\documentclass[twocolumn,aps,superscriptaddress,prx]{revtex4-2}
\setcitestyle{numbers,square}
\usepackage{amsmath,amssymb}
\usepackage{tabularx}
\usepackage{graphicx}
\usepackage{bmpsize}
\usepackage{bm}
\usepackage{color}
\usepackage{amssymb}
\usepackage[version=3]{mhchem}
\usepackage{newtxmath}
\usepackage{braket}
\usepackage[normalem]{ulem} 
\usepackage{float}
\usepackage{here}
\usepackage{comment}
\usepackage[T1]{fontenc}
\usepackage{booktabs}
\DeclareMathAlphabet{\mathpzc}{OT1}{pzc}{m}{it}

\newcommand{\blue}[1]{\textcolor{blue}{ #1}}
\newcommand{\mage}[1]{\textcolor{magenta}{ #1}}

\begin{document}
\title{One-Dimensional Electronic States in a Moir\'e Superlattice of Twisted Bilayer WTe$_2$}



\author{Takuto Kawakami}
\thanks{These authors contributed equally to this work.}
\affiliation{Department of Physics, Osaka University, Toyonaka, Osaka 560-0043, Japan}
\author{Hayato Tateishi}
\thanks{These authors contributed equally to this work.}
\affiliation{Department of Applied chemistry, Kyushu University, Motooka, Fukuoka 819-0395, Japan}
\author{Daiki Yoshida}
\affiliation{Department of Physics, Osaka University, Toyonaka, Osaka 560-0043, Japan}
\author{Xiaohan Yang}
\affiliation{Institute of Industrial Science, The University of Tokyo, Tokyo 153-8505, Japan}
\author{Naoto Nakatsuji}
\affiliation{Department of Physics and Astronomy, Stony Brook University, Stony Brook, New York 11794, USA}
\author{Limi Chen}
\affiliation{School of Materials Science, Japan Advanced Institute of Science and Technology, Ishikawa 923-1292, Japan}
\author{Kohei Aso}
\affiliation{School of Materials Science, Japan Advanced Institute of Science and Technology, Ishikawa 923-1292, Japan}
\author{Yukiko Yamada-Takamura}
\affiliation{School of Materials Science, Japan Advanced Institute of Science and Technology, Ishikawa 923-1292, Japan}
\author{Yoshifumi Oshima}
\affiliation{School of Materials Science, Japan Advanced Institute of Science and Technology, Ishikawa 923-1292, Japan}
\author{Yijin Zhang}
\affiliation{Institute of Industrial Science, The University of Tokyo, Tokyo 153-8505, Japan}
\affiliation{Department of Physics, The University of Tokyo, Tokyo 113-0022, Japan}
\author{Tomoki Machida}
\affiliation{Institute of Industrial Science, The University of Tokyo, Tokyo 153-8505, Japan}
\author{Koichiro Kato}
\thanks{Corresponding author: kato.koichiro.957@m.kyushu-u.ac.jp}
\affiliation{Department of Applied chemistry, Kyushu University, Motooka, Fukuoka 819-0395, Japan}
\author{Mikito Koshino}
\thanks{Corresponding author: koshino@phys.sci.osaka-u.ac.jp}
\affiliation{Department of Physics, Osaka University, Toyonaka, Osaka 560-0043, Japan}

\date{\today}

\begin{abstract}
One-dimensional (1D) moir\'e superlattices provide a route to engineering reduced-dimensional electronic states in van der Waals materials, yet their electronic structure and microscopic origin remain largely unexplored. 
Here, we investigate the structural relaxation and electronic properties of a 1D moir\'e superlattice formed in twisted bilayer 1T$'$-WTe$_2$ using density functional theory calculations, complemented by high-angle annular dark-field scanning transmission electron microscopy. 
We show that lattice relaxation strongly reconstructs the moir\'e stripes, leading to stacking-dependent stripe widths that are in excellent agreement with experimental observations. 
The relaxed structure hosts quasi-one-dimensional electronic bands near the Fermi level, characterized by strong dispersion along the stripe direction and nearly flat dispersion in the perpendicular direction. 
By comparing the full bilayer with isolated relaxed layers, we establish that these 1D electronic states are governed predominantly by an intralayer moir\'e potential induced by in-plane lattice relaxation, rather than by interlayer hybridization. 
We extract this position-dependent moir\'e potential directly from DFT calculations and construct an effective tight-binding model that reproduces both the band dispersion and the real-space localization of the electronic wave functions. 
Our results identify lattice relaxation as the key mechanism underlying 1D electronic states in 1D moir\'e superlattices.
The framework developed here provides a unified theoretical basis for realizing and exploring one-dimensional moir\'e physics in a broad class of anisotropic two-dimensional materials.
\end{abstract}

\maketitle
\section{introduction}

\begin{figure}
    \centering
    \includegraphics[width=85mm]{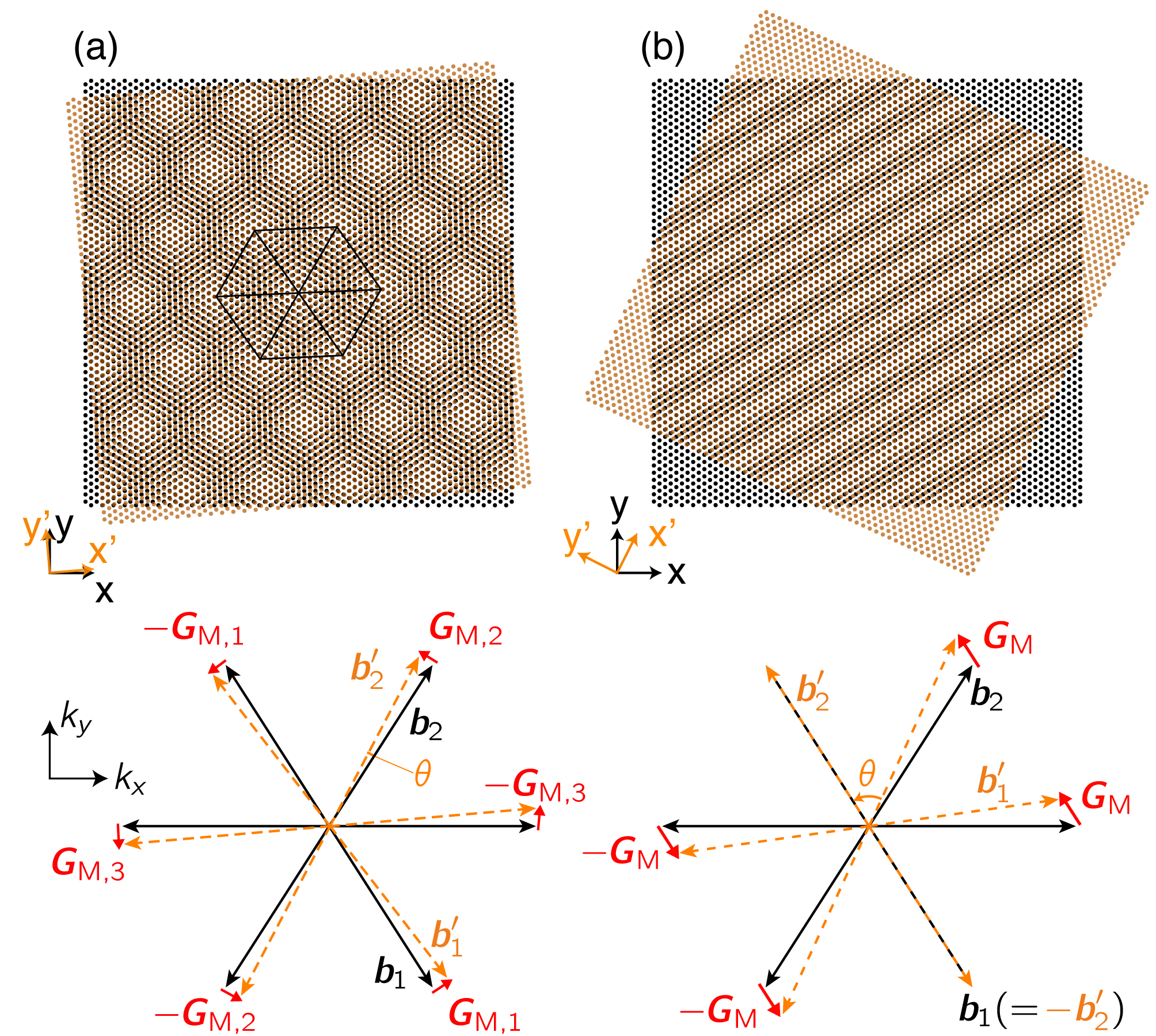}
    \caption{
Schematic illustration of a twisted bilayer composed of an anisotropic two-dimensional lattice stacked at (a) $\theta = 5^\circ$ and (b) $\theta \approx 65^\circ$.
The anisotropic monolayer lattice is constructed by uniaxially deforming an equilateral triangular lattice by a factor of 0.9 along the $x$ direction of the unrotated configuration.
While the low-angle stacking in (a) produces an anisotropic two-dimensional moir\'e pattern, the large-angle configuration in (b) gives rise to a purely one-dimensional moir\'e superlattice.
The lower panels show the corresponding momentum-space constructions; in the one-dimensional case, only a single moir\'e reciprocal lattice vector $\bm{G}_{\rm M}$ remains.
}   
    \label{fig:concept}
\end{figure}

Moir\'e superlattices in twisted two-dimensional (2D) materials have 
attracted considerable attention as a versatile platform for engineering electronic states~\cite{
andrei2020review, shah2024,gupta2025review,lopes2007graphene,morell2010,bistritzer2010,bistritzer2011moirepnas,moon2012,cclu2013,dean2013,cao2018_80,cao2018_43,shuang2018,koshino2018,yankowitz2019,wu2019,yu2019,carr2020,yasuda2021,xie2021,park2021,cai2023signatures,lu2024,xia2025,guo2025}. 
The long-wavelength interference between two slightly mismatched lattices can strongly reconstruct the electronic band structure and give rise to a variety of novel topological and strongly correlated phases.
In general, when two layers of the same 2D material are stacked with a small twist angle, 
the resulting moir\'e pattern can be regarded as a magnified and rotated image of the underlying atomic lattice. 
A representative example is twisted bilayer graphene, which hosts a two-dimensional moir\'e superlattice with sixfold rotational symmetry~\cite{lopes2007graphene,shallcross2010}. 
Likewise, uniaxially anisotropic monolayers with reduced rotational symmetry give rise to 
an anisotropic two-dimensional moir\'e pattern in their low-angle twisted bilayers~\cite{kennes2020, soltero2022, wang2022,fyuan2023,magorrian2024}, as illustrated in Fig.~\ref{fig:concept}(a).

In contrast, when anisotropic two-dimensional materials are stacked at specific large twist angles, 
the system can host a purely one-dimensional (1D) moir\'e superlattice, 
in which a given local stacking configuration extends infinitely along a single direction [see Fig.~\ref{fig:concept}(b)]~\cite{yang2025,an2025}. 
This situation arises when two inequivalent reciprocal lattice vectors of the monolayer are brought into exact alignment by twisting, 
as illustrated in the lower panel of Fig.~\ref{fig:concept}(b), leaving only a single moir\'e reciprocal lattice vector~\cite{yang2025}.
Such 1D moir\'e superlattices have been experimentally observed in twisted bilayer 1T$'$-WTe$_2$ at twist angles near $60^\circ$~\cite{yang2025} and in $\alpha$- and $\beta$-Sb~\cite{drozdz2024}, and have been theoretically predicted in twisted bilayers of black phosphorene~\cite{sousa2025} and PdSe$_2$~\cite{an2025}.
Moreover, the emergence of a 1D moir\'e pattern via reciprocal-lattice coincidence can also be achieved by applying heterostrain~\cite{sinner2023,escudero2024,hesp2024,boi2025,su2025}.

The resulting structural anisotropy provides a promising platform for realizing 1D electronic phenomena~\cite{sousa2025,an2025}. 
However, it remains an open question whether a 1D moir\'e pattern necessarily gives rise to intrinsically 1D electronic states. 
In this work, we perform density functional theory (DFT) calculations for the 1D moir\'e superlattice of twisted bilayer WTe$_2$~\cite{yang2025} and demonstrate the emergence of quasi-1D electronic states.
In moir\'e systems, lattice relaxation is known to have a profound impact on the electronic band structure~\cite{jonathan2013,uchida2014, sanjose2014,sanjose2014spontaneous,jung2015origin,wijk2015,dai2016twisted,PhysRevB.96.075311,jain2017structure,lucignano2019, yoo2019atomic,guinea2019, cantele2020, weston2020, rosenberger2020,sung2020,enaldiev2020,koshino2020,tsim2020,liang2020,zzhan2020, wu2021,li2021,vitale2021,leconte2022,engelke2023,van2023,mchugh2023moire,mchugh2023moire,xie2023,zhao2023,halbertal2023,zhang2024polarization,ceferino2024,ezzi2024,hagel2024,li2024, dang2025,yu2025,fu2025review}. 
Here, we carry out a full structural optimization of the one-dimensional moir\'e superlattice 
and show that moir\'e stripes with different local stacking configurations develop uneven widths depending on their energetic stability. 
We further experimentally probe the atomic structure of twisted 
bilayer WTe$_2$ at the corresponding twist angle using high-angle annular dark-field scanning transmission electron microscopy (HAADF-STEM), 
and confirm that the calculated atomic structure is in good agreement with the experimental images.

We then compute the electronic band structure of the optimized superlattice 
and demonstrate that the resulting bands exhibit a pronounced 1D character, 
with strong dispersion along the stripe direction and nearly flat dispersion in the perpendicular direction.
Notably, we find that the formation of 1D bands is driven by in-plane lattice relaxation. 
Specifically, structural relaxation induces a position-dependent shear strain that generates an on-site moir\'e potential that varies periodically in the direction perpendicular to the moir\'e stripes. 
The 1D bands emerge as bound states localized near the extrema of this potential.
We extract the moir\'e potential directly from DFT calculations of the locally distorted atomic structure and construct an effective tight-binding model incorporating this potential. 
The model successfully reproduces the essential features of the DFT band structure as well as the corresponding real-space wave functions.

We also introduce a general framework for constructing 1D moir\'e patterns, 
which is employed throughout this work. 
Specifically, we replace the original atomic structure of monolayer WTe$_2$ with an effective anisotropic triangular lattice. 
Within this description, the emergence of a 1D moir\'e pattern can be readily understood from the slight shape mismatch between the rotated triangular lattices of the two layers. 
This real-space picture is complementary to the reciprocal-space construction discussed above~\cite{yang2025}.
Although our analysis focuses on 1T$'$ transition-metal dichalcogenides (TMDCs), the method is broadly applicable to other systems exhibiting 1D moir\'e patterns. 
The general theoretical framework proposed here—combining one-dimensional moir\'e construction, lattice relaxation, and electronic structure analysis—is applicable to a broad class of anisotropic two-dimensional materials hosting one-dimensional moir\'e superlattices.

The remainder of this paper is organized as follows. 
In Sec.~\ref{sec:1dmoire}, we introduce a general framework for constructing one-dimensional moir\'e patterns and describe the geometry of the 1D moir\'e superlattice in twisted bilayer WTe$_2$. 
In Sec.~\ref{sec:relax}, we determine the relaxed atomic structure using DFT calculations 
and compare the results with experimental observations. 
Section~\ref{sec:electronic} discusses the resulting electronic band structure and presents an effective tight-binding model that captures its essential features. 
Finally, Sec.~\ref{sec:conclusion} summarizes our findings.

\begin{figure}[b]
    \centering
    \includegraphics[width=85mm]{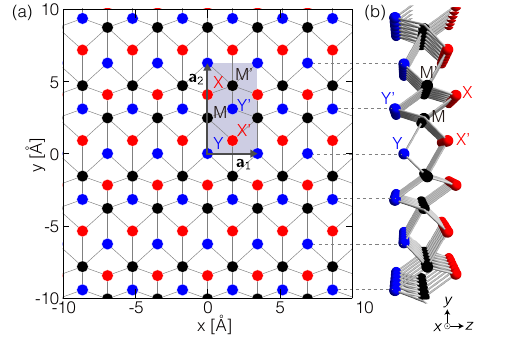}
    \caption{Atomic structure of a 1T$'$-WTe$_2$ monolayer. (a) Top view of the in-plane structure. (b) Side view in the $y$-$z$ plane. Black dots denote W atoms, while blue and red dots denote Te atoms. The shaded rectangle indicates the unit cell spanned by the primitive lattice vectors $\bm{a}_1$ and $\bm{a}_2$.}
    \label{fig:single}
\end{figure}

\begin{figure*}
    \centering
    \includegraphics[width=170mm]{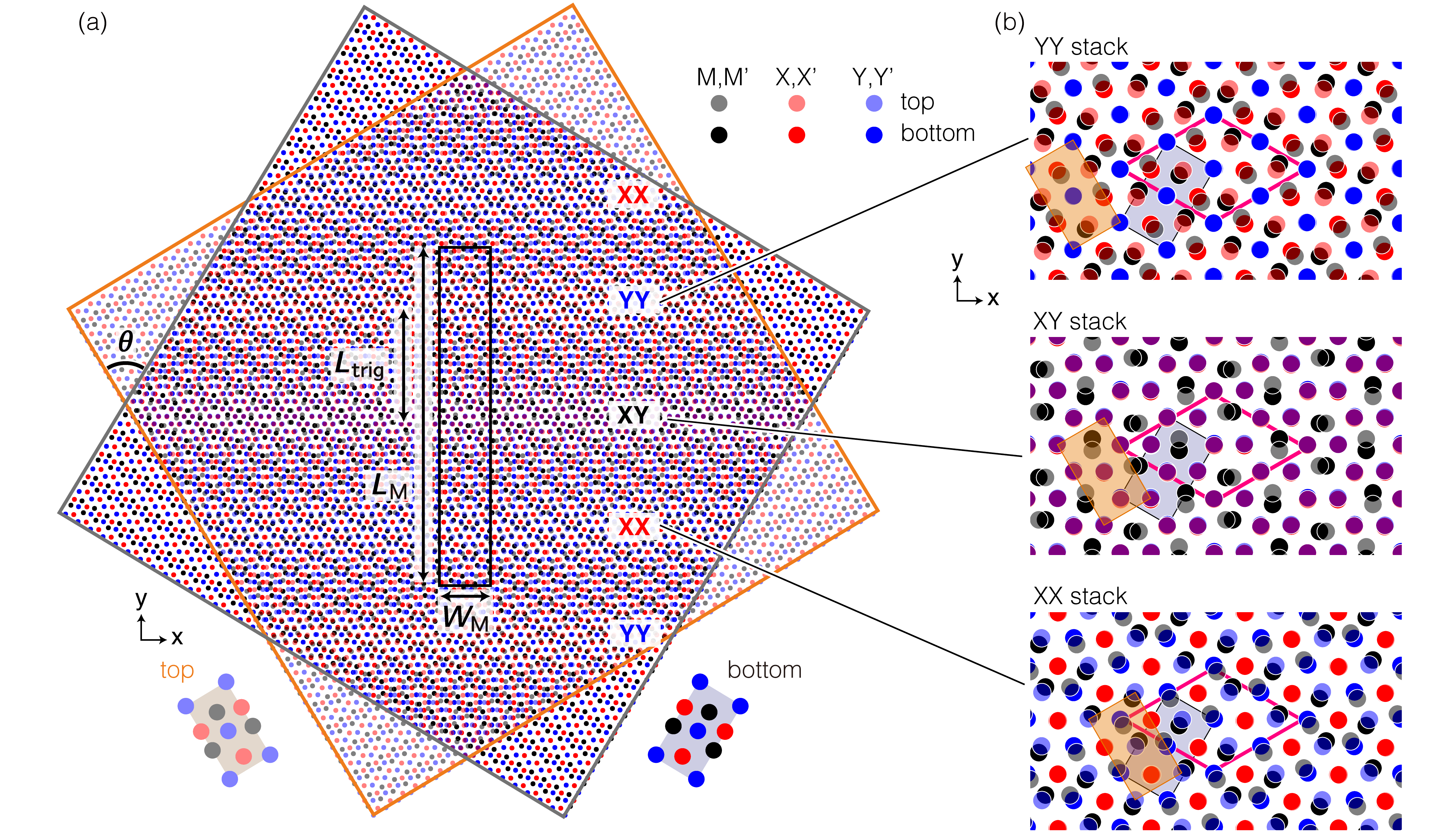}
    \caption{(a) One-dimensional moir\'e pattern formed by twisting two 1T$'$-WTe$_2$ monolayers. The top layer (transparent dots) is rotated by $\theta\approx62^\circ$ relative to the bottom layer (solid dots).
    $L_\mathrm{M}$ and $W_\mathrm{M}$ are the length and width of the moire unit cell of WTe$_2$, while $L_{\mathrm{trig}}$ denotes the moire wavelength of effective triangular lattice (see Sec.~\ref{sec:averaged} and Fig.~\ref{fig:triangle}).
    (b) Periodic stacking configuration corresponding to the local structure in (a), with the period indicated by the rhombus. Periodic interlayer sliding along the $x$ direction gives rise to the repeating XX, XY, and YY stacking configurations.}
    \label{fig:system}
\end{figure*}

\section{1D moir\'e pattern}\label{sec:1dmoire}




\subsection{Atomic structure}

We first describe the atomic structure of a single WTe$_2$ layer~\cite{qian2014,torun2016,muechler2016,fzheng2016,stang2017,fatemi2018,syxu2018,jyang2018,qzhang2019,shi2019,ok2019,czhao2020,wzhao2020,mhu2021,wzhao2021,bsun2022,maximenko2022,jlee2023,lwatson2025}, 
which serves as the basis for constructing the moir\'e pattern in the twisted bilayer system.
Figure~\ref{fig:single} shows its crystal structure. 
Black dots represent W atoms, while red and blue dots represent Te atoms
located above and below the W plane, respectively.
In contrast to the trigonal prismatic coordination in hexagonal TMDCs such as MoS$_2$, WTe$_2$ is characterized by a distorted octahedral coordination, referred to as 1T$'$ structure.
The in-plane structure, shown in Fig.~\ref{fig:single}(a),
has a rectangular unit cell containing six atoms:
two W atoms (labeled M and M$'$), two Te atoms above the W plane (X and X$'$),
and two Te atoms below the W plane (Y and Y$'$).
The in-plane lattice constants are $|\bm{a}_1|\approx 3.46$ \AA\ and $|\bm{a}_2|\approx6.26$ \AA, 
with directions $\bm{a}_1\parallel \hat{\bm{x}}$ and $\bm{a}_2\parallel \hat{\bm{y}}$.
This structure belongs to the layer group $P2_1/m11$,
generated by mirror reflection about the $yz$-plane and
screw operation along the $x$ direction.


We now consider a twisted bilayer 1T$'$-WTe$_2$,
where the top and bottom layers are rotated by 
$\pm \theta/2$. 
Figure \ref{fig:system}(a) illustrate the
atomic configuration for a twist angle of $\theta\sim62^\circ$.
We observe a one-dimensional moir\'e pattern, with the horizontal bright stripes indicating 1D moir\'e domains~\cite{yang2025}.
The local regions can be approximated as periodic stacking configurations, denoted as XX, XY, and YY, as shown in Fig.~\ref{fig:system}(b); their precise definitions are provided in the following section.
This classification is analogous to the AA, AB, and BA stackings commonly used for twisted bilayer graphene.
In the XY stacking, upper Te sites (X or X$'$) in one layer are vertically aligned with lower Te sites (Y or Y$'$) in the other layer.
The XX and YY stackings are defined analogously.

\begin{figure*}
    \centering
    \includegraphics[width=170mm]{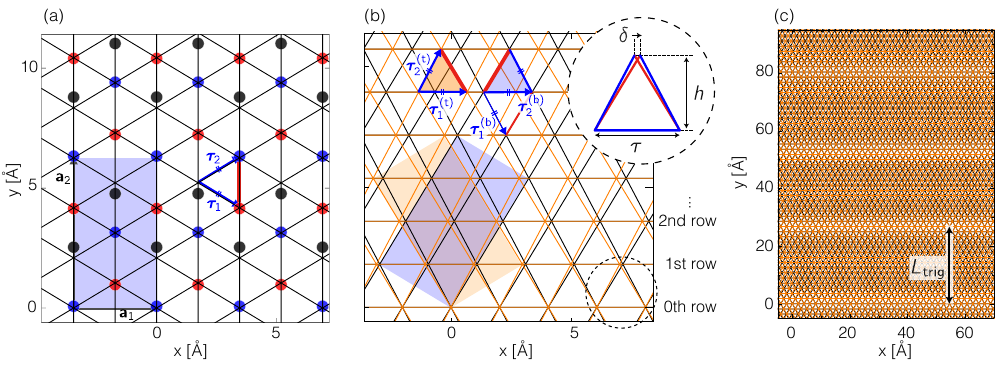}
    \caption{(a) Effective isoscales triangular lattice approximating the atomic structure of WTe$_2$. (b) Twisted stacking of the effective triangular lattices. Black and orange denotes the bottom and top layers, respectively. The two layers are rotated such that their lattice vectors parallel to the $x$ axis have the equal length but are not identical. The inset shows a magnified view of a single isoscales plaquette with an aligned baseline. (c) Zoom-out view of (b), showing a clear moir\'e period $L_\mathrm{trig}$.
    Atomic color codes are the same as in Fig.~\ref{fig:single}}
    \label{fig:triangle}
\end{figure*}
\subsection{Origin of the 1D moir\'e pattern}
\label{sec:averaged}

The formation of the 1D moir\'e pattern in  WTe$_2$ can be understood using
the concept of an effective triangular lattice,
introduced from a momentum-space perspective~\cite{yang2025}. 
Here we develop a complementary real-space description 
that provides an intuitive understanding of the
1D moir\'e structure in twisted bilayer 1T$'$-WTe$_2$.

The effective triangular lattice, shown by the solid lines in Fig.~\ref{fig:triangle}(a), is defined as a triangular grid commensurate with the monolayer unit cell of 1T$'$-WTe$_2$, where each vertex corresponds approximately to an atomic site of 1T$'$-WTe$_2$. 
The primitive lattice vectors of this effective lattice are given by
\begin{align}
   & \bm{\tau}_{1} =  \bm{a}_1/2 - \bm{a}_2/6\\
   & \bm{\tau}_{2} =  \bm{a}_1/2 + \bm{a}_2/6.
\end{align}
For each triangle plaquette, the mirror-reflection symmetry of the original monolayer enforces equal edge lengths, $|\boldsymbol{\tau}_{1}| = |\boldsymbol{\tau}_{2}| = \tau$,
while the deviation from an equilateral triangle
($|\boldsymbol{\tau}_{1} - \boldsymbol{\tau}_{2}| \neq \tau$) reflects the absence of threefold rotational symmetry in the 1T$'$ structure.
To match the high-symmety points of the effective lattice to those of the atomic structure, we fix the relative two-dimensional position of effective lattice such that
the inversion center of the atomic structure (the midpoint between the X and Y$'$ sites) coincides with the midpoint of an edge of the triangular lattice.

We consider a twisted bilayer of this triangular lattice, obtained by rotating the top and bottom layers by $\pm \theta/2$, respectively. The rotated lattice vectors are denoted as $\boldsymbol{\tau}_{i}^{(l)}$ with $l = \mathrm{t}, \mathrm{b}$ for the top and bottom layers. When $\theta$ is chosen as the relative angle between $\boldsymbol{\tau}_{1}$ and $\boldsymbol{\tau}_{2}$, the configuration shown in Fig.~\ref{fig:triangle}(b) is realized, in which $\boldsymbol{\tau}_{1}^{(\mathrm{t})}$ and $\boldsymbol{\tau}_{2}^{(\mathrm{b})}$ coincide along the $x$ direction.
In this arrangement, a one-dimensional moir\'e pattern naturally emerges.
Let $y=0$ denote the baseline, where the horizontal edges of the top and bottom triangular lattices coincide, defining a reference row [Fig.~\ref{fig:triangle}(b)].
In the first row above, the vertices are laterally shifted by a small amount $\delta$ along $x$, 
due to the non-equilateral geometry of the triangles, as shown in the inset of Fig.~\ref{fig:triangle}(b).
As the structure repeats row by row,
the displacement accumulates as $N\delta$ in the $N$-th row.
When the accumulated shift satisfies
    $N\delta \simeq \tau$, 
the vertices realign,
giving rise to a 1D moir\'e period [Fig.~\ref{fig:triangle}(c)].
The corresponding moir\'e wavelength is then 
\begin{equation}
L_{\mathrm{trig}}= Nh \simeq \frac{\tau h}{\delta},
\end{equation}
where $h$ is the height of the triangles in $y$-direction [see Fig.~\ref{fig:triangle}(b), inset].

The bright stripes observed in twisted bilayer of 1T$'$-WTe$_2$ 
[Fig.~\ref{fig:system}(a)] corresponds to the 1D moir\'e pattern in the effective triangular lattice.
However, these stripes do not share the same atomic arrangement; instead, the three distinct stacking patterns, XX, XY, and YY appear periodically, as discussed above.
Therefore, the true moir\'e period is given by
\begin{equation}
    L_{\mathrm{M}} = 3 L_{\mathrm{trig}}.
\end{equation}
This structural variation arises because the vertices of the effective triangular lattice 
correspond to different atomic species in the underlying WTe$_2$ lattice.
The periodic stacking structures for XX, XY, and YY, shown in Fig.~\ref{fig:system}(b),
are defined by linearly transforming the WTe$_2$ lattice so that its effective lattice becomes an equilateral triangular lattice, and then overlapping a pair of such layers with a twist angle of $60^\circ$.  
The resulting structure possesses a translational symmetry, with the unit cell indicated by the pink rhombus in Fig.~\ref{fig:system}(b), and the different stacking types (XX, XY, YY) correspond to different lateral shifts between the two layers.
In the original moir\'e twisted bilayer shown in Fig.~\ref{fig:system}(a),
a movement of length $L_{\rm trig}$ along the $y$-direction corresponds to a horizontal interlayer shift by one triangle side $(\tau)$ in the commensurate bilayer.  
It can be shown that the stacking sequence of the commensurate bilayer
changes periodically as XX, XY, YY, $\cdots$ for each shift by $\tau$,
which explains the evolution of the local stacking structures in the moir\'e bilayer
along the $y$-direction.

\subsection{Commensurate approximant}\label{sec:commensurate}
\begin{figure}
    \centering
    \includegraphics[width=85mm]{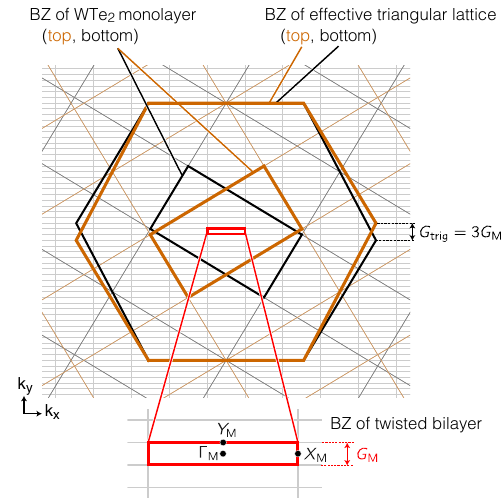}
    \caption{Momentum-space structure of twisted bilayer WTe$_2$ for a commensurate approximant. Three levels of Brilloiuin zones are indicated, corresponding to the effective triangular lattice, the WTe$_2$ monolayers, and the commensurate bilayer supercell.}
    \label{fig:bz}
\end{figure}

Although the long-range moir\'e pattern can be described as above,
the overall structure is not strictly periodic, due to the moir\'e period $L_M$ is not generally commensurate with the underlying WTe$_2$ lattice.
To obtain a commensurate structure,
we slightly deform the effective triangular lattice so that
its vertices realign exactly at $N= \lfloor \tau/\delta \rceil$ row,
where $\lfloor x \rceil$ denote the nearest integer to $x$.
To preserve the lattice symmetry,
we apply a uniaxial isochoric strain to the unrotated effective lattice, transforming $\bm{\tau}_i$ as
\begin{equation}\label{eq:epsilon}
\tilde{\bm{\tau}}_{i}=\left(\begin{array}{cc}
1+\varepsilon & 0 \\
0 & (1+\varepsilon)^{-1}
\end{array}\right)\bm{\tau}_{i},
\end{equation}
where $\varepsilon$ is determined to satisfy the condition $\tilde{\tau}/\tilde{\delta}=\lfloor \tau/\delta \rceil=N$.
Here, a tilde denotes the quantity in the commensurately adjusted system.
From geometric consideration (see Appendix~\ref{sec:derivepsilon}), the required strain magnitude is
\begin{equation}\label{eq:requiredepsilon}
    1+\varepsilon = \sqrt{\frac{|\bm{\tau}_1-\bm{\tau}_2|}{|\bm{\tau}_1+\bm{\tau}_2|}}
    \left(\frac{3N-1}{N+1}\right)^{1/4}.
\end{equation}
For WTe$_2$, we obtain $N=15$, and $\varepsilon\approx 3.8\times 10^{-5}$, using $\tau\approx2.02$ \AA\  and $\delta\approx 0.135$ \AA.
This deformation is sufficiently small that it does not significantly affect the
electronic structure.
Single-layer DFT calculations show that the resulting band shifts are approximately 1 meV.
The resulting moir\'e period is
$L_{\text{trig}}= N \tilde{h} \approx 26.8$ \AA,
and $L_{\text{M}}=3L_{\text{trig}}\approx 80.4$ \AA.

Using the adjusted lattice,
the moir\'e superlattice vectors are 
defined to satisfy
\begin{align}
    \bm{a}_{\mathrm{M},i} &= R(\theta/2)
    (m_{i1} \tilde{\bm{a}}_{1} + m_{i2} \tilde{\bm{a}}_{2}) \nonumber\\
    &= R(-\theta/2)
    (n_{i1} \tilde{\bm{a}}_{1} + n_{i2} \tilde{\bm{a}}_{2}).
\end{align}
The smallest integer solutions for the linearly independent $\bm{a}_{\mathrm{M},i}$ are
$(m_{11},m_{12},n_{11},n_{12})=(12,11,-12,11)$ and
$(m_{21},m_{22},n_{21},n_{22})=(-3,1,-3,-1)$, which yield the moir\'e lattice constants
\begin{gather}
    \bm{a}_\mathrm{M,1} = L_{\mathrm{M}} \hat{\bm{e}}_y \\
    \bm{a}_\mathrm{M,2} =-W_{\mathrm{M}} \hat{\bm{e}}_x
\end{gather}
with $L_{\mathrm{M}}=3L_{\mathrm{trig}}$ and $W_{\mathrm{M}}=6\tilde{\tau}$.
The unit cell spanned by $\bm{a}_\mathrm{M,1}$ and $\bm{a}_\mathrm{M,2}$ is shown a rectangle in Fig.~\ref{fig:system}(a).
We adopt this commensurate lattice to simulate the lattice relaxation and
the electronic properties in Sec.~\ref{sec:relax} and \ref{sec:electronic}.

Figure~\ref{fig:bz} illustrates the construction of the Brillouin zone (BZ) of the twisted bilayer system considered here.
The brown and black rectangles denote the BZs of the top and bottom WTe$_2$ monolayers, respectively, while the hexagons represent the BZs of the effective triangular lattice. 
The smallest red rectangle corresponds to the BZ of the moir\'e unit cell of the commensurate twisted bilayer system obtained above, with its extent along the $y$ direction given by $G_{\rm M} = 2\pi / L_{\rm M}$. 
The distance between the closest corners of the hexagonal Brillouin zones of the top and bottom layers defines $G_{\rm trig} = 2\pi / L_{\rm trig}$, which is the reciprocal lattice vector associated with the moir\'e pattern of the effective triangular lattice (Fig.~\ref{fig:triangle}). Here $G_{\rm trig} = 3 G_{\rm M}$, reflecting the relation $L_{\mathrm{M}} = 3 L_{\mathrm{trig}}$.

\section{Structural relaxation}\label{sec:relax}

\begin{figure*}
    \centering
    \includegraphics[width=170mm]{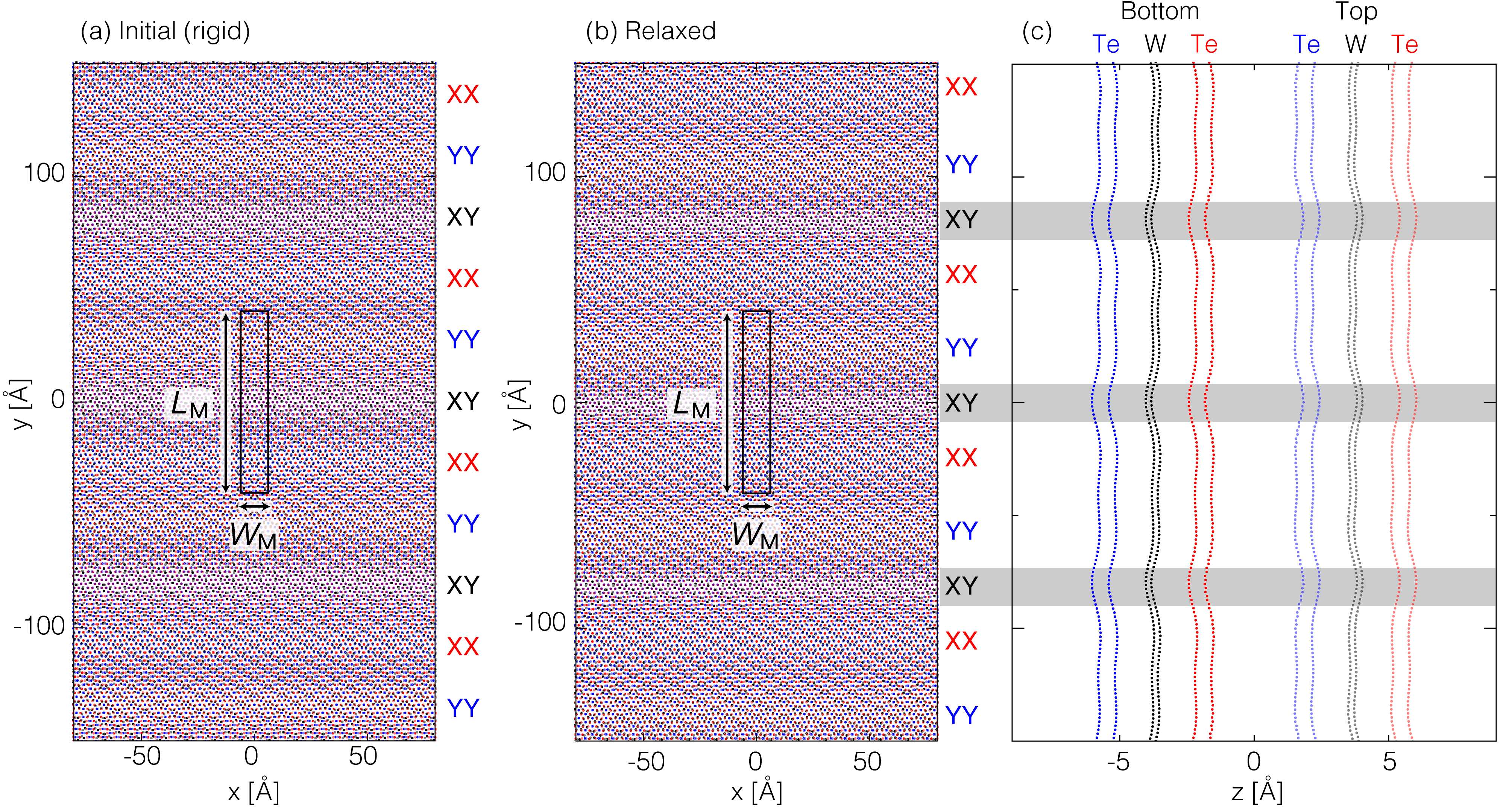}
    \caption{DFT structural relaxation of twisted bilayer WTe$_2$. 
    (a) Atomic structure of the rigid commensurate approximant before relaxation. The rectangle at the center denotes the unit cell. 
    Multiple neighboring unit cells are shown for clarity. 
    (b) Top view of the in-plane relaxed structure.
    The XY stacking stripe shrinks, while XX and YY stacking stripes broaden.
    (c) Side view of the relaxed structure with an enlarged $z$-axis scale. 
    The interlayer distance increases 
    in the XY stacking region. Atomic color codes are the same as  those used in Figs.~\ref{fig:single} and \ref{fig:system}.
    }
    \label{fig:relax}
\end{figure*}

In this section, we investigate the lattice relaxation of twisted bilayer WTe$_2$. Starting from the commensurate approximant of the rigid twisted bilayer described in Sec.~\ref{sec:commensurate}, we perform a full DFT structural optimization. We compare the calculated structures with HAADF-STEM measurements and demonstrate good agreement between theory and experiment.



\subsection{DFT structural analysis} 

We performed density functional theory (DFT) calculations by mainly employing two computational packages: Vienna Ab initio Simulation Package (VASP)~\cite{KRESSE199615, PhysRevB.54.11169} for structure optimization, OpenMX~\cite{PhysRevB.67.155108, PhysRevB.69.195113, PhysRevB.72.045121} for band calculation. The following calculations were conducted with Perdew-Burke-Ernzerhof (PBE)  generalized approximation (GGA)~\cite{PhysRevLett.77.3865} and DFT-D3 method of Grimme with zero-damping function~\cite{Grimme2010}. The projector augmented-wave (PAW) method~\cite{PhysRevB.50.17953} was employed for the pseudopotentials in VASP calculations. Also, norm-conserving Vanderbilt pseudopotential~\cite{PhysRevB.47.6728} and variationally optimizing numerical atomic orbitals~\cite{PhysRevB.72.045121} were used in OpenMX calculations.


We first optimized the primitive unit cell of monolayer 1T$'$-WTe$_2$ to determine its in-plane lattice constants ($|\bm{a}_1|$ and $|\bm{a}_2|$). The calculations employed a plane-wave cutoff energy of 300 eV and a $4\times 2\times 1$ Monkhorst–Pack $k$-point mesh for Brillouin-zone sampling~\cite{PhysRevB.13.5188}. The out-of-plane lattice constant was fixed at $c=20$ Å to suppress spurious interactions between periodic images.
After full relaxation of the atomic positions, cell shape, and volume, we obtained optimized lattice constants of approximately $|\bm{a}_1| = 3.46$ Å and $|\bm{a}_2| = 6.26$ Å.

We then constructed a commensurate 1D moir\'e system by stacking two structure-optimized monolayers at $\theta \approx 62^\circ$, following the commensuration procedure described in Sec.~\ref{sec:commensurate}. The resulting configuration, shown in Fig.~\ref{fig:relax}(a), is referred to as the rigid moir\'e structure. Starting from this structure, we performed a full structural optimization of the entire moir\'e supercell.
In these calculations, the out-of-plane lattice constant was fixed at $c = 25$ Å, the plane-wave cutoff energy was set to 300 eV, and only the $\Gamma$ point ($1 \times 1 \times 1$ $k$-point mesh) was used for Brillouin-zone sampling. The resulting total energy and geometric parameters are summarized in Appendix~\ref{sec:parameter}.


The optimized structure of the 1D moir\'e pattern, shown in Fig.~\ref{fig:relax}(b), exhibits uneven stripe widths, with the XX and YY regions becoming significantly wider than the XY region. 
In general, moir\'e structures tend to suppress energetically unfavorable stacking configurations while expanding more stable ones, much like in graphene where the AA-stacked regions shrink and the AB/BA regions are enlarged~\cite{jonathan2013,uchida2014,PhysRevB.96.075311}. 
In the present system, XY stacking is energetically unfavorable because the frontier Te atoms of the two layers approach each other too closely, resulting in excessive overlap of their filled out-of-plane $p_z$ orbitals. 
This behavior is consistent with the well-known instability of chalcogen–chalcogen stacking in bilayer TMDC systems~\cite{
he2014,weston2020,zhai2024review}.

In contrast, the XX and YY stackings adopt an interlocking arrangement in which a convex Te atom from one layer fits into a concave region of the opposing layer. 
Consequently, the system spontaneously reduces the XY-stacked domains while expanding the XX and YY regions through lateral atomic displacements. 
Figure~\ref{fig:relax}(c) shows a side view of the atomic structure in the $yz$ plane.
We find that the interlayer Te-Te distance varies periodically along the moir\'e pattern, ranging from 3.04 to 3.68~\AA. 
This variation is most pronounced in the XY-stacked regions, where the increased spacing reflects the near-contact configuration of the frontier Te atoms.  
A detailed analysis of the local stacking configurations is provided in Appendix~\ref{sec:localstacking}. 

\begin{figure}[b]
    \centering
    \includegraphics[width=85mm]{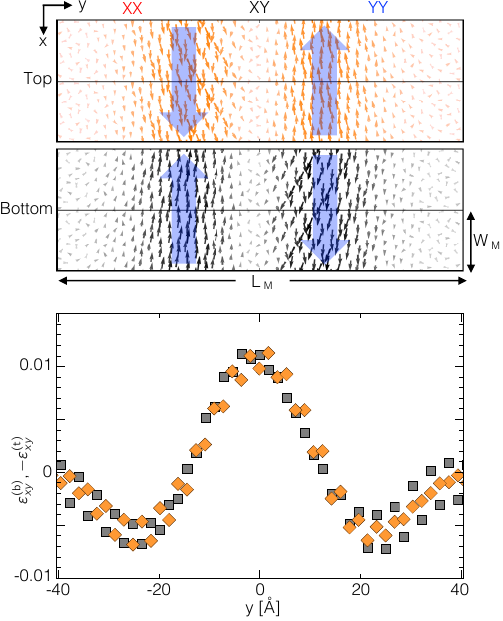}
    \caption{
(Top) Atomic displacements induced by lattice relaxation in top and bottom layers.  The arrow lengths are scaled to be 14 times the actual displacement amplitudes. 
Blue arrows schematically indicate the dominant displacement directions in each region.
(Bottom) Shear strain components, $\epsilon^{(b)}_{xy}, -\epsilon^{(t)}_{xy}$, plotted as a function of the unit-cell center position along the $y$ direction, perpendicular to the moir\'e stripes. Orange and black symbols denote the top and bottom layers, respectively.
}
    \label{fig:shear}
\end{figure}

We also investigate the local lattice distortions associated with structural relaxation. For each monolayer unit cell in the relaxed twist bilayer [illustrated in Fig.~\ref{fig:potential}(a) in a later section], we extract the local lattice vectors $\bm{a}_{i,\mathrm{relax}}^{(l)}$ ($i=1,2$) for layer $l={\rm t},{\rm b}$ (top and bottom), and characterize the local lattice distortion following the standard strain-rotation decomposition~\cite{chaikinbooksec6}.
\begin{equation} 
\bm{a}_{i, \mathrm{relax}}^{(l)}
=
\left(\begin{array}{cc} 1+\epsilon_{xx}^{(l)} & \epsilon_{xy}^{(l)}-\Omega^{(l)} \\ \epsilon_{xy}^{(l)}+\Omega^{(l)} & 1+\epsilon_{yy}^{(l)} \end{array}\right)\bm{a}_{i}^{(l)} 
\end{equation}
Here, $\bm{a}_{i}^{(l)} = R(\pm\theta/2)\bm{a}_{i}$ ($\pm$ for $l={\rm t},{\rm b}$) denotes the intrinsic lattice vectors of monolayer WTe$_2$ without the relaxation,
$\epsilon_{xx}^{(l)}$ and $\epsilon_{yy}^{(l)}$ represent the uniaxial strain components, $\epsilon_{xy}^{(l)}$ is the shear strain, and $\Omega^{(l)}$ corresponds to the local rotation.
Note that this decomposition is strictly valid to leading order in the rotation angle for $\Omega^{(l)}$.
Figure~\ref{fig:shear} shows the spatial distribution of the lattice distortions. The upper panels display vector maps of the atomic displacements in each layer, while the lower panels present the corresponding profiles of the shear strain $\epsilon_{xy}^{(l)}$ as functions of the unit-cell center position along the $y$ direction. 

In the vector maps, we observe relative interlayer sliding along the $x$ axis (vertical direction in Fig.~\ref{fig:shear}) on either side of the XY stacking region, with opposite sliding directions. This sliding motion is closely related to the narrowing of the XY stripe upon relaxation.
Specifically, interlayer sliding along the $x$ direction induces a shift of the moir\'e pattern along the $y$ direction, as can be understood from the effective triangular-lattice picture shown in Fig.~\ref{fig:triangle}(b). The opposite sliding directions on the left- and right-hand sides of the XY region lead to moir\'e pattern shifts toward $\pm y$, resulting in a reduction of the XY domain width. Correspondingly, $\epsilon_{xy}^{(l)}$ exhibits pronounced peaks in the XY stacking region, originating from opposite atomic displacements along the $x$ direction on either side of the region.

\subsection{Experimental validation} 

\begin{figure*}
    \centering
    \includegraphics[width=170mm]{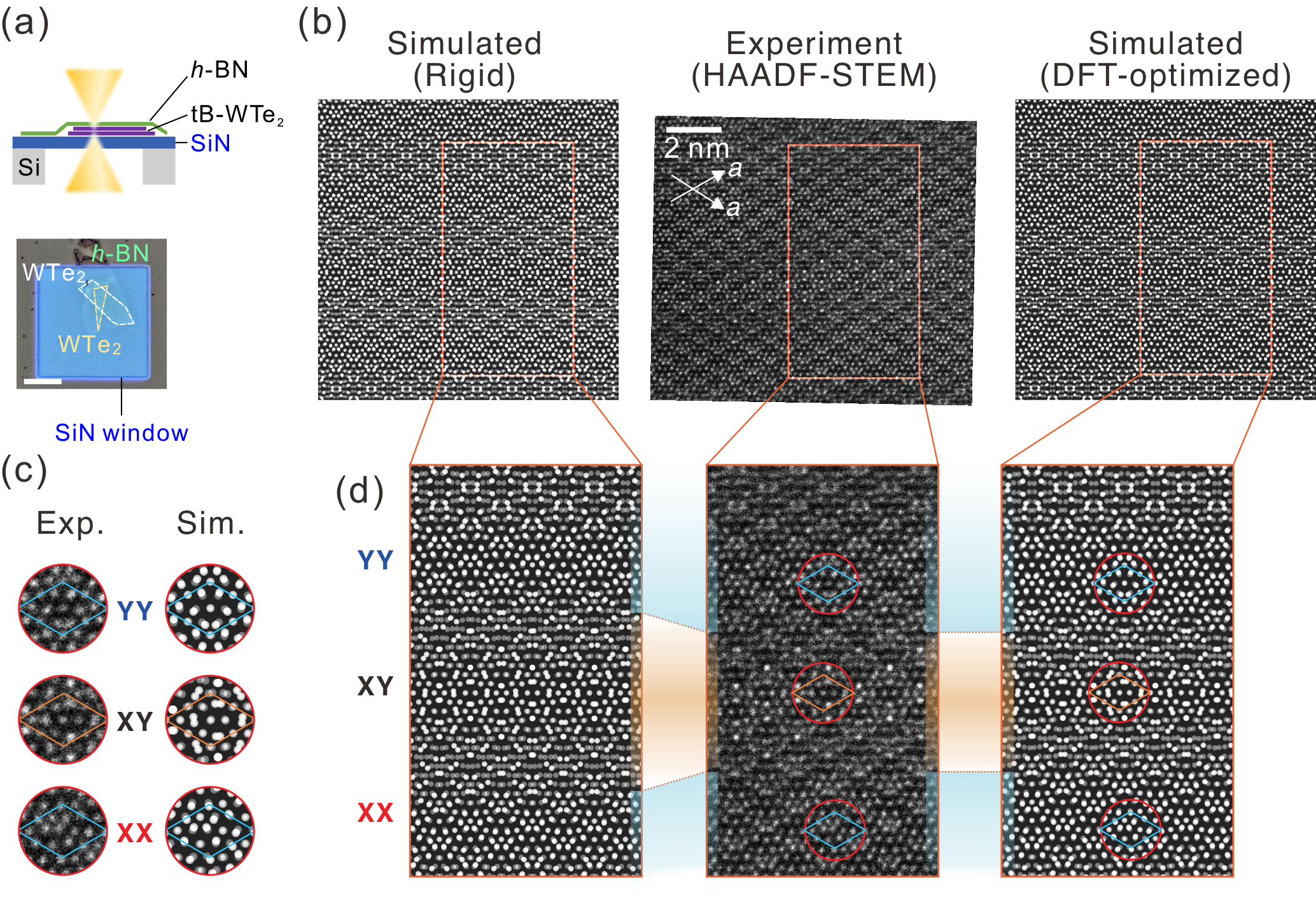}
    \caption{
    (a) Schematic and optical micrograph of the sample. The white scale bar: 10 $\mu$m. (b) Comparison of experimental HAADF-STEM image of 62.51$^\circ$ twisted bilayer 1T'-WTe$_2$ with simulated STEM images using (left) the rigid structure and (right) the DFT-optimized structure. (c) Atomic arrangements of three motifs. (d) Comparison of simulated STEM images with the experimental STEM image.}
    \label{fig:excomp}
\end{figure*}

We compare the DFT-optimized structure with the experimentally observed atomic arrangement of twisted bilayer 1T$'$-WTe$_2$, obtained using high-angle annular dark-field scanning transmission electron microscopy (HAADF-STEM)~\cite{yang2025}. The van der Waals structure was made from mechanically exfoliated monolayer 1T$'$-WTe$_2$ by the dry transfer method. The twisted bilayer 1T$'$-WTe$_2$ was made from a single monolayer 1T$'$-WTe$_2$ using the tear-and-stack method which provides good control of twist angle. As WTe$_2$ is highly air-sensitive, mechanical exfoliation and van der Waals assembly were carried out in the glove box filled with dry nitrogen. The twisted bilayer were covered by ultrahin {\it h}-BN as a protective layer before taking out from the glove box [Fig.~\ref{fig:excomp}(a)]. 

Figure~\ref{fig:excomp}(b) compares a HAADF-STEM image of a $\sim62^\circ$ twisted-bilayer 1T$'$-WTe$_2$ with two simulated images of the rigid and DFT-optimized structures. The latter two images were obtained simply by rendering the structures in Figs.~\ref{fig:relax}(a) and (b) in grayscale, respectively. Experimentally, we observed 1D moir\'e stripes running horizontally in the panel. Perpendicular to these stripes, the contrast alternates between relatively bright and dark stripes. The moir\'e period in our sample was ~8.2 nm and includes three bright and three dark stripes per period (orange rectangle). The three bright stripes from top to bottom correspond to YY, XY and XX regions.

As reported previously~\cite{yang2025}, the characteristic atomic configurations within each stripe are already well reproduced by the rigid model. This agreement is not altered for the DFT-optimized structure Fig.~\ref{fig:excomp}(c). Furthermore, we find that the widths of the stripes, which were not discussed in earlier work, agree more closely with the simulation based on the DFT-optimized structure. This comparison is highlighted in the zoomed images in Fig.~\ref{fig:excomp}(d), which show a single moir\'e period for the rigid structure, the STEM image, and the DFT-optimized structure extracted from Figs.~\ref{fig:excomp}(b), respectively. The DFT relaxation reduces the extent of the XY region (orange shading) while expanding the XX and YY regions (blue shading). These trends closely match the experimentally observed atomic arrangement of twisted bilayer 1T$'$-WTe$_2$, supporting the validity of our DFT-based approach for modeling lattice relaxation in 1D moir\'e systems.

\begin{figure}
    \centering
    \includegraphics[width=85mm]{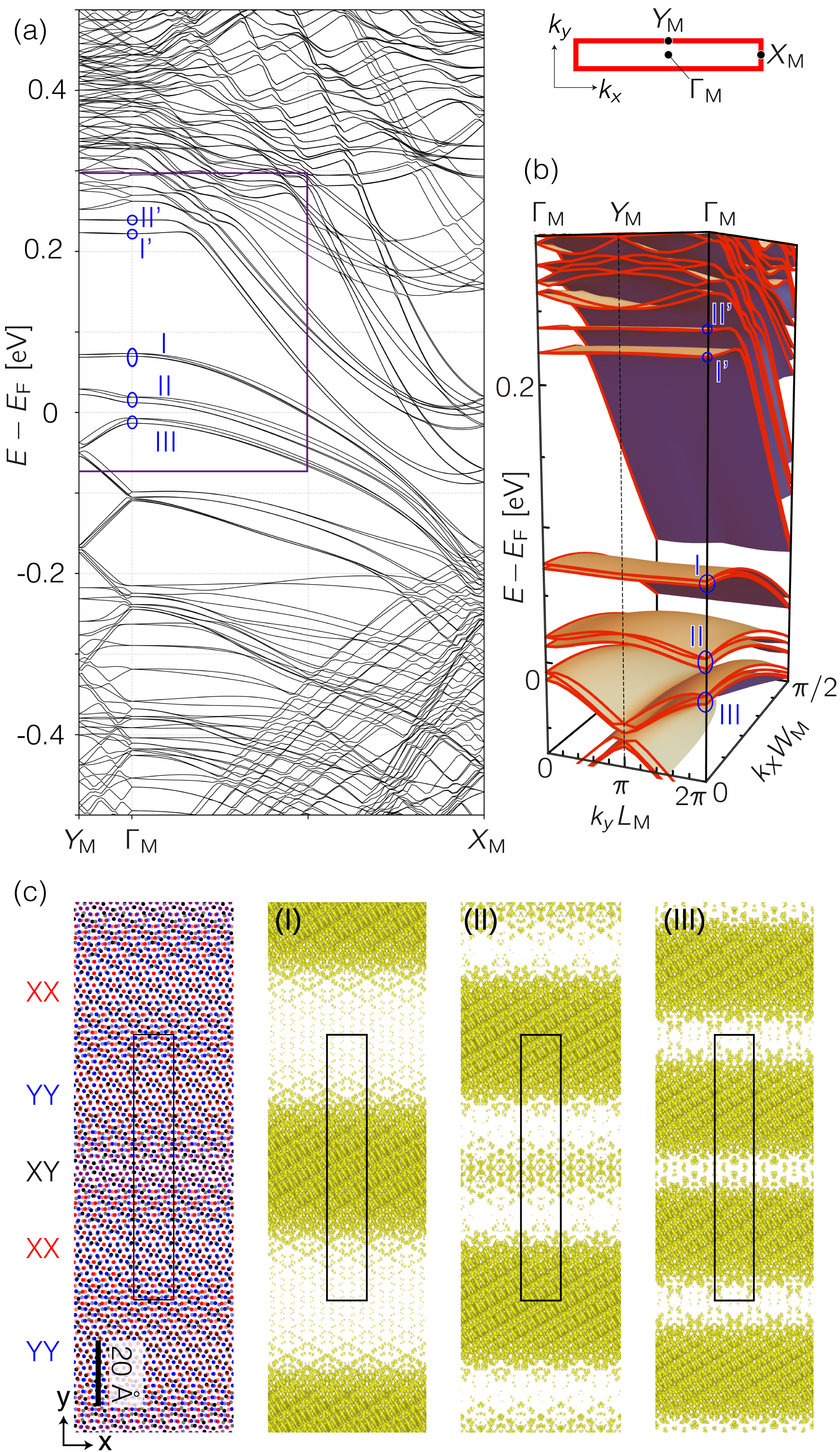}
    \caption{DFT electronic states. 
    (a) Electronic band structure of the moir\'e unit cell along the high symmetry paths in the moire Brillouin zone (see also Fig.~\ref{fig:bz}). (b) Three-dimensional band dispersion in the region enclosed by the purple line in (a). (c) From left to right: optimized 1D moir\'e structure, and real-space probability densities of the electronic states in group I, II, and III as indicated in the panels (a) and (b).}
    \label{fig:dft}
\end{figure}

\begin{figure}
    \centering
    \includegraphics[width=85mm]{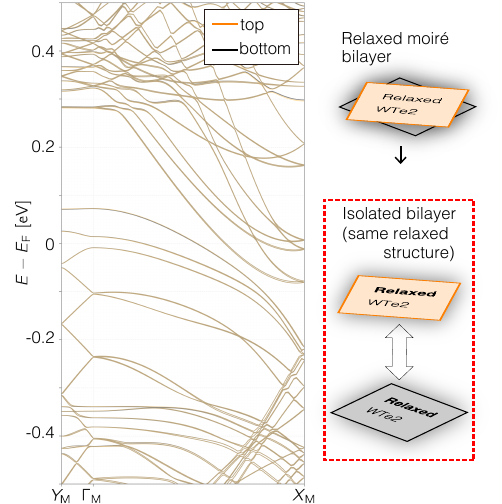}
    \caption{Electronic bands calculated for isolated layers extracted from the structurally optimized moir\'e bilayer. The orange and black curves denote the top and bottom layers, respectively. The bands of the top and bottom layers are superimposed throughout the entire energy and momentum range. The right panel shows a schematic illustration of isolated individual 1T$'$-WTe$_2$ layers under moir\'e-induced lattice relaxation.}
    \label{fig:dftlayer}
\end{figure}

\section{Electronic band structure}\label{sec:electronic}

We analyze the electronic properties of the relaxed twisted bilayer WTe$_2$ obtained in Sec.~\ref{sec:relax} using DFT calculations. 
As shown below, the system exhibits a pronounced 1D electronic character: the low-energy bands are strongly dispersive along the moir\'e stripes while remaining nearly flat in the perpendicular direction. 
We further find that this anisotropy arises primarily from lattice relaxation, 
which generates an in-plane potential modulation. 
The resulting low-energy states can be interpreted as confined modes within the moir\'e-induced potential landscape.

\subsection{DFT band structure}\label{sec:electronic:dft}
We performed band calculations for the optimized commensurate 1D moir\'e unit cell.
Charge density cutoff was set to 150 Ry, and Te7.0-s3p2d2f1 and W7.0-s3p2d2f1 were employed as pseudoatomic orbitals.
Spin-orbit coupling was included self-consistently~\cite{PhysRevB.93.224425}, and the Brillouin zone was sampled using a 1×4×1 grid of {$k$} points mesh.

The resulting band structure along the high-symmetry directions of the moir\'e Brillouin zone [Fig.~\ref{fig:bz}] is presented in Fig.~\ref{fig:dft}(a), where the zero of the energy axis is set at the Fermi energy.
Figure~\ref{fig:dft}(b) shows a three-dimensional plot of the band-energy surface for the region highlighted by the rectangle in Fig.~\ref{fig:dft}(a).
The bands near the valence-band maximum at the $\Gamma_\mathrm{M}$ point, labeled as groups I, II, and III, appear in fourfold groups. This reflects the two spin and two layer degrees of freedom, while the small splittings within each group arise from interlayer hybridization and inversion-symmetry breaking introduced by the twist.
Notably, the groups I is nearly flat along the $ \mathrm{Y_M} $-$ \mathrm{\Gamma_\mathrm{M}} $ direction (perpendicular to the moir\'e stripe), while showing significant dispersion along the $ \mathrm{\Gamma_\mathrm{M}}$-$\mathrm{X_M}$ direction (parallel to the stripe). 
This anisotropic structure suggests suppressed electron motion across the moir\'e stripes and enhanced transport along them, indicating a 1D electronic character. 
The dispersion along the $\mathrm{Y_M}$–$\Gamma_\mathrm{M}$ direction becomes progressively more pronounced for lower-energy groups (from I to III).
We also find that the lowest conduction band, labeled as I$'$ and II$'$, exhibit a similar one-dimensional dispersion.
The conduction band extends to lower energies and crosses the Fermi level, reflecting the semimetallic nature of monolayer WTe$_2$, where the lowest branch retains a quasi-one-dimensional character.

The one-dimensional character of the energy bands is also reflected in the electronic wave functions. 
Figure \ref{fig:dft}(c) shows the real-space distributions of the wave functions for groups I, II, and III (right three panels), obtained by summing the probability density over all states within each group. 
The leftmost panel shows the atomic configuration for reference.
The wave function of group I exhibits a pronounced stripe pattern, strongly localized around the XY stacking regions and separated by low-density areas, consistent with its 1D band structure, which disperses predominantly along the horizontal ($x$) direction. 
Groups II and III display similar stripe patterns but with increasing overlap between neighboring stripes, corresponding to their larger band dispersion along the perpendicular direction. 

To disentangle the respective roles of interlayer coupling and intralayer moir\'e strain in shaping the 1D electronic structure, we performed DFT band calculations for an isolated twisted bilayer, in which the two layers are separated while each layer retains the atomic configuration of the fully relaxed moir\'e bilayer. 
As shown in Fig.~\ref{fig:dftlayer}, in which the bands of the top and bottom layers are superimposed throughout the entire energy range, the resulting band structure, featuring nearly degenerate top and bottom layer bands, closely resembles that of the full bilayer [Fig.~\ref{fig:dft}(a)], including its 1D character, apart from minor splittings attributable to interlayer hybridization. 
This observation indicates that the electronic structure near the Fermi level is governed predominantly by the intralayer moir\'e potential arising from lattice relaxation, rather than by interlayer coupling. 

\subsection{Effective tight-binding model}\label{sec:electronic:eff}

\begin{figure}
    \centering
    \includegraphics[width=85mm]{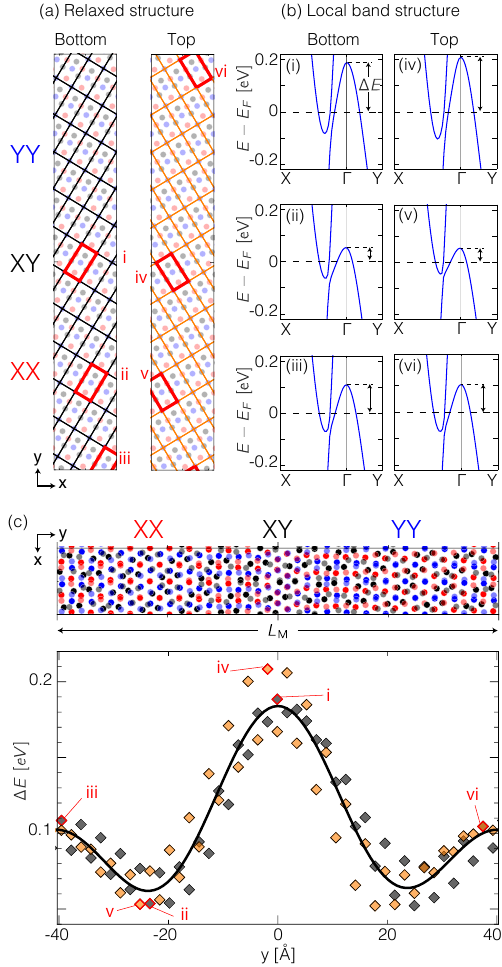}
    \caption{Electronic bands associated with the local atomic structures. (a) atomic structure of the relaxed twisted bilayer WTe$_2$, with the two layers separated. 
    Rectangles indicate local unit cells defined by connecting neighboring Y tellurium atoms (Fig.~\ref{fig:single}). 
    The local unit cells and internal atomic configurations exhibit small but finite modulation across the moir\'e pattern. 
    (b) Representative local band structures for unit cells labeled i-vi in panel (a).
    (c) Energy of the top of the hole-like band at $\Gamma$, relative to the Fermi energy, plotted as a function of unit-cell center position along the $y$ direction, perpendicular to moir\'e stripes. Orange and black symbols denote the top and bottom layers, respectively.
    }
    \label{fig:potential}
\end{figure}

To elucidate the microscopic origin of the observed electronic features, we construct an effective tight-binding model that reproduces the DFT band structure. As shown in the previous section, the emergence of the one-dimensional character is dominated by intralayer effects, while interlayer hopping leads only to relatively small band splittings. Motivated by this observation, we assume that the essential qualitative features can be captured by a tight-binding model incorporating an intralayer on-site moir\'e potential induced by lattice relaxation within each layer.

The derivation of the moir\'e potential proceeds as follows. We divide the relaxed twisted bilayer into monolayer unit cells, as illustrated in Fig.~\ref{fig:potential}(a), where the atomic configuration within each cell varies slightly from place to place due to moir\'e-induced lattice distortions. 
For each unit cell, we perform a monolayer DFT calculation assuming periodic boundary conditions for that cell. 
The resulting band structures, shown in Fig.~\ref{fig:potential}(b), exhibit a systematic dependence on position, reflecting the local variations in the atomic structure.
We define the energy difference between the Fermi level and the valence-band maximum as $\Delta E$, which serves as an effective local on-site potential for valence electrons. 

Figure~\ref{fig:potential}(c) plots $\Delta E$ as a function of the position of the unit-cell center projected onto the $y$ direction.
We find that $\Delta E$ varies nearly continuously across the moir\'e pattern, where the top and bottom layers exhibit nearly identical trends with maxima in the XY regions and minima in the XX and YY regions.
The intralayer moir\'e potential $V_{\rm eff}$ is then obtained by fitting all $\Delta E$ data points from both layers to a single function of the following form:
\[
V_{\rm eff}(\textbf{\textit{r}})=V_{0}+V_{1}\cos({\textbf{\textit{G}}_{\rm M} \cdot \textbf{\textit{r}}} + \varphi_{1}) + V_{2}\cos({2\textbf{\textit{G}}_{\rm M} \cdot \textbf{\textit{r}}} + \varphi_{2}),
\]
where $\textbf{\textit{G}}_{\rm M} = G_{\rm M} \mathbf{e}_y$ is the moir\'e reciprocal lattice vector.
The fitting parameters are obtained as $V_{0}\approx0.106$ eV, $V_{1}\approx0.0408$ eV, $V_{2}\approx0.0374$ eV, $\varphi_{1}\approx-0.0192$ rad, $\varphi_{2}\approx0.0105$ rad.
The inclusion of the second harmonic term reflects the anharmonic shape of the moir\'e potential, while the small phase shifts $\varphi_{1}$ and $\varphi_{2}$ indicate that the potential minima are nearly aligned with the high-symmetry stacking positions.
Notably, the spatial profile of the effective moir\'e potential closely resembles that of $\epsilon_{xy}^{(l)}$ shown in Fig.~\ref{fig:shear}, indicating that the intralayer moir\'e potential arises predominantly from shear strain induced by lattice relaxation.

\begin{figure}
    \centering
    \includegraphics[width=85mm]{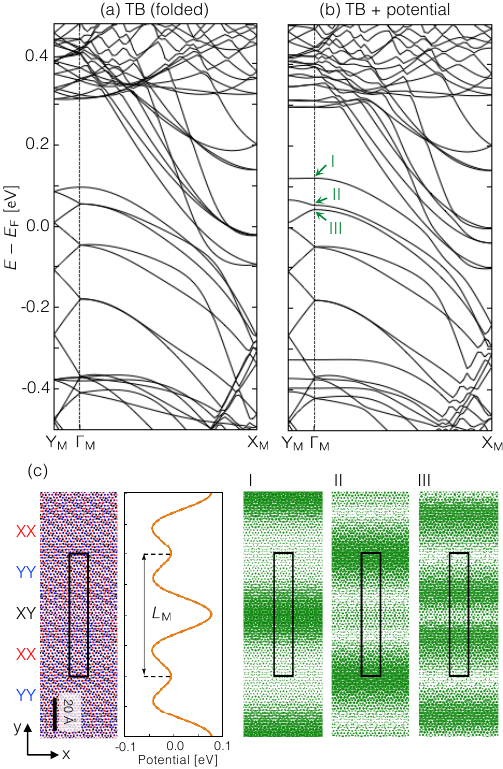}
    \caption{Band structure of the effective tight-binding model.
    (a) Electronic structure of single-layer WTe$_2$ folded into 
    the moir\'e Brillouin zone. 
    (b) Band structure including the potential induced by lattice relaxation.
    (c) spatial modulation of the potential, taken from the $\Gamma$-point energy variation in Fig.~\ref{fig:potential}, togather with 
    the wave-function amplitudes of states I-III indicated in panel (b)}
    \label{fig:tb}
\end{figure}

The effective tight-binding model is constructed by incorporating this moir\'e potential into the tight-binding Hamiltonian.
All monolayer unit cells are assumed to share identical hopping parameters,
which are obtained from DFT calculations of the intrinsic, undistorted WTe$_2$ monolayer.
We employ the $\{d_{xy}, d_{yz}, d_{zx}, d_{x^2-y^2}, d_{z^2}\}$ orbitals of W
and the $\{p_x, p_y, p_z\}$ orbitals of Te as the basis set.
By adding the position-dependent on-site potential $V_{\rm eff}(\mathbf{r})$
to all the atomic sites, the effective tight-binding model is fully specified.


Figure~\ref{fig:tb} presents the band structures obtained from the tight-binding model: (a) without and (b) with the effective moir\'e potential. In panel (b), pronounced one-dimensional features emerge near the valence-band maximum along the $\mathrm{Y}_{\rm M}$-$\Gamma_{\rm M}$ direction, in good agreement with the DFT results shown in Fig.~\ref{fig:dft}, apart from the small splittings induced by interlayer coupling, which are absent in the effective model. The wave functions at the $\Gamma_{\rm M}$ point, shown in Fig.~\ref{fig:tb}(c), also exhibit a one-dimensional spatial distribution aligned with the stripe pattern, consistent with the DFT wave-function profiles in Fig.~\ref{fig:dft}(c).
By comparison with the spatial profile of the effective potential $V_{\rm eff}$, we interpret the group-I and group-III states as the ground and second excited bound states, respectively, of holes confined near the potential maxima at the XY regions. The group-II states correspond to bound states associated with the secondary potential maxima at the XX and YY regions. From these considerations, we conclude that the emergence of one-dimensional energy bands near the Fermi level in twisted bilayer WTe$_2$ is primarily driven by the intralayer moir\'e potential induced by lattice relaxation.


\section{Conclusion}\label{sec:conclusion}

We have investigated the structural relaxation and electronic properties of a one-dimensional moir\'e superlattice formed in twisted bilayer 1T$'$-WTe$_2$. Using first-principles DFT calculations, supported by high-resolution HAADF-STEM measurements, we showed that lattice relaxation strongly reconstructs the moir\'e stripes, leading to stacking-dependent stripe widths in quantitative agreement with experiment. The relaxed structure hosts one-dimensional electronic bands near the Fermi level, characterized by strong dispersion along the stripe direction and weak dispersion in the perpendicular direction.

By disentangling interlayer and intralayer effects, we established that these 1D electronic states originate predominantly from an intralayer moir\'e potential induced by in-plane lattice relaxation. An effective tight-binding model incorporating this position-dependent potential successfully reproduces the DFT band structure and wave-function localization. Our results identify lattice relaxation as the key mechanism underlying one-dimensional electronic states in 1D moir\'e superlattices, and highlight twisted bilayer WTe$_2$ as a promising platform for exploring emergent one-dimensional moir\'e physics.

Unlike previously studied one-dimensional systems such as nanotubes or quantum wires, the 1D moir\'e superlattice naturally realizes an extended array of parallel one-dimensional channels with uniform geometry and tunable parameters inherited from the parent two-dimensional material. 
More generally, one-dimensional electronic channels embedded in two-dimensional materials
have been theoretically and experimentally explored, including domain-wall modes~\cite{sanjose2013,nam2017lattice,efimkin2018,walet2019emergence,tsim2020,hsu2023,nakatsuji2023,hcwang2024,yychang2025}, anisotropic van der Waals systems~\cite{wang2022}, and moir\'e structure arising from Brillouin-zone-edge states~\cite{fujimoto2022,cualuguaru2025}.

The simultaneous presence of many equivalent 1D channels is expected to substantially enhance one-dimensional signatures compared to isolated or sample-dependent realizations, providing a robust platform for exploring one-dimensional many-body physics—such as Tomonaga–Luttinger liquid behavior and other interaction-driven phenomena~\cite{haldane1981,egger1997, bockrath1999, yao1999}—as well as collective states emerging from coupled one-dimensional channels~\cite{emery2000,vishwanath2001}. 
However, in the present system, the one-dimensional valence-band channels intersect the conduction bands at the Fermi level, reflecting the intrinsic semimetallic nature of monolayer 1T$'$-WTe$_2$. 
This band overlap may influence the manifestation of one-dimensional many-body physics through additional screening or interband coupling effects. 
In contrast, semiconducting anisotropic materials, where such band overlap is absent, such as ReSe$_2$~\cite{yang2026}, are expected to provide a cleaner realization of isolated one-dimensional moir\'e channels. Exploring these systems therefore represents an important direction for future studies of one-dimensional moir\'e physics. Quantitative analysis of these spectra remains to be investigated.

In addition, the quasi-one-dimensional electronic structure is expected to give rise to experimentally accessible signatures. 
In particular, anisotropic features are expected to appear in energy-resolved probes such as optical conductivity, reflecting the one-dimensional character of the valence states. 
The spatially anisotropic wave functions may also be detectable in local probes such as scanning tunneling microscopy and spectroscopy (STM/STS), 
providing a direct signature of the one-dimensional electronic states. 
A quantitative analysis of these spectra remains to be carried out.

The general construction method for one-dimensional moir\'e patterns based on an effective anisotropic triangular lattice, which is adopted throughout this paper, predicts the emergence of 1D moir\'e superlattices in a wide class of anisotropic two-dimensional materials, including twisted bilayers of black phosphorene~\cite{sousa2025} PdSe$_2$~\cite{an2025}, and ReSe$_2$~\cite{yang2026} as well as other low-symmetry transition-metal dichalcogenides. 
The theoretical approach presented here—including the one-dimensional moir\'e description, structural optimization, and electronic band analysis—provides a unified framework for exploring one-dimensional moir\'e physics across a broad range of materials beyond twisted bilayer WTe$_2$.

We note that a one-dimensional moir\'e pattern has also been experimentally observed at $\theta=58^\circ$~\cite{yang2025}.
While the present framework is, in principle, applicable to this case, the stacking configuration is expected to be qualitatively different from the 62$^\circ$ structures feature alternating XX-XY-YY domains, 
where the presence of XY regions drives shear strain and generates an intralayer moir\'e potential. 
In contrast, the 58$^\circ$ configuration is expected to lack such XY regions, and associated strain modulation may therefore be significantly reduced.
As a result, the mechanism underlying the quasi-one-dimensional electronic structure may differ in the $58^\circ$ case. 
A detailed investigation of this regime is left for future work.


\begin{acknowledgements}
This work was supported by JSPS KAKENHI (Grants Nos. 
JP21H05232, 
JP21H05234, 
JP21H05236, 
JP22K18317, JP24K01293, 
JP24K06921, 
JP24H01198, 
JP24K21195, JP25H00602, 
JP25K00938);  
JST-CREST and JST-PRESTO 
(Grant Nos. 
JPMJCR20T3, 
JPMJCR20B4, 
JPMJPR24H8); 
Advanced Research Infrastructure for Materials and Nanotechnology in Japan
(ARIM), MEXT (Grant Numbers JPMXP1223JI0033, JPMXP1224JI0026);
JSPS Overseas Research Fellowship; 
JST SPRING (Grant Nos. JPMJSP2108, JPMJSP2102);
the World-Leading Innovative Graduate Study Program for Advanced Basic Science Course at the University of Tokyo. 
The DFT calculations were carried out using the computer resource offered under the category of General Projects by Research Institute for Information Technology, Kyushu University.
\end{acknowledgements}

\appendix
\begin{table*}
\centering
\caption{Structural  parameters and energy decomposition of the monolayer and 1D moir\'e WTe$_2$ systems. 
All energies are given per formula unit (per WTe$_2$).
In particular, the moir\'e energies are given per moir\'e supercell containing $N_{\mathrm{u.c.}}$ primitive unit cells per layer. 
$E_{\mathrm{iso}}$ is defined as the total energy of the twisted bilayer with interlayer interactions switched off, while the intralayer atomic geometry is kept identical to that of the relaxed 1D moir\'e structure (see Fig.~\ref{fig:dftlayer} for definition). }
\label{tab:parameters}

\begin{tabular}{c@{\hspace{12pt}}c@{\hspace{12pt}}l}
\hline
\hline
Quantity & Value & Expression \\
\hline
$a_1$ & 3.461\AA & Single layer lattice constant \\
$a_2$ & 6.261\AA & Single layer lattice constant \\
$\theta$ & 62.18$^\circ$ & Twist angle \\
$L_{\mathrm{M}}$ & 80.61\AA & Moir\'e supercell lattice constant \\
$W_{\mathrm{M}}$ & 12.12\AA & Moir\'e supercell width\\
$N_{\mathrm{u.c.}}$ & 45 & Number of unit cells per layer in 1D moir\'e cell\\
$\langle d \rangle$ & 3.240 \AA & Average interlayer spacing (relaxed 1D moir\'e structure) \\
$d_0$ & 3.19 \AA & Interlayer spacing of untwisted bilayer \\
\addlinespace
\hline
$E_{\mathrm{mono}}$ & $-20.21$ eV & Energy of single layer (reference energy) 
\\
$E_{\mathrm{bi}}$ & $-20.34$ eV & Energy of untwisted bilayer (1T$'$ stacking, $180^\circ$ rotated)
\\
$E_{\mathrm{moire}}$ & $-20.32$ eV & Energy of relaxed 1D moir\'e bilayer \\
$E_{\mathrm{bind}}(=E_{\mathrm{moire}}-E_{\mathrm{mono}})$& $-0.11$ eV& Binding energy of 1D moir\'e bilayer \\
$E_{\mathrm{iso}}$ & $-20.19$ eV & Energy of isolated 1D moir\'e bilayer \\
$E_{\mathrm{strain}}(=E_{\mathrm{iso}}-E_{\mathrm{mono}})$ & 0.021 eV &   Strain energy induced by relaxation \\
$E_{\mathrm{adh}}(=E_{\mathrm{moire}}-E_{\mathrm{iso}})$ & $-0.13$ eV & Adhesion energy\\
\hline
\hline
\end{tabular}
\end{table*}

\section{Derivation for commensurate approximant}\label{sec:derivepsilon}

In this appendix, we derive the strain magnitude Eq.~\eqref{eq:requiredepsilon} required to satisfy the commensurate condition. 
From the isosceles triangular geometry shown in Fig.~\ref{fig:triangle}(b), with $|\bm{\tau}_1|=|\bm{\tau}_2|=\tau$, the shift $\delta$ (see the inset of Fig.~\ref{fig:triangle}(b)) is given by
\begin{equation}
\delta=2(\bm{\tau}_2-\bm{\tau}_1)\cdot\frac{\bm{\tau}_2}{\tau}-\tau.
\end{equation}
Using the isosceles condition  
$(\bm{\tau}_2+\bm{\tau}_1)\cdot(\bm{\tau}_2-\bm{\tau}_1)=0$, 
the ratio $\tau/\delta$ can be expressed in terms of the parameter 
$\eta=|\bm{\tau}_2+\bm{\tau}_1|/|\bm{\tau}_2-\bm{\tau}_1|$ 
as
\begin{equation}
\frac{\tau}{\delta}=\frac{1+\eta^2}{3-\eta^2}
\end{equation}
Under an area-conserving uniaxial strain [Eq.~\eqref{eq:epsilon}], 
, where the $x$ and $y$ directions are scaled as $(1+\varepsilon)$ and $(1+\varepsilon)^{-1}$, respectively, 
the parameter transforms as $\eta\to \tilde{\eta}=(\varepsilon+1)^2\eta$, since $\bm{\tau}_1+\bm{\tau_2}\parallel\hat{\bm{x}}$ and 
$\bm{\tau}_1-\bm{\tau_2}\parallel\hat{\bm{y}}$. 
Imposing the commensurate condition,  
\begin{equation}
\frac{\tilde\tau}{\tilde\delta} = \frac{1+(1+\varepsilon)^4\eta^2}{3-(1+\varepsilon)^4\eta^2} = N,
\end{equation}
we obtain the required strain magnitude given in Eq.~\eqref{eq:requiredepsilon}.

\section{Structural parameters and total energy decomposition}~\label{sec:parameter}
We summarize in Table~\ref{tab:parameters} the structural parameters and total energies obtained from DFT calculations for the monolayer, untwisted bilayer, and 1D moir\'e WTe$_2$ systems. The lattice constants $a_1$, $a_2$, and twist angle $\theta$ are consistent with those introduced in the main text and are listed here for completeness. 
The moir\'e supercell parameters ($L_{\mathrm{M}}$, $W_{\mathrm{M}}$, $N_{\mathrm{u.c.}}$) are also provided explicitly.

The average interlayer spacing $\langle d \rangle$ in the relaxed 1D moir\'e structure is defined as the average out-of-plane ($z$ direction) distance between the nearest Te atoms (X site of the bottom layer and Y$'$ site of the top layer; see also Fig.~\ref{fig:relax}(c)). The reference spacing $d_0$ is defined analogously for the untwisted bilayer (1T$'$ stacking with $180^\circ$ rotation), corresponding to the natural stacking realized in bulk WTe$_2$. 
The moir\'e structure exhibits a slightly larger average interlayer spacing than the untwisted case.

We further list the total energies and their energy decomposition. 
In particular, $E_{\mathrm{moire}}$ is the total energy of the structure shown in Fig.~\ref{fig:relax}(b) and $E_{\mathrm{iso}}$ is the total energy for the structure discussed in Fig.~\ref{fig:dftlayer}. 
These energies are given per formula unit (per WTe$_2$). 
The reference energy is defined as $E_{\mathrm{mono}}$ (per-formula-unit energy of monolayer).  
The binding energy $E_{\mathrm{bind}}$, 
strain energy $E_{\mathrm{strain}}$, 
and adhesion energy $E_{\mathrm{adh}}$ 
are defined in Table~\ref{tab:parameters} and provide a decomposition of the total energy difference 
between the relaxed moir\'e bilayer and the reference state.

\section{Electronic states of local stacking configurations}~\label{sec:localstacking}
As shown in Fig.~\ref{fig:system}, 
the local stacking structures of the one-dimensional moir\'e in 62$^\circ$ twisted bilayer WTe$_2$ are characterized 
by a set of periodic stacking configurations (Fig.~\ref{fig:system}(b)).
These configurations are constructed from a 60$^\circ$ stacking of uniaxially expanded WTe$_2$, such that the ratio between the lattice vectors satisfies $|\bm{a}_2|/|\bm{a}_1|\to\sqrt{3}$.
In this appendix, we calculate the electronic structures and energetics of these periodic stacking configurations, by using DFT calculation.

\begin{figure}[h]
    \centering
    \includegraphics[width=85mm]{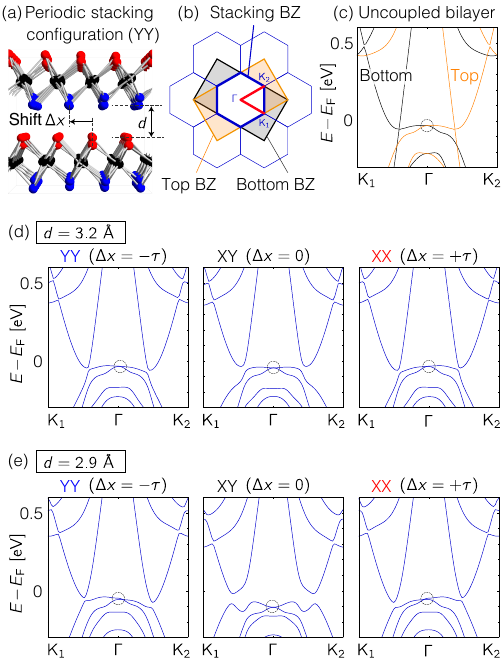}
    \caption{Electronic structure of the periodic stacking configurations shown in Fig.~\ref{fig:system}(b). 
    (a) The side view of YY stacking in Fig.~\ref{fig:system}(b), showing the definition of the shift $\Delta x$ and the interlayer spacing $d$.
    (b) Brillouin zone of the periodic stacking configuration.
    (c) Band structure of decoupled bilayer. Bands originating from top and bottom layers are colored orange and black, respectively.
    (d) Band structure of $YY$, $XY$, and $XX$ stackings for an interlayer spacing $d=3.2$ \AA. 
    (e) The same as (d), but smaller interlayer spacing $d=2.9$ \AA.
    }
    \label{fig:localband}
\end{figure}

Figure~\ref{fig:localband}(a) shows the side view of the YY stacking presented in Fig.~\ref{fig:system}(b). 
By introducing a lateral interlayer shift along the $x$ direction, we generate the local stacking structures of the one-dimensional moir\'e system. 
Specifically, we define the shift $\Delta x$ in Fig.~\ref{fig:localband}(a) such that $\Delta x=-\tau$
corresponds to YY stacking, $\Delta x=0$ to XY stacking, and $\Delta x=+\tau$ to XX stacking. Here $\tau$ is the lattice constant of the effective triangular lattice defined in Sec.~\ref{sec:averaged}.

As shown in Fig.~\ref{fig:system}(b), 
the unit cell of the periodic stacking configuration has twice the area of the monolayer unit cell. 
Correspondingly, in momentum space, 
the Brillouin zone of this configuration is reduced 
to half of the monolayer BZ, as illustrated in Fig. 13(b). 
Figure~\ref{fig:localband} (c) displays the band structure in the decoupled limit, 
where the monolayer bands are folded into the stacking BZ. 
Here, the paths K$_1\Gamma$ and K$_2\Gamma$ correspond to the screw axis of the top and bottom layers, respectively, 
along which band crossings occur~\cite{mhu2021}. 
In the decoupled bilayer, the top and bottom bands are degenerate at the $\Gamma$ point.

Figure~\ref{fig:localband}(d) shows the band structures of the periodic bilayers corresponding to YY, XY, and XX stackings. 
In this calculation, we set the interlayer spacing to $d=3.2$~\AA, which is slightly larger than the value for the non-moir\'e 1T$'$ bilayer (180$^\circ$ stacking). 
As highlighted by the dashed circle, 
the valence bands at $\Gamma$, just below the Fermi energy, remain degenerate even in the presence of finite interlayer coupling. 
This indicates that the interlayer coupling is ineffective,
consistent with the results in the main text. 
Note that this degeneracy is not symmetry-protected: the present bilayer has only $C_{2y}$ symmetry, 
which does not support degenerate irreducible representations. 
Therefore, the degeneracy likely originates 
from the detailed orbital character of the states at the $\Gamma$ point and phase cancellation in the interlayer hopping.

Figure~\ref{fig:localband}(e) shows the band structures for reduced interlayer spacing. 
Although the interlayer coupling is enhanced at smaller spacing, the band splitting of the top valence band at $\Gamma$ remains very small. 
This indicates that the interlayer coupling remains ineffective even under pressure.

\begin{figure}[b]
    \centering
    \includegraphics[width=85mm]{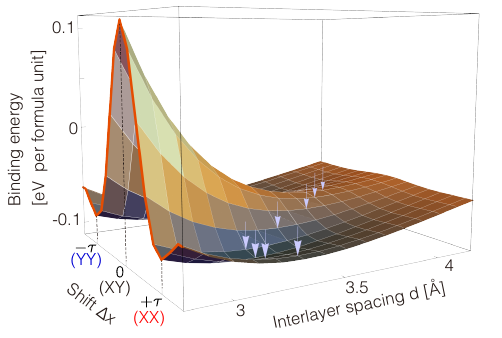}
    \caption{
    Binding energy (per unit cell) landscape of periodic stacking configurations shown in Fig.~\ref{fig:system}(b), as a function of the interlayer shift $\Delta x$ and interlayer spacing $d$ (see Fig.~\ref{fig:localband}(a) for the definition). 
    Arrows indicate the minima along the $d$ direction at fixed $\Delta x$. 
    The landscape is symmetric with respect to $\Delta x=0$}
    \label{fig:binding}
\end{figure}

On the other hand, moir\'e relaxation, 
which is the primary factor determining the band structure of 62$^\circ$ twisted WTe$_2$, 
as discussed in the main text, can be further enhanced under pressure. 
To analyze this effect, we evaluate the binding energy landscape of the periodic stacking configurations, 
defined by the relative total energy $E_\mathrm{tot}-E_{\mathrm{mono}}$ in per formula unit (per WTe$_2$), 
as a function of the lateral shift $\Delta x$ and interlayer spacing $d$ (see Fig.~\ref{fig:binding}).
The XY stacking is clearly less stable than the XX and YY stackings. 
As a result, in the moir\'e system, the XY regions shrink while the XX and YY regions expand, as shown in the main text. 
Moreover, the energy contrast between stacking configurations becomes more pronounced at smaller interlayer spacing. 
It is therefore natural that the structural modulation is enhanced under pressure, 
leading to a stronger impact on the electronic structure.


\clearpage
\bibliography{reference}

\begin{thebibliography}{129}%
\makeatletter
\providecommand \@ifxundefined [1]{%
 \@ifx{#1\undefined}
}%
\providecommand \@ifnum [1]{%
 \ifnum #1\expandafter \@firstoftwo
 \else \expandafter \@secondoftwo
 \fi
}%
\providecommand \@ifx [1]{%
 \ifx #1\expandafter \@firstoftwo
 \else \expandafter \@secondoftwo
 \fi
}%
\providecommand \natexlab [1]{#1}%
\providecommand \enquote  [1]{``#1''}%
\providecommand \bibnamefont  [1]{#1}%
\providecommand \bibfnamefont [1]{#1}%
\providecommand \citenamefont [1]{#1}%
\providecommand \href@noop [0]{\@secondoftwo}%
\providecommand \href [0]{\begingroup \@sanitize@url \@href}%
\providecommand \@href[1]{\@@startlink{#1}\@@href}%
\providecommand \@@href[1]{\endgroup#1\@@endlink}%
\providecommand \@sanitize@url [0]{\catcode `\\12\catcode `\$12\catcode
  `\&12\catcode `\#12\catcode `\^12\catcode `\_12\catcode `\%12\relax}%
\providecommand \@@startlink[1]{}%
\providecommand \@@endlink[0]{}%
\providecommand \url  [0]{\begingroup\@sanitize@url \@url }%
\providecommand \@url [1]{\endgroup\@href {#1}{\urlprefix }}%
\providecommand \urlprefix  [0]{URL }%
\providecommand \Eprint [0]{\href }%
\providecommand \doibase [0]{https://doi.org/}%
\providecommand \selectlanguage [0]{\@gobble}%
\providecommand \bibinfo  [0]{\@secondoftwo}%
\providecommand \bibfield  [0]{\@secondoftwo}%
\providecommand \translation [1]{[#1]}%
\providecommand \BibitemOpen [0]{}%
\providecommand \bibitemStop [0]{}%
\providecommand \bibitemNoStop [0]{.\EOS\space}%
\providecommand \EOS [0]{\spacefactor3000\relax}%
\providecommand \BibitemShut  [1]{\csname bibitem#1\endcsname}%
\let\auto@bib@innerbib\@empty
\bibitem [{\citenamefont {Andrei}\ and\ \citenamefont
  {MacDonald}(2020)}]{andrei2020review}%
  \BibitemOpen
  \bibfield  {author} {\bibinfo {author} {\bibfnamefont {E.~Y.}\ \bibnamefont
  {Andrei}}\ and\ \bibinfo {author} {\bibfnamefont {A.~H.}\ \bibnamefont
  {MacDonald}},\ }\bibfield  {title} {\bibinfo {title} {Graphene bilayers with
  a twist},\ }\href@noop {} {\bibfield  {journal} {\bibinfo  {journal} {Nature
  materials}\ }\textbf {\bibinfo {volume} {19}},\ \bibinfo {pages} {1265}
  (\bibinfo {year} {2020})}\BibitemShut {NoStop}%
\bibitem [{\citenamefont {Shah}\ \emph {et~al.}(2024)\citenamefont {Shah},
  \citenamefont {Chen}, \citenamefont {Xie}, \citenamefont {Oyang},
  \citenamefont {Ouyang}, \citenamefont {Liu}, \citenamefont {Wang},
  \citenamefont {He},\ and\ \citenamefont {Liu}}]{shah2024}%
  \BibitemOpen
  \bibfield  {author} {\bibinfo {author} {\bibfnamefont {S.~J.}\ \bibnamefont
  {Shah}}, \bibinfo {author} {\bibfnamefont {J.}~\bibnamefont {Chen}}, \bibinfo
  {author} {\bibfnamefont {X.}~\bibnamefont {Xie}}, \bibinfo {author}
  {\bibfnamefont {X.}~\bibnamefont {Oyang}}, \bibinfo {author} {\bibfnamefont
  {F.}~\bibnamefont {Ouyang}}, \bibinfo {author} {\bibfnamefont
  {Z.}~\bibnamefont {Liu}}, \bibinfo {author} {\bibfnamefont {J.-T.}\
  \bibnamefont {Wang}}, \bibinfo {author} {\bibfnamefont {J.}~\bibnamefont
  {He}},\ and\ \bibinfo {author} {\bibfnamefont {Y.}~\bibnamefont {Liu}},\
  }\bibfield  {title} {\bibinfo {title} {Progress and prospects of moir{\'e}
  superlattices in twisted tmd heterostructures},\ }\href@noop {} {\bibfield
  {journal} {\bibinfo  {journal} {Nano Research}\ }\textbf {\bibinfo {volume}
  {17}},\ \bibinfo {pages} {10134} (\bibinfo {year} {2024})}\BibitemShut
  {NoStop}%
\bibitem [{\citenamefont {Gupta}\ \emph {et~al.}(2025)\citenamefont {Gupta},
  \citenamefont {Zhang}, \citenamefont {Lei}, \citenamefont {Yu}, \citenamefont
  {Liu}, \citenamefont {Zou},\ and\ \citenamefont
  {Yakobson}}]{gupta2025review}%
  \BibitemOpen
  \bibfield  {author} {\bibinfo {author} {\bibfnamefont {S.}~\bibnamefont
  {Gupta}}, \bibinfo {author} {\bibfnamefont {J.-J.}\ \bibnamefont {Zhang}},
  \bibinfo {author} {\bibfnamefont {J.}~\bibnamefont {Lei}}, \bibinfo {author}
  {\bibfnamefont {H.}~\bibnamefont {Yu}}, \bibinfo {author} {\bibfnamefont
  {M.}~\bibnamefont {Liu}}, \bibinfo {author} {\bibfnamefont {X.}~\bibnamefont
  {Zou}},\ and\ \bibinfo {author} {\bibfnamefont {B.~I.}\ \bibnamefont
  {Yakobson}},\ }\bibfield  {title} {\bibinfo {title} {Two-dimensional
  transition metal dichalcogenides: A theory and simulation perspective},\
  }\href@noop {} {\bibfield  {journal} {\bibinfo  {journal} {Chemical Reviews}\
  }\textbf {\bibinfo {volume} {125}},\ \bibinfo {pages} {786} (\bibinfo {year}
  {2025})}\BibitemShut {NoStop}%
\bibitem [{\citenamefont {Lopes~dos Santos}\ \emph {et~al.}(2007)\citenamefont
  {Lopes~dos Santos}, \citenamefont {Peres},\ and\ \citenamefont
  {Castro~Neto}}]{lopes2007graphene}%
  \BibitemOpen
  \bibfield  {author} {\bibinfo {author} {\bibfnamefont {J.}~\bibnamefont
  {Lopes~dos Santos}}, \bibinfo {author} {\bibfnamefont {N.}~\bibnamefont
  {Peres}},\ and\ \bibinfo {author} {\bibfnamefont {A.}~\bibnamefont
  {Castro~Neto}},\ }\bibfield  {title} {\bibinfo {title} {Graphene bilayer with
  a twist: Electronic structure},\ }\href@noop {} {\bibfield  {journal}
  {\bibinfo  {journal} {Phys. Rev. Lett.}\ }\textbf {\bibinfo {volume} {99}},\
  \bibinfo {pages} {256802} (\bibinfo {year} {2007})}\BibitemShut {NoStop}%
\bibitem [{\citenamefont {Su\'arez~Morell}\ \emph {et~al.}(2010)\citenamefont
  {Su\'arez~Morell}, \citenamefont {Correa}, \citenamefont {Vargas},
  \citenamefont {Pacheco},\ and\ \citenamefont {Barticevic}}]{morell2010}%
  \BibitemOpen
  \bibfield  {author} {\bibinfo {author} {\bibfnamefont {E.}~\bibnamefont
  {Su\'arez~Morell}}, \bibinfo {author} {\bibfnamefont {J.~D.}\ \bibnamefont
  {Correa}}, \bibinfo {author} {\bibfnamefont {P.}~\bibnamefont {Vargas}},
  \bibinfo {author} {\bibfnamefont {M.}~\bibnamefont {Pacheco}},\ and\ \bibinfo
  {author} {\bibfnamefont {Z.}~\bibnamefont {Barticevic}},\ }\bibfield  {title}
  {\bibinfo {title} {Flat bands in slightly twisted bilayer graphene:
  Tight-binding calculations},\ }\href
  {https://doi.org/10.1103/PhysRevB.82.121407} {\bibfield  {journal} {\bibinfo
  {journal} {Phys. Rev. B}\ }\textbf {\bibinfo {volume} {82}},\ \bibinfo
  {pages} {121407} (\bibinfo {year} {2010})}\BibitemShut {NoStop}%
\bibitem [{\citenamefont {Bistritzer}\ and\ \citenamefont
  {MacDonald}(2010)}]{bistritzer2010}%
  \BibitemOpen
  \bibfield  {author} {\bibinfo {author} {\bibfnamefont {R.}~\bibnamefont
  {Bistritzer}}\ and\ \bibinfo {author} {\bibfnamefont {A.~H.}\ \bibnamefont
  {MacDonald}},\ }\bibfield  {title} {\bibinfo {title} {Transport between
  twisted graphene layers},\ }\href
  {https://doi.org/10.1103/PhysRevB.81.245412} {\bibfield  {journal} {\bibinfo
  {journal} {Phys. Rev. B}\ }\textbf {\bibinfo {volume} {81}},\ \bibinfo
  {pages} {245412} (\bibinfo {year} {2010})}\BibitemShut {NoStop}%
\bibitem [{\citenamefont {Bistritzer}\ and\ \citenamefont
  {MacDonald}(2011)}]{bistritzer2011moirepnas}%
  \BibitemOpen
  \bibfield  {author} {\bibinfo {author} {\bibfnamefont {R.}~\bibnamefont
  {Bistritzer}}\ and\ \bibinfo {author} {\bibfnamefont {A.}~\bibnamefont
  {MacDonald}},\ }\bibfield  {title} {\bibinfo {title} {Moir{\'e} bands in
  twisted double-layer graphene},\ }\href@noop {} {\bibfield  {journal}
  {\bibinfo  {journal} {Proc. Natl. Acad. Sci.}\ }\textbf {\bibinfo {volume}
  {108}},\ \bibinfo {pages} {12233} (\bibinfo {year} {2011})}\BibitemShut
  {NoStop}%
\bibitem [{\citenamefont {Moon}\ and\ \citenamefont
  {Koshino}(2012)}]{moon2012}%
  \BibitemOpen
  \bibfield  {author} {\bibinfo {author} {\bibfnamefont {P.}~\bibnamefont
  {Moon}}\ and\ \bibinfo {author} {\bibfnamefont {M.}~\bibnamefont {Koshino}},\
  }\bibfield  {title} {\bibinfo {title} {Energy spectrum and quantum hall
  effect in twisted bilayer graphene},\ }\href@noop {} {\bibfield  {journal}
  {\bibinfo  {journal} {Phys. Rev. B}\ }\textbf {\bibinfo {volume} {85}},\
  \bibinfo {pages} {195458} (\bibinfo {year} {2012})}\BibitemShut {NoStop}%
\bibitem [{\citenamefont {Lu}\ \emph {et~al.}(2013)\citenamefont {Lu},
  \citenamefont {Lin}, \citenamefont {Liu}, \citenamefont {Yeh}, \citenamefont
  {Suenaga},\ and\ \citenamefont {Chiu}}]{cclu2013}%
  \BibitemOpen
  \bibfield  {author} {\bibinfo {author} {\bibfnamefont {C.-C.}\ \bibnamefont
  {Lu}}, \bibinfo {author} {\bibfnamefont {Y.-C.}\ \bibnamefont {Lin}},
  \bibinfo {author} {\bibfnamefont {Z.}~\bibnamefont {Liu}}, \bibinfo {author}
  {\bibfnamefont {C.-H.}\ \bibnamefont {Yeh}}, \bibinfo {author} {\bibfnamefont
  {K.}~\bibnamefont {Suenaga}},\ and\ \bibinfo {author} {\bibfnamefont {P.-W.}\
  \bibnamefont {Chiu}},\ }\bibfield  {title} {\bibinfo {title} {Twisting
  bilayer graphene superlattices},\ }\href {https://doi.org/10.1021/nn3059828}
  {\bibfield  {journal} {\bibinfo  {journal} {ACS Nano}\ }\textbf {\bibinfo
  {volume} {7}},\ \bibinfo {pages} {2587} (\bibinfo {year} {2013})},\ \bibinfo
  {note} {pMID: 23448165}\BibitemShut {NoStop}%
\bibitem [{\citenamefont {Dean}\ \emph {et~al.}(2013)\citenamefont {Dean},
  \citenamefont {Wang}, \citenamefont {Maher}, \citenamefont {Forsythe},
  \citenamefont {Ghahari}, \citenamefont {Gao}, \citenamefont {Katoch},
  \citenamefont {Ishigami}, \citenamefont {Moon}, \citenamefont {Koshino} \emph
  {et~al.}}]{dean2013}%
  \BibitemOpen
  \bibfield  {author} {\bibinfo {author} {\bibfnamefont {C.~R.}\ \bibnamefont
  {Dean}}, \bibinfo {author} {\bibfnamefont {L.}~\bibnamefont {Wang}}, \bibinfo
  {author} {\bibfnamefont {P.}~\bibnamefont {Maher}}, \bibinfo {author}
  {\bibfnamefont {C.}~\bibnamefont {Forsythe}}, \bibinfo {author}
  {\bibfnamefont {F.}~\bibnamefont {Ghahari}}, \bibinfo {author} {\bibfnamefont
  {Y.}~\bibnamefont {Gao}}, \bibinfo {author} {\bibfnamefont {J.}~\bibnamefont
  {Katoch}}, \bibinfo {author} {\bibfnamefont {M.}~\bibnamefont {Ishigami}},
  \bibinfo {author} {\bibfnamefont {P.}~\bibnamefont {Moon}}, \bibinfo {author}
  {\bibfnamefont {M.}~\bibnamefont {Koshino}}, \emph {et~al.},\ }\bibfield
  {title} {\bibinfo {title} {Hofstadter’s butterfly and the fractal quantum
  hall effect in moir{\'e} superlattices},\ }\href@noop {} {\bibfield
  {journal} {\bibinfo  {journal} {Nature}\ }\textbf {\bibinfo {volume} {497}},\
  \bibinfo {pages} {598} (\bibinfo {year} {2013})}\BibitemShut {NoStop}%
\bibitem [{\citenamefont {Cao}\ \emph {et~al.}(2018{\natexlab{a}})\citenamefont
  {Cao}, \citenamefont {Fatemi}, \citenamefont {Demir}, \citenamefont {Fang},
  \citenamefont {Tomarken}, \citenamefont {Luo}, \citenamefont
  {Sanchez-Yamagishi}, \citenamefont {Watanabe}, \citenamefont {Taniguchi},
  \citenamefont {Kaxiras}, \citenamefont {Ashoori},\ and\ \citenamefont
  {Jarillo-Herrero}}]{cao2018_80}%
  \BibitemOpen
  \bibfield  {author} {\bibinfo {author} {\bibfnamefont {Y.}~\bibnamefont
  {Cao}}, \bibinfo {author} {\bibfnamefont {V.}~\bibnamefont {Fatemi}},
  \bibinfo {author} {\bibfnamefont {A.}~\bibnamefont {Demir}}, \bibinfo
  {author} {\bibfnamefont {S.}~\bibnamefont {Fang}}, \bibinfo {author}
  {\bibfnamefont {S.~L.}\ \bibnamefont {Tomarken}}, \bibinfo {author}
  {\bibfnamefont {J.~Y.}\ \bibnamefont {Luo}}, \bibinfo {author} {\bibfnamefont
  {J.~D.}\ \bibnamefont {Sanchez-Yamagishi}}, \bibinfo {author} {\bibfnamefont
  {K.}~\bibnamefont {Watanabe}}, \bibinfo {author} {\bibfnamefont
  {T.}~\bibnamefont {Taniguchi}}, \bibinfo {author} {\bibfnamefont
  {E.}~\bibnamefont {Kaxiras}}, \bibinfo {author} {\bibfnamefont {R.~C.}\
  \bibnamefont {Ashoori}},\ and\ \bibinfo {author} {\bibfnamefont
  {P.}~\bibnamefont {Jarillo-Herrero}},\ }\bibfield  {title} {\bibinfo {title}
  {Correlated insulator behaviour at half-filling in magic-angle graphene
  superlattices},\ }\href {https://doi.org/10.1038/nature26154} {\bibfield
  {journal} {\bibinfo  {journal} {Nature}\ }\textbf {\bibinfo {volume} {556}},\
  \bibinfo {pages} {80} (\bibinfo {year} {2018}{\natexlab{a}})}\BibitemShut
  {NoStop}%
\bibitem [{\citenamefont {Cao}\ \emph {et~al.}(2018{\natexlab{b}})\citenamefont
  {Cao}, \citenamefont {Fatemi}, \citenamefont {Fang}, \citenamefont
  {Watanabe}, \citenamefont {Taniguchi}, \citenamefont {Kaxiras},\ and\
  \citenamefont {Jarillo-Herrero}}]{cao2018_43}%
  \BibitemOpen
  \bibfield  {author} {\bibinfo {author} {\bibfnamefont {Y.}~\bibnamefont
  {Cao}}, \bibinfo {author} {\bibfnamefont {V.}~\bibnamefont {Fatemi}},
  \bibinfo {author} {\bibfnamefont {S.}~\bibnamefont {Fang}}, \bibinfo {author}
  {\bibfnamefont {K.}~\bibnamefont {Watanabe}}, \bibinfo {author}
  {\bibfnamefont {T.}~\bibnamefont {Taniguchi}}, \bibinfo {author}
  {\bibfnamefont {E.}~\bibnamefont {Kaxiras}},\ and\ \bibinfo {author}
  {\bibfnamefont {P.}~\bibnamefont {Jarillo-Herrero}},\ }\bibfield  {title}
  {\bibinfo {title} {Unconventional superconductivity in magic-angle graphene
  superlattices},\ }\href {https://doi.org/10.1038/nature26160} {\bibfield
  {journal} {\bibinfo  {journal} {Nature}\ }\textbf {\bibinfo {volume} {556}},\
  \bibinfo {pages} {43} (\bibinfo {year} {2018}{\natexlab{b}})}\BibitemShut
  {NoStop}%
\bibitem [{\citenamefont {Huang}\ \emph {et~al.}(2018)\citenamefont {Huang},
  \citenamefont {Kim}, \citenamefont {Efimkin}, \citenamefont {Lovorn},
  \citenamefont {Taniguchi}, \citenamefont {Watanabe}, \citenamefont
  {MacDonald}, \citenamefont {Tutuc},\ and\ \citenamefont
  {LeRoy}}]{shuang2018}%
  \BibitemOpen
  \bibfield  {author} {\bibinfo {author} {\bibfnamefont {S.}~\bibnamefont
  {Huang}}, \bibinfo {author} {\bibfnamefont {K.}~\bibnamefont {Kim}}, \bibinfo
  {author} {\bibfnamefont {D.~K.}\ \bibnamefont {Efimkin}}, \bibinfo {author}
  {\bibfnamefont {T.}~\bibnamefont {Lovorn}}, \bibinfo {author} {\bibfnamefont
  {T.}~\bibnamefont {Taniguchi}}, \bibinfo {author} {\bibfnamefont
  {K.}~\bibnamefont {Watanabe}}, \bibinfo {author} {\bibfnamefont {A.~H.}\
  \bibnamefont {MacDonald}}, \bibinfo {author} {\bibfnamefont {E.}~\bibnamefont
  {Tutuc}},\ and\ \bibinfo {author} {\bibfnamefont {B.~J.}\ \bibnamefont
  {LeRoy}},\ }\bibfield  {title} {\bibinfo {title} {Topologically protected
  helical states in minimally twisted bilayer graphene},\ }\href
  {https://doi.org/10.1103/PhysRevLett.121.037702} {\bibfield  {journal}
  {\bibinfo  {journal} {Phys. Rev. Lett.}\ }\textbf {\bibinfo {volume} {121}},\
  \bibinfo {pages} {037702} (\bibinfo {year} {2018})}\BibitemShut {NoStop}%
\bibitem [{\citenamefont {Koshino}\ \emph {et~al.}(2018)\citenamefont
  {Koshino}, \citenamefont {Yuan}, \citenamefont {Koretsune}, \citenamefont
  {Ochi}, \citenamefont {Kuroki},\ and\ \citenamefont {Fu}}]{koshino2018}%
  \BibitemOpen
  \bibfield  {author} {\bibinfo {author} {\bibfnamefont {M.}~\bibnamefont
  {Koshino}}, \bibinfo {author} {\bibfnamefont {N.~F.~Q.}\ \bibnamefont
  {Yuan}}, \bibinfo {author} {\bibfnamefont {T.}~\bibnamefont {Koretsune}},
  \bibinfo {author} {\bibfnamefont {M.}~\bibnamefont {Ochi}}, \bibinfo {author}
  {\bibfnamefont {K.}~\bibnamefont {Kuroki}},\ and\ \bibinfo {author}
  {\bibfnamefont {L.}~\bibnamefont {Fu}},\ }\bibfield  {title} {\bibinfo
  {title} {Maximally localized wannier orbitals and the extended hubbard model
  for twisted bilayer graphene},\ }\href
  {https://doi.org/10.1103/PhysRevX.8.031087} {\bibfield  {journal} {\bibinfo
  {journal} {Phys. Rev. X}\ }\textbf {\bibinfo {volume} {8}},\ \bibinfo {pages}
  {031087} (\bibinfo {year} {2018})}\BibitemShut {NoStop}%
\bibitem [{\citenamefont {Yankowitz}\ \emph {et~al.}(2019)\citenamefont
  {Yankowitz}, \citenamefont {Chen}, \citenamefont {Polshyn}, \citenamefont
  {Zhang}, \citenamefont {Watanabe}, \citenamefont {Taniguchi}, \citenamefont
  {Graf}, \citenamefont {Young},\ and\ \citenamefont {Dean}}]{yankowitz2019}%
  \BibitemOpen
  \bibfield  {author} {\bibinfo {author} {\bibfnamefont {M.}~\bibnamefont
  {Yankowitz}}, \bibinfo {author} {\bibfnamefont {S.}~\bibnamefont {Chen}},
  \bibinfo {author} {\bibfnamefont {H.}~\bibnamefont {Polshyn}}, \bibinfo
  {author} {\bibfnamefont {Y.}~\bibnamefont {Zhang}}, \bibinfo {author}
  {\bibfnamefont {K.}~\bibnamefont {Watanabe}}, \bibinfo {author}
  {\bibfnamefont {T.}~\bibnamefont {Taniguchi}}, \bibinfo {author}
  {\bibfnamefont {D.}~\bibnamefont {Graf}}, \bibinfo {author} {\bibfnamefont
  {A.~F.}\ \bibnamefont {Young}},\ and\ \bibinfo {author} {\bibfnamefont
  {C.~R.}\ \bibnamefont {Dean}},\ }\bibfield  {title} {\bibinfo {title} {Tuning
  superconductivity in twisted bilayer graphene},\ }\href
  {https://doi.org/10.1126/science.aav1910} {\bibfield  {journal} {\bibinfo
  {journal} {Science}\ }\textbf {\bibinfo {volume} {363}},\ \bibinfo {pages}
  {1059} (\bibinfo {year} {2019})}\BibitemShut {NoStop}%
\bibitem [{\citenamefont {Wu}\ \emph {et~al.}(2019)\citenamefont {Wu},
  \citenamefont {Lovorn}, \citenamefont {Tutuc}, \citenamefont {Martin},\ and\
  \citenamefont {MacDonald}}]{wu2019}%
  \BibitemOpen
  \bibfield  {author} {\bibinfo {author} {\bibfnamefont {F.}~\bibnamefont
  {Wu}}, \bibinfo {author} {\bibfnamefont {T.}~\bibnamefont {Lovorn}}, \bibinfo
  {author} {\bibfnamefont {E.}~\bibnamefont {Tutuc}}, \bibinfo {author}
  {\bibfnamefont {I.}~\bibnamefont {Martin}},\ and\ \bibinfo {author}
  {\bibfnamefont {A.~H.}\ \bibnamefont {MacDonald}},\ }\bibfield  {title}
  {\bibinfo {title} {Topological insulators in twisted transition metal
  dichalcogenide homobilayers},\ }\href
  {https://doi.org/10.1103/PhysRevLett.122.086402} {\bibfield  {journal}
  {\bibinfo  {journal} {Phys. Rev. Lett.}\ }\textbf {\bibinfo {volume} {122}},\
  \bibinfo {pages} {086402} (\bibinfo {year} {2019})}\BibitemShut {NoStop}%
\bibitem [{\citenamefont {Yu}\ \emph {et~al.}(2019)\citenamefont {Yu},
  \citenamefont {Chen},\ and\ \citenamefont {Yao}}]{yu2019}%
  \BibitemOpen
  \bibfield  {author} {\bibinfo {author} {\bibfnamefont {H.}~\bibnamefont
  {Yu}}, \bibinfo {author} {\bibfnamefont {M.}~\bibnamefont {Chen}},\ and\
  \bibinfo {author} {\bibfnamefont {W.}~\bibnamefont {Yao}},\ }\bibfield
  {title} {\bibinfo {title} {Giant magnetic field from moir{\'e} induced berry
  phase in homobilayer semiconductors},\ }\href
  {https://doi.org/10.1093/nsr/nwz117} {\bibfield  {journal} {\bibinfo
  {journal} {National Science Review}\ }\textbf {\bibinfo {volume} {7}},\
  \bibinfo {pages} {12} (\bibinfo {year} {2019})}\BibitemShut {NoStop}%
\bibitem [{\citenamefont {Carr}\ \emph {et~al.}(2020)\citenamefont {Carr},
  \citenamefont {Fang},\ and\ \citenamefont {Kaxiras}}]{carr2020}%
  \BibitemOpen
  \bibfield  {author} {\bibinfo {author} {\bibfnamefont {S.}~\bibnamefont
  {Carr}}, \bibinfo {author} {\bibfnamefont {S.}~\bibnamefont {Fang}},\ and\
  \bibinfo {author} {\bibfnamefont {E.}~\bibnamefont {Kaxiras}},\ }\bibfield
  {title} {\bibinfo {title} {Electronic-structure methods for twisted moir{\'e}
  layers},\ }\href@noop {} {\bibfield  {journal} {\bibinfo  {journal} {Nature
  Reviews Materials}\ }\textbf {\bibinfo {volume} {5}},\ \bibinfo {pages} {748}
  (\bibinfo {year} {2020})}\BibitemShut {NoStop}%
\bibitem [{\citenamefont {Yasuda}\ \emph {et~al.}(2021)\citenamefont {Yasuda},
  \citenamefont {Wang}, \citenamefont {Watanabe}, \citenamefont {Taniguchi},\
  and\ \citenamefont {Jarillo-Herrero}}]{yasuda2021}%
  \BibitemOpen
  \bibfield  {author} {\bibinfo {author} {\bibfnamefont {K.}~\bibnamefont
  {Yasuda}}, \bibinfo {author} {\bibfnamefont {X.}~\bibnamefont {Wang}},
  \bibinfo {author} {\bibfnamefont {K.}~\bibnamefont {Watanabe}}, \bibinfo
  {author} {\bibfnamefont {T.}~\bibnamefont {Taniguchi}},\ and\ \bibinfo
  {author} {\bibfnamefont {P.}~\bibnamefont {Jarillo-Herrero}},\ }\bibfield
  {title} {\bibinfo {title} {Stacking-engineered ferroelectricity in bilayer
  boron nitride},\ }\href@noop {} {\bibfield  {journal} {\bibinfo  {journal}
  {Science}\ }\textbf {\bibinfo {volume} {372}},\ \bibinfo {pages} {1458}
  (\bibinfo {year} {2021})}\BibitemShut {NoStop}%
\bibitem [{\citenamefont {Xie}\ \emph {et~al.}(2021)\citenamefont {Xie},
  \citenamefont {Pierce}, \citenamefont {Park}, \citenamefont {Parker},
  \citenamefont {Khalaf}, \citenamefont {Ledwith}, \citenamefont {Cao},
  \citenamefont {Lee}, \citenamefont {Chen}, \citenamefont {Forrester} \emph
  {et~al.}}]{xie2021}%
  \BibitemOpen
  \bibfield  {author} {\bibinfo {author} {\bibfnamefont {Y.}~\bibnamefont
  {Xie}}, \bibinfo {author} {\bibfnamefont {A.~T.}\ \bibnamefont {Pierce}},
  \bibinfo {author} {\bibfnamefont {J.~M.}\ \bibnamefont {Park}}, \bibinfo
  {author} {\bibfnamefont {D.~E.}\ \bibnamefont {Parker}}, \bibinfo {author}
  {\bibfnamefont {E.}~\bibnamefont {Khalaf}}, \bibinfo {author} {\bibfnamefont
  {P.}~\bibnamefont {Ledwith}}, \bibinfo {author} {\bibfnamefont
  {Y.}~\bibnamefont {Cao}}, \bibinfo {author} {\bibfnamefont {S.~H.}\
  \bibnamefont {Lee}}, \bibinfo {author} {\bibfnamefont {S.}~\bibnamefont
  {Chen}}, \bibinfo {author} {\bibfnamefont {P.~R.}\ \bibnamefont {Forrester}},
  \emph {et~al.},\ }\bibfield  {title} {\bibinfo {title} {Fractional chern
  insulators in magic-angle twisted bilayer graphene},\ }\href@noop {}
  {\bibfield  {journal} {\bibinfo  {journal} {Nature}\ }\textbf {\bibinfo
  {volume} {600}},\ \bibinfo {pages} {439} (\bibinfo {year}
  {2021})}\BibitemShut {NoStop}%
\bibitem [{\citenamefont {Park}\ \emph {et~al.}(2021)\citenamefont {Park},
  \citenamefont {Cao}, \citenamefont {Watanabe}, \citenamefont {Taniguchi},\
  and\ \citenamefont {Jarillo-Herrero}}]{park2021}%
  \BibitemOpen
  \bibfield  {author} {\bibinfo {author} {\bibfnamefont {J.~M.}\ \bibnamefont
  {Park}}, \bibinfo {author} {\bibfnamefont {Y.}~\bibnamefont {Cao}}, \bibinfo
  {author} {\bibfnamefont {K.}~\bibnamefont {Watanabe}}, \bibinfo {author}
  {\bibfnamefont {T.}~\bibnamefont {Taniguchi}},\ and\ \bibinfo {author}
  {\bibfnamefont {P.}~\bibnamefont {Jarillo-Herrero}},\ }\bibfield  {title}
  {\bibinfo {title} {Tunable strongly coupled superconductivity in magic-angle
  twisted trilayer graphene},\ }\href@noop {} {\bibfield  {journal} {\bibinfo
  {journal} {Nature}\ }\textbf {\bibinfo {volume} {590}},\ \bibinfo {pages}
  {249} (\bibinfo {year} {2021})}\BibitemShut {NoStop}%
\bibitem [{\citenamefont {Cai}\ \emph {et~al.}(2023)\citenamefont {Cai},
  \citenamefont {Anderson}, \citenamefont {Wang}, \citenamefont {Zhang},
  \citenamefont {Liu}, \citenamefont {Holtzmann}, \citenamefont {Zhang},
  \citenamefont {Fan}, \citenamefont {Taniguchi}, \citenamefont {Watanabe}
  \emph {et~al.}}]{cai2023signatures}%
  \BibitemOpen
  \bibfield  {author} {\bibinfo {author} {\bibfnamefont {J.}~\bibnamefont
  {Cai}}, \bibinfo {author} {\bibfnamefont {E.}~\bibnamefont {Anderson}},
  \bibinfo {author} {\bibfnamefont {C.}~\bibnamefont {Wang}}, \bibinfo {author}
  {\bibfnamefont {X.}~\bibnamefont {Zhang}}, \bibinfo {author} {\bibfnamefont
  {X.}~\bibnamefont {Liu}}, \bibinfo {author} {\bibfnamefont {W.}~\bibnamefont
  {Holtzmann}}, \bibinfo {author} {\bibfnamefont {Y.}~\bibnamefont {Zhang}},
  \bibinfo {author} {\bibfnamefont {F.}~\bibnamefont {Fan}}, \bibinfo {author}
  {\bibfnamefont {T.}~\bibnamefont {Taniguchi}}, \bibinfo {author}
  {\bibfnamefont {K.}~\bibnamefont {Watanabe}}, \emph {et~al.},\ }\bibfield
  {title} {\bibinfo {title} {Signatures of fractional quantum anomalous hall
  states in twisted {MoTe$_2$}},\ }\href@noop {} {\bibfield  {journal}
  {\bibinfo  {journal} {Nature}\ }\textbf {\bibinfo {volume} {622}},\ \bibinfo
  {pages} {63} (\bibinfo {year} {2023})}\BibitemShut {NoStop}%
\bibitem [{\citenamefont {Lu}\ \emph {et~al.}(2024)\citenamefont {Lu},
  \citenamefont {Han}, \citenamefont {Yao}, \citenamefont {Reddy},
  \citenamefont {Yang}, \citenamefont {Seo}, \citenamefont {Watanabe},
  \citenamefont {Taniguchi}, \citenamefont {Fu},\ and\ \citenamefont
  {Ju}}]{lu2024}%
  \BibitemOpen
  \bibfield  {author} {\bibinfo {author} {\bibfnamefont {Z.}~\bibnamefont
  {Lu}}, \bibinfo {author} {\bibfnamefont {T.}~\bibnamefont {Han}}, \bibinfo
  {author} {\bibfnamefont {Y.}~\bibnamefont {Yao}}, \bibinfo {author}
  {\bibfnamefont {A.~P.}\ \bibnamefont {Reddy}}, \bibinfo {author}
  {\bibfnamefont {J.}~\bibnamefont {Yang}}, \bibinfo {author} {\bibfnamefont
  {J.}~\bibnamefont {Seo}}, \bibinfo {author} {\bibfnamefont {K.}~\bibnamefont
  {Watanabe}}, \bibinfo {author} {\bibfnamefont {T.}~\bibnamefont {Taniguchi}},
  \bibinfo {author} {\bibfnamefont {L.}~\bibnamefont {Fu}},\ and\ \bibinfo
  {author} {\bibfnamefont {L.}~\bibnamefont {Ju}},\ }\bibfield  {title}
  {\bibinfo {title} {Fractional quantum anomalous hall effect in multilayer
  graphene},\ }\href@noop {} {\bibfield  {journal} {\bibinfo  {journal}
  {Nature}\ }\textbf {\bibinfo {volume} {626}},\ \bibinfo {pages} {759}
  (\bibinfo {year} {2024})}\BibitemShut {NoStop}%
\bibitem [{\citenamefont {Xia}\ \emph {et~al.}(2025)\citenamefont {Xia},
  \citenamefont {Han}, \citenamefont {Watanabe}, \citenamefont {Taniguchi},
  \citenamefont {Shan},\ and\ \citenamefont {Mak}}]{xia2025}%
  \BibitemOpen
  \bibfield  {author} {\bibinfo {author} {\bibfnamefont {Y.}~\bibnamefont
  {Xia}}, \bibinfo {author} {\bibfnamefont {Z.}~\bibnamefont {Han}}, \bibinfo
  {author} {\bibfnamefont {K.}~\bibnamefont {Watanabe}}, \bibinfo {author}
  {\bibfnamefont {T.}~\bibnamefont {Taniguchi}}, \bibinfo {author}
  {\bibfnamefont {J.}~\bibnamefont {Shan}},\ and\ \bibinfo {author}
  {\bibfnamefont {K.~F.}\ \bibnamefont {Mak}},\ }\bibfield  {title} {\bibinfo
  {title} {Superconductivity in twisted bilayer wse2},\ }\href@noop {}
  {\bibfield  {journal} {\bibinfo  {journal} {Nature}\ }\textbf {\bibinfo
  {volume} {637}},\ \bibinfo {pages} {833} (\bibinfo {year}
  {2025})}\BibitemShut {NoStop}%
\bibitem [{\citenamefont {Guo}\ \emph {et~al.}(2025)\citenamefont {Guo},
  \citenamefont {Pack}, \citenamefont {Swann}, \citenamefont {Holtzman},
  \citenamefont {Cothrine}, \citenamefont {Watanabe}, \citenamefont
  {Taniguchi}, \citenamefont {Mandrus}, \citenamefont {Barmak}, \citenamefont
  {Hone} \emph {et~al.}}]{guo2025}%
  \BibitemOpen
  \bibfield  {author} {\bibinfo {author} {\bibfnamefont {Y.}~\bibnamefont
  {Guo}}, \bibinfo {author} {\bibfnamefont {J.}~\bibnamefont {Pack}}, \bibinfo
  {author} {\bibfnamefont {J.}~\bibnamefont {Swann}}, \bibinfo {author}
  {\bibfnamefont {L.}~\bibnamefont {Holtzman}}, \bibinfo {author}
  {\bibfnamefont {M.}~\bibnamefont {Cothrine}}, \bibinfo {author}
  {\bibfnamefont {K.}~\bibnamefont {Watanabe}}, \bibinfo {author}
  {\bibfnamefont {T.}~\bibnamefont {Taniguchi}}, \bibinfo {author}
  {\bibfnamefont {D.~G.}\ \bibnamefont {Mandrus}}, \bibinfo {author}
  {\bibfnamefont {K.}~\bibnamefont {Barmak}}, \bibinfo {author} {\bibfnamefont
  {J.}~\bibnamefont {Hone}}, \emph {et~al.},\ }\bibfield  {title} {\bibinfo
  {title} {Superconductivity in 5.0° twisted bilayer wse2},\ }\href@noop {}
  {\bibfield  {journal} {\bibinfo  {journal} {Nature}\ }\textbf {\bibinfo
  {volume} {637}},\ \bibinfo {pages} {839} (\bibinfo {year}
  {2025})}\BibitemShut {NoStop}%
\bibitem [{\citenamefont {Shallcross}\ \emph {et~al.}(2010)\citenamefont
  {Shallcross}, \citenamefont {Sharma}, \citenamefont {Kandelaki},\ and\
  \citenamefont {Pankratov}}]{shallcross2010}%
  \BibitemOpen
  \bibfield  {author} {\bibinfo {author} {\bibfnamefont {S.}~\bibnamefont
  {Shallcross}}, \bibinfo {author} {\bibfnamefont {S.}~\bibnamefont {Sharma}},
  \bibinfo {author} {\bibfnamefont {E.}~\bibnamefont {Kandelaki}},\ and\
  \bibinfo {author} {\bibfnamefont {O.~A.}\ \bibnamefont {Pankratov}},\
  }\bibfield  {title} {\bibinfo {title} {Electronic structure of turbostratic
  graphene},\ }\href {https://doi.org/10.1103/PhysRevB.81.165105} {\bibfield
  {journal} {\bibinfo  {journal} {Phys. Rev. B}\ }\textbf {\bibinfo {volume}
  {81}},\ \bibinfo {pages} {165105} (\bibinfo {year} {2010})}\BibitemShut
  {NoStop}%
\bibitem [{\citenamefont {Kennes}\ \emph {et~al.}(2020)\citenamefont {Kennes},
  \citenamefont {Xian}, \citenamefont {Claassen},\ and\ \citenamefont
  {Rubio}}]{kennes2020}%
  \BibitemOpen
  \bibfield  {author} {\bibinfo {author} {\bibfnamefont {D.~M.}\ \bibnamefont
  {Kennes}}, \bibinfo {author} {\bibfnamefont {L.}~\bibnamefont {Xian}},
  \bibinfo {author} {\bibfnamefont {M.}~\bibnamefont {Claassen}},\ and\
  \bibinfo {author} {\bibfnamefont {A.}~\bibnamefont {Rubio}},\ }\bibfield
  {title} {\bibinfo {title} {One-dimensional flat bands in twisted bilayer
  germanium selenide},\ }\href@noop {} {\bibfield  {journal} {\bibinfo
  {journal} {Nature communications}\ }\textbf {\bibinfo {volume} {11}},\
  \bibinfo {pages} {1124} (\bibinfo {year} {2020})}\BibitemShut {NoStop}%
\bibitem [{\citenamefont {Soltero}\ \emph {et~al.}(2022)\citenamefont
  {Soltero}, \citenamefont {Guerrero-S{\'a}nchez}, \citenamefont {Mireles},\
  and\ \citenamefont {Ruiz-Tijerina}}]{soltero2022}%
  \BibitemOpen
  \bibfield  {author} {\bibinfo {author} {\bibfnamefont {I.}~\bibnamefont
  {Soltero}}, \bibinfo {author} {\bibfnamefont {J.}~\bibnamefont
  {Guerrero-S{\'a}nchez}}, \bibinfo {author} {\bibfnamefont {F.}~\bibnamefont
  {Mireles}},\ and\ \bibinfo {author} {\bibfnamefont {D.~A.}\ \bibnamefont
  {Ruiz-Tijerina}},\ }\bibfield  {title} {\bibinfo {title} {Moir{\'e} band
  structures of twisted phosphorene bilayers},\ }\href@noop {} {\bibfield
  {journal} {\bibinfo  {journal} {Phys. Rev. B}\ }\textbf {\bibinfo {volume}
  {105}},\ \bibinfo {pages} {235421} (\bibinfo {year} {2022})}\BibitemShut
  {NoStop}%
\bibitem [{\citenamefont {Wang}\ \emph {et~al.}(2022)\citenamefont {Wang},
  \citenamefont {Yu}, \citenamefont {Kwan}, \citenamefont {Jia}, \citenamefont
  {Lei}, \citenamefont {Klemenz}, \citenamefont {Cevallos}, \citenamefont
  {Singha}, \citenamefont {Devakul}, \citenamefont {Watanabe} \emph
  {et~al.}}]{wang2022}%
  \BibitemOpen
  \bibfield  {author} {\bibinfo {author} {\bibfnamefont {P.}~\bibnamefont
  {Wang}}, \bibinfo {author} {\bibfnamefont {G.}~\bibnamefont {Yu}}, \bibinfo
  {author} {\bibfnamefont {Y.~H.}\ \bibnamefont {Kwan}}, \bibinfo {author}
  {\bibfnamefont {Y.}~\bibnamefont {Jia}}, \bibinfo {author} {\bibfnamefont
  {S.}~\bibnamefont {Lei}}, \bibinfo {author} {\bibfnamefont {S.}~\bibnamefont
  {Klemenz}}, \bibinfo {author} {\bibfnamefont {F.~A.}\ \bibnamefont
  {Cevallos}}, \bibinfo {author} {\bibfnamefont {R.}~\bibnamefont {Singha}},
  \bibinfo {author} {\bibfnamefont {T.}~\bibnamefont {Devakul}}, \bibinfo
  {author} {\bibfnamefont {K.}~\bibnamefont {Watanabe}}, \emph {et~al.},\
  }\bibfield  {title} {\bibinfo {title} {One-dimensional luttinger liquids in a
  two-dimensional moir{\'e} lattice},\ }\href@noop {} {\bibfield  {journal}
  {\bibinfo  {journal} {Nature}\ }\textbf {\bibinfo {volume} {605}},\ \bibinfo
  {pages} {57} (\bibinfo {year} {2022})}\BibitemShut {NoStop}%
\bibitem [{\citenamefont {Yuan}\ \emph {et~al.}(2023)\citenamefont {Yuan},
  \citenamefont {Jia}, \citenamefont {Cheng}, \citenamefont {Singha},
  \citenamefont {Lei}, \citenamefont {Yao}, \citenamefont {Wu},\ and\
  \citenamefont {Schoop}}]{fyuan2023}%
  \BibitemOpen
  \bibfield  {author} {\bibinfo {author} {\bibfnamefont {F.}~\bibnamefont
  {Yuan}}, \bibinfo {author} {\bibfnamefont {Y.}~\bibnamefont {Jia}}, \bibinfo
  {author} {\bibfnamefont {G.}~\bibnamefont {Cheng}}, \bibinfo {author}
  {\bibfnamefont {R.}~\bibnamefont {Singha}}, \bibinfo {author} {\bibfnamefont
  {S.}~\bibnamefont {Lei}}, \bibinfo {author} {\bibfnamefont {N.}~\bibnamefont
  {Yao}}, \bibinfo {author} {\bibfnamefont {S.}~\bibnamefont {Wu}},\ and\
  \bibinfo {author} {\bibfnamefont {L.~M.}\ \bibnamefont {Schoop}},\ }\bibfield
   {title} {\bibinfo {title} {Atomic resolution imaging of highly air-sensitive
  monolayer and twisted-bilayer {WTe$_2$}},\ }\href@noop {} {\bibfield
  {journal} {\bibinfo  {journal} {Nano Letters}\ }\textbf {\bibinfo {volume}
  {23}},\ \bibinfo {pages} {6868} (\bibinfo {year} {2023})}\BibitemShut
  {NoStop}%
\bibitem [{\citenamefont {Magorrian}\ and\ \citenamefont
  {Hine}(2024)}]{magorrian2024}%
  \BibitemOpen
  \bibfield  {author} {\bibinfo {author} {\bibfnamefont {S.~J.}\ \bibnamefont
  {Magorrian}}\ and\ \bibinfo {author} {\bibfnamefont {N.~D.}\ \bibnamefont
  {Hine}},\ }\bibfield  {title} {\bibinfo {title} {Strain-dependent
  one-dimensional confinement channels in twisted bilayer {1T$'$-WTe$_2$}},\
  }\href@noop {} {\bibfield  {journal} {\bibinfo  {journal} {Phys. Rev. B}\
  }\textbf {\bibinfo {volume} {110}},\ \bibinfo {pages} {045410} (\bibinfo
  {year} {2024})}\BibitemShut {NoStop}%
\bibitem [{\citenamefont {Yang}\ \emph {et~al.}(2025)\citenamefont {Yang},
  \citenamefont {Zhang}, \citenamefont {Chen}, \citenamefont {Aso},
  \citenamefont {Yamamori}, \citenamefont {Moriya}, \citenamefont {Watanabe},
  \citenamefont {Taniguchi}, \citenamefont {Sasagawa}, \citenamefont
  {Nakatsuji}, \citenamefont {Koshino}, \citenamefont {Yamada-Takamura},
  \citenamefont {Oshima},\ and\ \citenamefont {Machida}}]{yang2025}%
  \BibitemOpen
  \bibfield  {author} {\bibinfo {author} {\bibfnamefont {X.}~\bibnamefont
  {Yang}}, \bibinfo {author} {\bibfnamefont {Y.}~\bibnamefont {Zhang}},
  \bibinfo {author} {\bibfnamefont {L.}~\bibnamefont {Chen}}, \bibinfo {author}
  {\bibfnamefont {K.}~\bibnamefont {Aso}}, \bibinfo {author} {\bibfnamefont
  {W.}~\bibnamefont {Yamamori}}, \bibinfo {author} {\bibfnamefont
  {R.}~\bibnamefont {Moriya}}, \bibinfo {author} {\bibfnamefont
  {K.}~\bibnamefont {Watanabe}}, \bibinfo {author} {\bibfnamefont
  {T.}~\bibnamefont {Taniguchi}}, \bibinfo {author} {\bibfnamefont
  {T.}~\bibnamefont {Sasagawa}}, \bibinfo {author} {\bibfnamefont
  {N.}~\bibnamefont {Nakatsuji}}, \bibinfo {author} {\bibfnamefont
  {M.}~\bibnamefont {Koshino}}, \bibinfo {author} {\bibfnamefont
  {Y.}~\bibnamefont {Yamada-Takamura}}, \bibinfo {author} {\bibfnamefont
  {Y.}~\bibnamefont {Oshima}},\ and\ \bibinfo {author} {\bibfnamefont
  {T.}~\bibnamefont {Machida}},\ }\bibfield  {title} {\bibinfo {title}
  {Intrinsic one-dimensional moir{\'e} superlattice in large-angle twisted
  bilayer {WTe$_2$}},\ }\href {https://doi.org/10.1021/acsnano.4c17317}
  {\bibfield  {journal} {\bibinfo  {journal} {ACS Nano}\ }\textbf {\bibinfo
  {volume} {19}},\ \bibinfo {pages} {13007} (\bibinfo {year} {2025})},\
  \bibinfo {note} {pMID: 40145593}\BibitemShut {NoStop}%
\bibitem [{\citenamefont {An}\ \emph {et~al.}(2025)\citenamefont {An},
  \citenamefont {Zhang}, \citenamefont {Xu}, \citenamefont {Guo}, \citenamefont
  {Rehman}, \citenamefont {Kennes}, \citenamefont {Rubio}, \citenamefont
  {Wang},\ and\ \citenamefont {Xian}}]{an2025}%
  \BibitemOpen
  \bibfield  {author} {\bibinfo {author} {\bibfnamefont {D.}~\bibnamefont
  {An}}, \bibinfo {author} {\bibfnamefont {T.}~\bibnamefont {Zhang}}, \bibinfo
  {author} {\bibfnamefont {Q.}~\bibnamefont {Xu}}, \bibinfo {author}
  {\bibfnamefont {H.}~\bibnamefont {Guo}}, \bibinfo {author} {\bibfnamefont
  {M.~U.}\ \bibnamefont {Rehman}}, \bibinfo {author} {\bibfnamefont {D.~M.}\
  \bibnamefont {Kennes}}, \bibinfo {author} {\bibfnamefont {A.}~\bibnamefont
  {Rubio}}, \bibinfo {author} {\bibfnamefont {L.}~\bibnamefont {Wang}},\ and\
  \bibinfo {author} {\bibfnamefont {L.}~\bibnamefont {Xian}},\ }\bibfield
  {title} {\bibinfo {title} {Critical angles and one-dimensional moir{\'e}
  physics in twisted rectangular lattices},\ }\href@noop {} {\bibfield
  {journal} {\bibinfo  {journal} {arXiv preprint arXiv:2507.14435}\ } (\bibinfo
  {year} {2025})}\BibitemShut {NoStop}%
\bibitem [{\citenamefont {Dr{\'o}{\.z}d{\.z}}\ \emph
  {et~al.}(2024)\citenamefont {Dr{\'o}{\.z}d{\.z}}, \citenamefont
  {Go{\l}{\k{e}}biowski},\ and\ \citenamefont {Zdyb}}]{drozdz2024}%
  \BibitemOpen
  \bibfield  {author} {\bibinfo {author} {\bibfnamefont {P.}~\bibnamefont
  {Dr{\'o}{\.z}d{\.z}}}, \bibinfo {author} {\bibfnamefont {M.}~\bibnamefont
  {Go{\l}{\k{e}}biowski}},\ and\ \bibinfo {author} {\bibfnamefont
  {R.}~\bibnamefont {Zdyb}},\ }\bibfield  {title} {\bibinfo {title} {Quasi-1d
  moir{\'e} superlattices in self-twisted two-allotropic antimonene
  heterostructures},\ }\href@noop {} {\bibfield  {journal} {\bibinfo  {journal}
  {Nanoscale}\ }\textbf {\bibinfo {volume} {16}},\ \bibinfo {pages} {15960}
  (\bibinfo {year} {2024})}\BibitemShut {NoStop}%
\bibitem [{\citenamefont {de~Sousa}\ \emph {et~al.}(2026)\citenamefont
  {de~Sousa}, \citenamefont {Lee}, \citenamefont {Guinea},\ and\ \citenamefont
  {Low}}]{sousa2025}%
  \BibitemOpen
  \bibfield  {author} {\bibinfo {author} {\bibfnamefont {D.~J.}\ \bibnamefont
  {de~Sousa}}, \bibinfo {author} {\bibfnamefont {S.}~\bibnamefont {Lee}},
  \bibinfo {author} {\bibfnamefont {F.}~\bibnamefont {Guinea}},\ and\ \bibinfo
  {author} {\bibfnamefont {T.}~\bibnamefont {Low}},\ }\bibfield  {title}
  {\bibinfo {title} {Moir{\'e} collapse and luttinger liquids in twisted
  anisotropic homobilayers},\ }\href@noop {} {\bibfield  {journal} {\bibinfo
  {journal} {Proc. Natl. Acad. Sci. USA}\ }\textbf {\bibinfo {volume} {123}},\
  \bibinfo {pages} {e2527371123} (\bibinfo {year} {2026})}\BibitemShut
  {NoStop}%
\bibitem [{\citenamefont {Sinner}\ \emph {et~al.}(2023)\citenamefont {Sinner},
  \citenamefont {Pantale{\'o}n},\ and\ \citenamefont {Guinea}}]{sinner2023}%
  \BibitemOpen
  \bibfield  {author} {\bibinfo {author} {\bibfnamefont {A.}~\bibnamefont
  {Sinner}}, \bibinfo {author} {\bibfnamefont {P.~A.}\ \bibnamefont
  {Pantale{\'o}n}},\ and\ \bibinfo {author} {\bibfnamefont {F.}~\bibnamefont
  {Guinea}},\ }\bibfield  {title} {\bibinfo {title} {Strain-induced quasi-1d
  channels in twisted moir{\'e} lattices},\ }\href@noop {} {\bibfield
  {journal} {\bibinfo  {journal} {Phys. Rev. Lett.}\ }\textbf {\bibinfo
  {volume} {131}},\ \bibinfo {pages} {166402} (\bibinfo {year}
  {2023})}\BibitemShut {NoStop}%
\bibitem [{\citenamefont {Escudero}\ \emph {et~al.}(2024)\citenamefont
  {Escudero}, \citenamefont {Sinner}, \citenamefont {Zhan}, \citenamefont
  {Pantale{\'o}n},\ and\ \citenamefont {Guinea}}]{escudero2024}%
  \BibitemOpen
  \bibfield  {author} {\bibinfo {author} {\bibfnamefont {F.}~\bibnamefont
  {Escudero}}, \bibinfo {author} {\bibfnamefont {A.}~\bibnamefont {Sinner}},
  \bibinfo {author} {\bibfnamefont {Z.}~\bibnamefont {Zhan}}, \bibinfo {author}
  {\bibfnamefont {P.~A.}\ \bibnamefont {Pantale{\'o}n}},\ and\ \bibinfo
  {author} {\bibfnamefont {F.}~\bibnamefont {Guinea}},\ }\bibfield  {title}
  {\bibinfo {title} {Designing moir{\'e} patterns by strain},\ }\href@noop {}
  {\bibfield  {journal} {\bibinfo  {journal} {Physical Review Research}\
  }\textbf {\bibinfo {volume} {6}},\ \bibinfo {pages} {023203} (\bibinfo {year}
  {2024})}\BibitemShut {NoStop}%
\bibitem [{\citenamefont {Hesp}\ \emph {et~al.}(2024)\citenamefont {Hesp},
  \citenamefont {Batlle-Porro}, \citenamefont {Krishna~Kumar}, \citenamefont
  {Agarwal}, \citenamefont {Barcons~Ruiz}, \citenamefont {Herzig~Sheinfux},
  \citenamefont {Watanabe}, \citenamefont {Taniguchi}, \citenamefont
  {Stepanov},\ and\ \citenamefont {Koppens}}]{hesp2024}%
  \BibitemOpen
  \bibfield  {author} {\bibinfo {author} {\bibfnamefont {N.~C.}\ \bibnamefont
  {Hesp}}, \bibinfo {author} {\bibfnamefont {S.}~\bibnamefont {Batlle-Porro}},
  \bibinfo {author} {\bibfnamefont {R.}~\bibnamefont {Krishna~Kumar}}, \bibinfo
  {author} {\bibfnamefont {H.}~\bibnamefont {Agarwal}}, \bibinfo {author}
  {\bibfnamefont {D.}~\bibnamefont {Barcons~Ruiz}}, \bibinfo {author}
  {\bibfnamefont {H.}~\bibnamefont {Herzig~Sheinfux}}, \bibinfo {author}
  {\bibfnamefont {K.}~\bibnamefont {Watanabe}}, \bibinfo {author}
  {\bibfnamefont {T.}~\bibnamefont {Taniguchi}}, \bibinfo {author}
  {\bibfnamefont {P.}~\bibnamefont {Stepanov}},\ and\ \bibinfo {author}
  {\bibfnamefont {F.~H.}\ \bibnamefont {Koppens}},\ }\bibfield  {title}
  {\bibinfo {title} {Cryogenic nano-imaging of second-order moir{\'e}
  superlattices},\ }\href@noop {} {\bibfield  {journal} {\bibinfo  {journal}
  {Nature Materials}\ }\textbf {\bibinfo {volume} {23}},\ \bibinfo {pages}
  {1664} (\bibinfo {year} {2024})}\BibitemShut {NoStop}%
\bibitem [{\citenamefont {Boi}\ \emph {et~al.}(2025)\citenamefont {Boi},
  \citenamefont {Odunmbaku}, \citenamefont {Taallah},\ and\ \citenamefont
  {Wang}}]{boi2025}%
  \BibitemOpen
  \bibfield  {author} {\bibinfo {author} {\bibfnamefont {F.~S.}\ \bibnamefont
  {Boi}}, \bibinfo {author} {\bibfnamefont {O.}~\bibnamefont {Odunmbaku}},
  \bibinfo {author} {\bibfnamefont {A.}~\bibnamefont {Taallah}},\ and\ \bibinfo
  {author} {\bibfnamefont {S.}~\bibnamefont {Wang}},\ }\bibfield  {title}
  {\bibinfo {title} {Quasi-1d, rectangular-like and hexagonal moir{\'e}
  superlattices in exfoliated highly oriented pyrolytic graphite},\ }\href@noop
  {} {\bibfield  {journal} {\bibinfo  {journal} {Diamond and Related
  Materials}\ }\textbf {\bibinfo {volume} {151}},\ \bibinfo {pages} {111843}
  (\bibinfo {year} {2025})}\BibitemShut {NoStop}%
\bibitem [{\citenamefont {Su}\ \emph {et~al.}(2025)\citenamefont {Su},
  \citenamefont {Gao}, \citenamefont {Chen}, \citenamefont {Farooq},
  \citenamefont {Xian},\ and\ \citenamefont {Huang}}]{su2025}%
  \BibitemOpen
  \bibfield  {author} {\bibinfo {author} {\bibfnamefont {L.}~\bibnamefont
  {Su}}, \bibinfo {author} {\bibfnamefont {Y.}~\bibnamefont {Gao}}, \bibinfo
  {author} {\bibfnamefont {Y.}~\bibnamefont {Chen}}, \bibinfo {author}
  {\bibfnamefont {M.~U.}\ \bibnamefont {Farooq}}, \bibinfo {author}
  {\bibfnamefont {L.}~\bibnamefont {Xian}},\ and\ \bibinfo {author}
  {\bibfnamefont {L.}~\bibnamefont {Huang}},\ }\bibfield  {title} {\bibinfo
  {title} {Domain-selective 1d moir{\'e} engineering and topological
  transitions in bilayer graphene},\ }\href@noop {} {\bibfield  {journal}
  {\bibinfo  {journal} {Nano Letters}\ }\textbf {\bibinfo {volume} {25}},\
  \bibinfo {pages} {14060} (\bibinfo {year} {2025})}\BibitemShut {NoStop}%
\bibitem [{\citenamefont {Alden}\ \emph {et~al.}(2013)\citenamefont {Alden},
  \citenamefont {Tsen}, \citenamefont {Huang}, \citenamefont {Hovden},
  \citenamefont {Brown}, \citenamefont {Park}, \citenamefont {Muller},\ and\
  \citenamefont {McEuen}}]{jonathan2013}%
  \BibitemOpen
  \bibfield  {author} {\bibinfo {author} {\bibfnamefont {J.~S.}\ \bibnamefont
  {Alden}}, \bibinfo {author} {\bibfnamefont {A.~W.}\ \bibnamefont {Tsen}},
  \bibinfo {author} {\bibfnamefont {P.~Y.}\ \bibnamefont {Huang}}, \bibinfo
  {author} {\bibfnamefont {R.}~\bibnamefont {Hovden}}, \bibinfo {author}
  {\bibfnamefont {L.}~\bibnamefont {Brown}}, \bibinfo {author} {\bibfnamefont
  {J.}~\bibnamefont {Park}}, \bibinfo {author} {\bibfnamefont {D.~A.}\
  \bibnamefont {Muller}},\ and\ \bibinfo {author} {\bibfnamefont {P.~L.}\
  \bibnamefont {McEuen}},\ }\bibfield  {title} {\bibinfo {title} {Strain
  solitons and topological defects in bilayer graphene},\ }\href
  {https://doi.org/10.1073/pnas.1309394110} {\bibfield  {journal} {\bibinfo
  {journal} {Proceedings of the National Academy of Sciences}\ }\textbf
  {\bibinfo {volume} {110}},\ \bibinfo {pages} {11256} (\bibinfo {year}
  {2013})}\BibitemShut {NoStop}%
\bibitem [{\citenamefont {Uchida}\ \emph {et~al.}(2014)\citenamefont {Uchida},
  \citenamefont {Furuya}, \citenamefont {Iwata},\ and\ \citenamefont
  {Oshiyama}}]{uchida2014}%
  \BibitemOpen
  \bibfield  {author} {\bibinfo {author} {\bibfnamefont {K.}~\bibnamefont
  {Uchida}}, \bibinfo {author} {\bibfnamefont {S.}~\bibnamefont {Furuya}},
  \bibinfo {author} {\bibfnamefont {J.-I.}\ \bibnamefont {Iwata}},\ and\
  \bibinfo {author} {\bibfnamefont {A.}~\bibnamefont {Oshiyama}},\ }\bibfield
  {title} {\bibinfo {title} {Atomic corrugation and electron localization due
  to moir{\'e} patterns in twisted bilayer graphenes},\ }\href
  {https://doi.org/10.1103/PhysRevB.90.155451} {\bibfield  {journal} {\bibinfo
  {journal} {Phys. Rev. B}\ }\textbf {\bibinfo {volume} {90}},\ \bibinfo
  {pages} {155451} (\bibinfo {year} {2014})}\BibitemShut {NoStop}%
\bibitem [{\citenamefont {San-Jose}\ \emph
  {et~al.}(2014{\natexlab{a}})\citenamefont {San-Jose}, \citenamefont
  {Guti\'errez-Rubio}, \citenamefont {Sturla},\ and\ \citenamefont
  {Guinea}}]{sanjose2014}%
  \BibitemOpen
  \bibfield  {author} {\bibinfo {author} {\bibfnamefont {P.}~\bibnamefont
  {San-Jose}}, \bibinfo {author} {\bibfnamefont {A.}~\bibnamefont
  {Guti\'errez-Rubio}}, \bibinfo {author} {\bibfnamefont {M.}~\bibnamefont
  {Sturla}},\ and\ \bibinfo {author} {\bibfnamefont {F.}~\bibnamefont
  {Guinea}},\ }\bibfield  {title} {\bibinfo {title} {Electronic structure of
  spontaneously strained graphene on hexagonal boron nitride},\ }\href
  {https://doi.org/10.1103/PhysRevB.90.115152} {\bibfield  {journal} {\bibinfo
  {journal} {Phys. Rev. B}\ }\textbf {\bibinfo {volume} {90}},\ \bibinfo
  {pages} {115152} (\bibinfo {year} {2014}{\natexlab{a}})}\BibitemShut
  {NoStop}%
\bibitem [{\citenamefont {San-Jose}\ \emph
  {et~al.}(2014{\natexlab{b}})\citenamefont {San-Jose}, \citenamefont
  {Guti\'errez-Rubio}, \citenamefont {Sturla},\ and\ \citenamefont
  {Guinea}}]{sanjose2014spontaneous}%
  \BibitemOpen
  \bibfield  {author} {\bibinfo {author} {\bibfnamefont {P.}~\bibnamefont
  {San-Jose}}, \bibinfo {author} {\bibfnamefont {A.}~\bibnamefont
  {Guti\'errez-Rubio}}, \bibinfo {author} {\bibfnamefont {M.}~\bibnamefont
  {Sturla}},\ and\ \bibinfo {author} {\bibfnamefont {F.}~\bibnamefont
  {Guinea}},\ }\bibfield  {title} {\bibinfo {title} {Spontaneous strains and
  gap in graphene on boron nitride},\ }\href@noop {} {\bibfield  {journal}
  {\bibinfo  {journal} {Physical Review B}\ }\textbf {\bibinfo {volume} {90}},\
  \bibinfo {pages} {075428} (\bibinfo {year} {2014}{\natexlab{b}})}\BibitemShut
  {NoStop}%
\bibitem [{\citenamefont {Jung}\ \emph {et~al.}(2015)\citenamefont {Jung},
  \citenamefont {DaSilva}, \citenamefont {MacDonald},\ and\ \citenamefont
  {Adam}}]{jung2015origin}%
  \BibitemOpen
  \bibfield  {author} {\bibinfo {author} {\bibfnamefont {J.}~\bibnamefont
  {Jung}}, \bibinfo {author} {\bibfnamefont {A.~M.}\ \bibnamefont {DaSilva}},
  \bibinfo {author} {\bibfnamefont {A.~H.}\ \bibnamefont {MacDonald}},\ and\
  \bibinfo {author} {\bibfnamefont {S.}~\bibnamefont {Adam}},\ }\bibfield
  {title} {\bibinfo {title} {Origin of band gaps in graphene on hexagonal boron
  nitride},\ }\href@noop {} {\bibfield  {journal} {\bibinfo  {journal} {Nat.
  Commun.}\ }\textbf {\bibinfo {volume} {6}},\ \bibinfo {pages} {6308}
  (\bibinfo {year} {2015})}\BibitemShut {NoStop}%
\bibitem [{\citenamefont {Wijk}\ \emph {et~al.}(2015)\citenamefont {Wijk},
  \citenamefont {Schuring}, \citenamefont {Katsnelson},\ and\ \citenamefont
  {Fasolino}}]{wijk2015}%
  \BibitemOpen
  \bibfield  {author} {\bibinfo {author} {\bibfnamefont {M.~v.}\ \bibnamefont
  {Wijk}}, \bibinfo {author} {\bibfnamefont {A.}~\bibnamefont {Schuring}},
  \bibinfo {author} {\bibfnamefont {M.}~\bibnamefont {Katsnelson}},\ and\
  \bibinfo {author} {\bibfnamefont {A.}~\bibnamefont {Fasolino}},\ }\bibfield
  {title} {\bibinfo {title} {Relaxation of moir{\'e} patterns for slightly
  misaligned identical lattices: graphene on graphite},\ }\href@noop {}
  {\bibfield  {journal} {\bibinfo  {journal} {2D Materials}\ }\textbf {\bibinfo
  {volume} {2}},\ \bibinfo {pages} {034010} (\bibinfo {year}
  {2015})}\BibitemShut {NoStop}%
\bibitem [{\citenamefont {Dai}\ \emph {et~al.}(2016)\citenamefont {Dai},
  \citenamefont {Xiang},\ and\ \citenamefont {Srolovitz}}]{dai2016twisted}%
  \BibitemOpen
  \bibfield  {author} {\bibinfo {author} {\bibfnamefont {S.}~\bibnamefont
  {Dai}}, \bibinfo {author} {\bibfnamefont {Y.}~\bibnamefont {Xiang}},\ and\
  \bibinfo {author} {\bibfnamefont {D.~J.}\ \bibnamefont {Srolovitz}},\
  }\bibfield  {title} {\bibinfo {title} {Twisted bilayer graphene: Moir{\'e}
  with a twist},\ }\href@noop {} {\bibfield  {journal} {\bibinfo  {journal}
  {Nano Lett.}\ }\textbf {\bibinfo {volume} {16}},\ \bibinfo {pages} {5923}
  (\bibinfo {year} {2016})}\BibitemShut {NoStop}%
\bibitem [{\citenamefont {Nam}\ and\ \citenamefont
  {Koshino}(2017{\natexlab{a}})}]{PhysRevB.96.075311}%
  \BibitemOpen
  \bibfield  {author} {\bibinfo {author} {\bibfnamefont {N.~N.~T.}\
  \bibnamefont {Nam}}\ and\ \bibinfo {author} {\bibfnamefont {M.}~\bibnamefont
  {Koshino}},\ }\bibfield  {title} {\bibinfo {title} {Lattice relaxation and
  energy band modulation in twisted bilayer graphene},\ }\href
  {https://doi.org/10.1103/PhysRevB.96.075311} {\bibfield  {journal} {\bibinfo
  {journal} {Phys. Rev. B}\ }\textbf {\bibinfo {volume} {96}},\ \bibinfo
  {pages} {075311} (\bibinfo {year} {2017}{\natexlab{a}})}\BibitemShut
  {NoStop}%
\bibitem [{\citenamefont {Jain}\ \emph {et~al.}(2017)\citenamefont {Jain},
  \citenamefont {Juri{\v{c}}i{\'c}},\ and\ \citenamefont
  {Barkema}}]{jain2017structure}%
  \BibitemOpen
  \bibfield  {author} {\bibinfo {author} {\bibfnamefont {S.~K.}\ \bibnamefont
  {Jain}}, \bibinfo {author} {\bibfnamefont {V.}~\bibnamefont
  {Juri{\v{c}}i{\'c}}},\ and\ \bibinfo {author} {\bibfnamefont {G.~T.}\
  \bibnamefont {Barkema}},\ }\bibfield  {title} {\bibinfo {title} {Structure of
  twisted and buckled bilayer graphene},\ }\href@noop {} {\bibfield  {journal}
  {\bibinfo  {journal} {2D Mater.}\ }\textbf {\bibinfo {volume} {4}},\ \bibinfo
  {pages} {015018} (\bibinfo {year} {2017})}\BibitemShut {NoStop}%
\bibitem [{\citenamefont {Lucignano}\ \emph {et~al.}(2019)\citenamefont
  {Lucignano}, \citenamefont {Alf{\`e}}, \citenamefont {Cataudella},
  \citenamefont {Ninno},\ and\ \citenamefont {Cantele}}]{lucignano2019}%
  \BibitemOpen
  \bibfield  {author} {\bibinfo {author} {\bibfnamefont {P.}~\bibnamefont
  {Lucignano}}, \bibinfo {author} {\bibfnamefont {D.}~\bibnamefont {Alf{\`e}}},
  \bibinfo {author} {\bibfnamefont {V.}~\bibnamefont {Cataudella}}, \bibinfo
  {author} {\bibfnamefont {D.}~\bibnamefont {Ninno}},\ and\ \bibinfo {author}
  {\bibfnamefont {G.}~\bibnamefont {Cantele}},\ }\bibfield  {title} {\bibinfo
  {title} {Crucial role of atomic corrugation on the flat bands and energy gaps
  of twisted bilayer graphene at the magic angle {$\theta\sim 1. 08^\circ$}},\
  }\href@noop {} {\bibfield  {journal} {\bibinfo  {journal} {Physical Review
  B}\ }\textbf {\bibinfo {volume} {99}},\ \bibinfo {pages} {195419} (\bibinfo
  {year} {2019})}\BibitemShut {NoStop}%
\bibitem [{\citenamefont {Yoo}\ \emph {et~al.}(2019)\citenamefont {Yoo},
  \citenamefont {Engelke}, \citenamefont {Carr}, \citenamefont {Fang},
  \citenamefont {Zhang}, \citenamefont {Cazeaux}, \citenamefont {Sung},
  \citenamefont {Hovden}, \citenamefont {Tsen}, \citenamefont {Taniguchi},
  \citenamefont {Watanabe}, \citenamefont {Kim}, \citenamefont {Mitchell},
  \citenamefont {Tadmor}, \citenamefont {Kaxiras},\ and\ \citenamefont
  {Kim}}]{yoo2019atomic}%
  \BibitemOpen
  \bibfield  {author} {\bibinfo {author} {\bibfnamefont {H.}~\bibnamefont
  {Yoo}}, \bibinfo {author} {\bibfnamefont {R.}~\bibnamefont {Engelke}},
  \bibinfo {author} {\bibfnamefont {S.}~\bibnamefont {Carr}}, \bibinfo {author}
  {\bibfnamefont {S.}~\bibnamefont {Fang}}, \bibinfo {author} {\bibfnamefont
  {K.}~\bibnamefont {Zhang}}, \bibinfo {author} {\bibfnamefont
  {P.}~\bibnamefont {Cazeaux}}, \bibinfo {author} {\bibfnamefont {S.~H.}\
  \bibnamefont {Sung}}, \bibinfo {author} {\bibfnamefont {R.}~\bibnamefont
  {Hovden}}, \bibinfo {author} {\bibfnamefont {A.~W.}\ \bibnamefont {Tsen}},
  \bibinfo {author} {\bibfnamefont {T.}~\bibnamefont {Taniguchi}}, \bibinfo
  {author} {\bibfnamefont {G.-C.}\ \bibnamefont {Watanabe}, \bibfnamefont
  {Kenji~Yi}}, \bibinfo {author} {\bibfnamefont {M.}~\bibnamefont {Kim}},
  \bibinfo {author} {\bibfnamefont {L.}~\bibnamefont {Mitchell}}, \bibinfo
  {author} {\bibfnamefont {E.~B.}\ \bibnamefont {Tadmor}}, \bibinfo {author}
  {\bibfnamefont {E.}~\bibnamefont {Kaxiras}},\ and\ \bibinfo {author}
  {\bibfnamefont {P.}~\bibnamefont {Kim}},\ }\bibfield  {title} {\bibinfo
  {title} {Atomic and electronic reconstruction at the van der waals interface
  in twisted bilayer graphene},\ }\href@noop {} {\bibfield  {journal} {\bibinfo
   {journal} {Nat. Mater.}\ }\textbf {\bibinfo {volume} {18}},\ \bibinfo
  {pages} {448} (\bibinfo {year} {2019})}\BibitemShut {NoStop}%
\bibitem [{\citenamefont {Guinea}\ and\ \citenamefont
  {Walet}(2019)}]{guinea2019}%
  \BibitemOpen
  \bibfield  {author} {\bibinfo {author} {\bibfnamefont {F.}~\bibnamefont
  {Guinea}}\ and\ \bibinfo {author} {\bibfnamefont {N.~R.}\ \bibnamefont
  {Walet}},\ }\bibfield  {title} {\bibinfo {title} {Continuum models for
  twisted bilayer graphene: Effect of lattice deformation and hopping
  parameters},\ }\href {https://doi.org/10.1103/PhysRevB.99.205134} {\bibfield
  {journal} {\bibinfo  {journal} {Phys. Rev. B}\ }\textbf {\bibinfo {volume}
  {99}},\ \bibinfo {pages} {205134} (\bibinfo {year} {2019})}\BibitemShut
  {NoStop}%
\bibitem [{\citenamefont {Cantele}\ \emph {et~al.}(2020)\citenamefont
  {Cantele}, \citenamefont {Alf{\`e}}, \citenamefont {Conte}, \citenamefont
  {Cataudella}, \citenamefont {Ninno},\ and\ \citenamefont
  {Lucignano}}]{cantele2020}%
  \BibitemOpen
  \bibfield  {author} {\bibinfo {author} {\bibfnamefont {G.}~\bibnamefont
  {Cantele}}, \bibinfo {author} {\bibfnamefont {D.}~\bibnamefont {Alf{\`e}}},
  \bibinfo {author} {\bibfnamefont {F.}~\bibnamefont {Conte}}, \bibinfo
  {author} {\bibfnamefont {V.}~\bibnamefont {Cataudella}}, \bibinfo {author}
  {\bibfnamefont {D.}~\bibnamefont {Ninno}},\ and\ \bibinfo {author}
  {\bibfnamefont {P.}~\bibnamefont {Lucignano}},\ }\bibfield  {title} {\bibinfo
  {title} {Structural relaxation and low-energy properties of twisted bilayer
  graphene},\ }\href@noop {} {\bibfield  {journal} {\bibinfo  {journal}
  {Physical Review Research}\ }\textbf {\bibinfo {volume} {2}},\ \bibinfo
  {pages} {043127} (\bibinfo {year} {2020})}\BibitemShut {NoStop}%
\bibitem [{\citenamefont {Weston}\ \emph {et~al.}(2020)\citenamefont {Weston},
  \citenamefont {Zou}, \citenamefont {Enaldiev}, \citenamefont {Summerfield},
  \citenamefont {Clark}, \citenamefont {Z{\'o}lyomi}, \citenamefont {Graham},
  \citenamefont {Yelgel}, \citenamefont {Magorrian}, \citenamefont {Zhou} \emph
  {et~al.}}]{weston2020}%
  \BibitemOpen
  \bibfield  {author} {\bibinfo {author} {\bibfnamefont {A.}~\bibnamefont
  {Weston}}, \bibinfo {author} {\bibfnamefont {Y.}~\bibnamefont {Zou}},
  \bibinfo {author} {\bibfnamefont {V.}~\bibnamefont {Enaldiev}}, \bibinfo
  {author} {\bibfnamefont {A.}~\bibnamefont {Summerfield}}, \bibinfo {author}
  {\bibfnamefont {N.}~\bibnamefont {Clark}}, \bibinfo {author} {\bibfnamefont
  {V.}~\bibnamefont {Z{\'o}lyomi}}, \bibinfo {author} {\bibfnamefont
  {A.}~\bibnamefont {Graham}}, \bibinfo {author} {\bibfnamefont
  {C.}~\bibnamefont {Yelgel}}, \bibinfo {author} {\bibfnamefont
  {S.}~\bibnamefont {Magorrian}}, \bibinfo {author} {\bibfnamefont
  {M.}~\bibnamefont {Zhou}}, \emph {et~al.},\ }\bibfield  {title} {\bibinfo
  {title} {Atomic reconstruction in twisted bilayers of transition metal
  dichalcogenides},\ }\href@noop {} {\bibfield  {journal} {\bibinfo  {journal}
  {Nature nanotechnology}\ }\textbf {\bibinfo {volume} {15}},\ \bibinfo {pages}
  {592} (\bibinfo {year} {2020})}\BibitemShut {NoStop}%
\bibitem [{\citenamefont {Rosenberger}\ \emph {et~al.}(2020)\citenamefont
  {Rosenberger}, \citenamefont {Chuang}, \citenamefont {Phillips},
  \citenamefont {Oleshko}, \citenamefont {McCreary}, \citenamefont {Sivaram},
  \citenamefont {Hellberg},\ and\ \citenamefont {Jonker}}]{rosenberger2020}%
  \BibitemOpen
  \bibfield  {author} {\bibinfo {author} {\bibfnamefont {M.~R.}\ \bibnamefont
  {Rosenberger}}, \bibinfo {author} {\bibfnamefont {H.-J.}\ \bibnamefont
  {Chuang}}, \bibinfo {author} {\bibfnamefont {M.}~\bibnamefont {Phillips}},
  \bibinfo {author} {\bibfnamefont {V.~P.}\ \bibnamefont {Oleshko}}, \bibinfo
  {author} {\bibfnamefont {K.~M.}\ \bibnamefont {McCreary}}, \bibinfo {author}
  {\bibfnamefont {S.~V.}\ \bibnamefont {Sivaram}}, \bibinfo {author}
  {\bibfnamefont {C.~S.}\ \bibnamefont {Hellberg}},\ and\ \bibinfo {author}
  {\bibfnamefont {B.~T.}\ \bibnamefont {Jonker}},\ }\bibfield  {title}
  {\bibinfo {title} {Twist angle-dependent atomic reconstruction and moir{\'e}
  patterns in transition metal dichalcogenide heterostructures},\ }\href
  {https://doi.org/10.1021/acsnano.0c00088} {\bibfield  {journal} {\bibinfo
  {journal} {ACS Nano}\ }\textbf {\bibinfo {volume} {14}},\ \bibinfo {pages}
  {4550} (\bibinfo {year} {2020})},\ \bibinfo {note} {pMID: 32167748},\ \Eprint
  {https://arxiv.org/abs/https://doi.org/10.1021/acsnano.0c00088}
  {https://doi.org/10.1021/acsnano.0c00088} \BibitemShut {NoStop}%
\bibitem [{\citenamefont {Sung}\ \emph {et~al.}(2020)\citenamefont {Sung},
  \citenamefont {Zhou}, \citenamefont {Scuri}, \citenamefont {Z{\'o}lyomi},
  \citenamefont {Andersen}, \citenamefont {Yoo}, \citenamefont {Wild},
  \citenamefont {Joe}, \citenamefont {Gelly}, \citenamefont {Heo} \emph
  {et~al.}}]{sung2020}%
  \BibitemOpen
  \bibfield  {author} {\bibinfo {author} {\bibfnamefont {J.}~\bibnamefont
  {Sung}}, \bibinfo {author} {\bibfnamefont {Y.}~\bibnamefont {Zhou}}, \bibinfo
  {author} {\bibfnamefont {G.}~\bibnamefont {Scuri}}, \bibinfo {author}
  {\bibfnamefont {V.}~\bibnamefont {Z{\'o}lyomi}}, \bibinfo {author}
  {\bibfnamefont {T.~I.}\ \bibnamefont {Andersen}}, \bibinfo {author}
  {\bibfnamefont {H.}~\bibnamefont {Yoo}}, \bibinfo {author} {\bibfnamefont
  {D.~S.}\ \bibnamefont {Wild}}, \bibinfo {author} {\bibfnamefont {A.~Y.}\
  \bibnamefont {Joe}}, \bibinfo {author} {\bibfnamefont {R.~J.}\ \bibnamefont
  {Gelly}}, \bibinfo {author} {\bibfnamefont {H.}~\bibnamefont {Heo}}, \emph
  {et~al.},\ }\bibfield  {title} {\bibinfo {title} {Broken mirror symmetry in
  excitonic response of reconstructed domains in twisted mose2/mose2
  bilayers},\ }\href@noop {} {\bibfield  {journal} {\bibinfo  {journal} {Nature
  nanotechnology}\ }\textbf {\bibinfo {volume} {15}},\ \bibinfo {pages} {750}
  (\bibinfo {year} {2020})}\BibitemShut {NoStop}%
\bibitem [{\citenamefont {Enaldiev}\ \emph {et~al.}(2020)\citenamefont
  {Enaldiev}, \citenamefont {Z{\'o}lyomi}, \citenamefont {Yelgel},
  \citenamefont {Magorrian},\ and\ \citenamefont {Fal’ko}}]{enaldiev2020}%
  \BibitemOpen
  \bibfield  {author} {\bibinfo {author} {\bibfnamefont {V.~V.}\ \bibnamefont
  {Enaldiev}}, \bibinfo {author} {\bibfnamefont {V.}~\bibnamefont
  {Z{\'o}lyomi}}, \bibinfo {author} {\bibfnamefont {C.}~\bibnamefont {Yelgel}},
  \bibinfo {author} {\bibfnamefont {S.}~\bibnamefont {Magorrian}},\ and\
  \bibinfo {author} {\bibfnamefont {V.}~\bibnamefont {Fal’ko}},\ }\bibfield
  {title} {\bibinfo {title} {Stacking domains and dislocation networks in
  marginally twisted bilayers of transition metal dichalcogenides},\
  }\href@noop {} {\bibfield  {journal} {\bibinfo  {journal} {Physical review
  letters}\ }\textbf {\bibinfo {volume} {124}},\ \bibinfo {pages} {206101}
  (\bibinfo {year} {2020})}\BibitemShut {NoStop}%
\bibitem [{\citenamefont {Koshino}\ and\ \citenamefont
  {Nam}(2020)}]{koshino2020}%
  \BibitemOpen
  \bibfield  {author} {\bibinfo {author} {\bibfnamefont {M.}~\bibnamefont
  {Koshino}}\ and\ \bibinfo {author} {\bibfnamefont {N.~N.~T.}\ \bibnamefont
  {Nam}},\ }\bibfield  {title} {\bibinfo {title} {Effective continuum model for
  relaxed twisted bilayer graphene and moir\'e electron-phonon interaction},\
  }\href {https://doi.org/10.1103/PhysRevB.101.195425} {\bibfield  {journal}
  {\bibinfo  {journal} {Phys. Rev. B}\ }\textbf {\bibinfo {volume} {101}},\
  \bibinfo {pages} {195425} (\bibinfo {year} {2020})}\BibitemShut {NoStop}%
\bibitem [{\citenamefont {Tsim}\ \emph {et~al.}(2020)\citenamefont {Tsim},
  \citenamefont {Nam},\ and\ \citenamefont {Koshino}}]{tsim2020}%
  \BibitemOpen
  \bibfield  {author} {\bibinfo {author} {\bibfnamefont {B.}~\bibnamefont
  {Tsim}}, \bibinfo {author} {\bibfnamefont {N.~N.~T.}\ \bibnamefont {Nam}},\
  and\ \bibinfo {author} {\bibfnamefont {M.}~\bibnamefont {Koshino}},\
  }\bibfield  {title} {\bibinfo {title} {Perfect one-dimensional chiral states
  in biased twisted bilayer graphene},\ }\href
  {https://doi.org/10.1103/PhysRevB.101.125409} {\bibfield  {journal} {\bibinfo
   {journal} {Phys. Rev. B}\ }\textbf {\bibinfo {volume} {101}},\ \bibinfo
  {pages} {125409} (\bibinfo {year} {2020})}\BibitemShut {NoStop}%
\bibitem [{\citenamefont {Liang}\ \emph {et~al.}(2020)\citenamefont {Liang},
  \citenamefont {Goodwin}, \citenamefont {Vitale}, \citenamefont {Corsetti},
  \citenamefont {Mostofi},\ and\ \citenamefont {Lischner}}]{liang2020}%
  \BibitemOpen
  \bibfield  {author} {\bibinfo {author} {\bibfnamefont {X.}~\bibnamefont
  {Liang}}, \bibinfo {author} {\bibfnamefont {Z.~A.}\ \bibnamefont {Goodwin}},
  \bibinfo {author} {\bibfnamefont {V.}~\bibnamefont {Vitale}}, \bibinfo
  {author} {\bibfnamefont {F.}~\bibnamefont {Corsetti}}, \bibinfo {author}
  {\bibfnamefont {A.~A.}\ \bibnamefont {Mostofi}},\ and\ \bibinfo {author}
  {\bibfnamefont {J.}~\bibnamefont {Lischner}},\ }\bibfield  {title} {\bibinfo
  {title} {Effect of bilayer stacking on the atomic and electronic structure of
  twisted double bilayer graphene},\ }\href@noop {} {\bibfield  {journal}
  {\bibinfo  {journal} {Physical Review B}\ }\textbf {\bibinfo {volume}
  {102}},\ \bibinfo {pages} {155146} (\bibinfo {year} {2020})}\BibitemShut
  {NoStop}%
\bibitem [{\citenamefont {Zhan}\ \emph {et~al.}(2020)\citenamefont {Zhan},
  \citenamefont {Zhang}, \citenamefont {Lv}, \citenamefont {Zhong},
  \citenamefont {Yu}, \citenamefont {Guinea}, \citenamefont
  {Silva-Guill{\'e}n},\ and\ \citenamefont {Yuan}}]{zzhan2020}%
  \BibitemOpen
  \bibfield  {author} {\bibinfo {author} {\bibfnamefont {Z.}~\bibnamefont
  {Zhan}}, \bibinfo {author} {\bibfnamefont {Y.}~\bibnamefont {Zhang}},
  \bibinfo {author} {\bibfnamefont {P.}~\bibnamefont {Lv}}, \bibinfo {author}
  {\bibfnamefont {H.}~\bibnamefont {Zhong}}, \bibinfo {author} {\bibfnamefont
  {G.}~\bibnamefont {Yu}}, \bibinfo {author} {\bibfnamefont {F.}~\bibnamefont
  {Guinea}}, \bibinfo {author} {\bibfnamefont {J.~{\'A}.}\ \bibnamefont
  {Silva-Guill{\'e}n}},\ and\ \bibinfo {author} {\bibfnamefont
  {S.}~\bibnamefont {Yuan}},\ }\bibfield  {title} {\bibinfo {title} {Tunability
  of multiple ultraflat bands and effect of spin-orbit coupling in twisted
  bilayer transition metal dichalcogenides},\ }\href@noop {} {\bibfield
  {journal} {\bibinfo  {journal} {Physical Review B}\ }\textbf {\bibinfo
  {volume} {102}},\ \bibinfo {pages} {241106} (\bibinfo {year}
  {2020})}\BibitemShut {NoStop}%
\bibitem [{\citenamefont {Wu}\ \emph {et~al.}(2021)\citenamefont {Wu},
  \citenamefont {Zhan},\ and\ \citenamefont {Yuan}}]{wu2021}%
  \BibitemOpen
  \bibfield  {author} {\bibinfo {author} {\bibfnamefont {Z.}~\bibnamefont
  {Wu}}, \bibinfo {author} {\bibfnamefont {Z.}~\bibnamefont {Zhan}},\ and\
  \bibinfo {author} {\bibfnamefont {S.}~\bibnamefont {Yuan}},\ }\bibfield
  {title} {\bibinfo {title} {Lattice relaxation, mirror symmetry and magnetic
  field effects on ultraflat bands in twisted trilayer graphene},\ }\href@noop
  {} {\bibfield  {journal} {\bibinfo  {journal} {Science China Physics,
  Mechanics \& Astronomy}\ }\textbf {\bibinfo {volume} {64}},\ \bibinfo {pages}
  {267811} (\bibinfo {year} {2021})}\BibitemShut {NoStop}%
\bibitem [{\citenamefont {Li}\ \emph {et~al.}(2021)\citenamefont {Li},
  \citenamefont {Li}, \citenamefont {Naik}, \citenamefont {Xie}, \citenamefont
  {Li}, \citenamefont {Wang}, \citenamefont {Regan}, \citenamefont {Wang},
  \citenamefont {Zhao}, \citenamefont {Zhao} \emph {et~al.}}]{li2021}%
  \BibitemOpen
  \bibfield  {author} {\bibinfo {author} {\bibfnamefont {H.}~\bibnamefont
  {Li}}, \bibinfo {author} {\bibfnamefont {S.}~\bibnamefont {Li}}, \bibinfo
  {author} {\bibfnamefont {M.~H.}\ \bibnamefont {Naik}}, \bibinfo {author}
  {\bibfnamefont {J.}~\bibnamefont {Xie}}, \bibinfo {author} {\bibfnamefont
  {X.}~\bibnamefont {Li}}, \bibinfo {author} {\bibfnamefont {J.}~\bibnamefont
  {Wang}}, \bibinfo {author} {\bibfnamefont {E.}~\bibnamefont {Regan}},
  \bibinfo {author} {\bibfnamefont {D.}~\bibnamefont {Wang}}, \bibinfo {author}
  {\bibfnamefont {W.}~\bibnamefont {Zhao}}, \bibinfo {author} {\bibfnamefont
  {S.}~\bibnamefont {Zhao}}, \emph {et~al.},\ }\bibfield  {title} {\bibinfo
  {title} {Imaging moir{\'e} flat bands in three-dimensional reconstructed
  wse2/ws2 superlattices},\ }\href@noop {} {\bibfield  {journal} {\bibinfo
  {journal} {Nature materials}\ }\textbf {\bibinfo {volume} {20}},\ \bibinfo
  {pages} {945} (\bibinfo {year} {2021})}\BibitemShut {NoStop}%
\bibitem [{\citenamefont {Vitale}\ \emph {et~al.}(2021)\citenamefont {Vitale},
  \citenamefont {Atalar}, \citenamefont {Mostofi},\ and\ \citenamefont
  {Lischner}}]{vitale2021}%
  \BibitemOpen
  \bibfield  {author} {\bibinfo {author} {\bibfnamefont {V.}~\bibnamefont
  {Vitale}}, \bibinfo {author} {\bibfnamefont {K.}~\bibnamefont {Atalar}},
  \bibinfo {author} {\bibfnamefont {A.~A.}\ \bibnamefont {Mostofi}},\ and\
  \bibinfo {author} {\bibfnamefont {J.}~\bibnamefont {Lischner}},\ }\bibfield
  {title} {\bibinfo {title} {Flat band properties of twisted transition metal
  dichalcogenide homo-and heterobilayers of mos2, mose2, ws2 and wse2},\
  }\href@noop {} {\bibfield  {journal} {\bibinfo  {journal} {2D Materials}\
  }\textbf {\bibinfo {volume} {8}},\ \bibinfo {pages} {045010} (\bibinfo {year}
  {2021})}\BibitemShut {NoStop}%
\bibitem [{\citenamefont {Leconte}\ \emph {et~al.}(2022)\citenamefont
  {Leconte}, \citenamefont {Javvaji}, \citenamefont {An}, \citenamefont
  {Samudrala},\ and\ \citenamefont {Jung}}]{leconte2022}%
  \BibitemOpen
  \bibfield  {author} {\bibinfo {author} {\bibfnamefont {N.}~\bibnamefont
  {Leconte}}, \bibinfo {author} {\bibfnamefont {S.}~\bibnamefont {Javvaji}},
  \bibinfo {author} {\bibfnamefont {J.}~\bibnamefont {An}}, \bibinfo {author}
  {\bibfnamefont {A.}~\bibnamefont {Samudrala}},\ and\ \bibinfo {author}
  {\bibfnamefont {J.}~\bibnamefont {Jung}},\ }\bibfield  {title} {\bibinfo
  {title} {Relaxation effects in twisted bilayer graphene: A multiscale
  approach},\ }\href {https://doi.org/10.1103/PhysRevB.106.115410} {\bibfield
  {journal} {\bibinfo  {journal} {Phys. Rev. B}\ }\textbf {\bibinfo {volume}
  {106}},\ \bibinfo {pages} {115410} (\bibinfo {year} {2022})}\BibitemShut
  {NoStop}%
\bibitem [{\citenamefont {Engelke}\ \emph {et~al.}(2023)\citenamefont
  {Engelke}, \citenamefont {Yoo}, \citenamefont {Carr}, \citenamefont {Xu},
  \citenamefont {Cazeaux}, \citenamefont {Allen}, \citenamefont {Valdivia},
  \citenamefont {Luskin}, \citenamefont {Kaxiras}, \citenamefont {Kim} \emph
  {et~al.}}]{engelke2023}%
  \BibitemOpen
  \bibfield  {author} {\bibinfo {author} {\bibfnamefont {R.}~\bibnamefont
  {Engelke}}, \bibinfo {author} {\bibfnamefont {H.}~\bibnamefont {Yoo}},
  \bibinfo {author} {\bibfnamefont {S.}~\bibnamefont {Carr}}, \bibinfo {author}
  {\bibfnamefont {K.}~\bibnamefont {Xu}}, \bibinfo {author} {\bibfnamefont
  {P.}~\bibnamefont {Cazeaux}}, \bibinfo {author} {\bibfnamefont
  {R.}~\bibnamefont {Allen}}, \bibinfo {author} {\bibfnamefont {A.~M.}\
  \bibnamefont {Valdivia}}, \bibinfo {author} {\bibfnamefont {M.}~\bibnamefont
  {Luskin}}, \bibinfo {author} {\bibfnamefont {E.}~\bibnamefont {Kaxiras}},
  \bibinfo {author} {\bibfnamefont {M.}~\bibnamefont {Kim}}, \emph {et~al.},\
  }\bibfield  {title} {\bibinfo {title} {Topological nature of dislocation
  networks in two-dimensional moir{\'e} materials},\ }\href@noop {} {\bibfield
  {journal} {\bibinfo  {journal} {Physical Review B}\ }\textbf {\bibinfo
  {volume} {107}},\ \bibinfo {pages} {125413} (\bibinfo {year}
  {2023})}\BibitemShut {NoStop}%
\bibitem [{\citenamefont {Van~Winkle}\ \emph {et~al.}(2023)\citenamefont
  {Van~Winkle}, \citenamefont {Craig}, \citenamefont {Carr}, \citenamefont
  {Dandu}, \citenamefont {Bustillo}, \citenamefont {Ciston}, \citenamefont
  {Ophus}, \citenamefont {Taniguchi}, \citenamefont {Watanabe}, \citenamefont
  {Raja} \emph {et~al.}}]{van2023}%
  \BibitemOpen
  \bibfield  {author} {\bibinfo {author} {\bibfnamefont {M.}~\bibnamefont
  {Van~Winkle}}, \bibinfo {author} {\bibfnamefont {I.~M.}\ \bibnamefont
  {Craig}}, \bibinfo {author} {\bibfnamefont {S.}~\bibnamefont {Carr}},
  \bibinfo {author} {\bibfnamefont {M.}~\bibnamefont {Dandu}}, \bibinfo
  {author} {\bibfnamefont {K.~C.}\ \bibnamefont {Bustillo}}, \bibinfo {author}
  {\bibfnamefont {J.}~\bibnamefont {Ciston}}, \bibinfo {author} {\bibfnamefont
  {C.}~\bibnamefont {Ophus}}, \bibinfo {author} {\bibfnamefont
  {T.}~\bibnamefont {Taniguchi}}, \bibinfo {author} {\bibfnamefont
  {K.}~\bibnamefont {Watanabe}}, \bibinfo {author} {\bibfnamefont
  {A.}~\bibnamefont {Raja}}, \emph {et~al.},\ }\bibfield  {title} {\bibinfo
  {title} {Rotational and dilational reconstruction in transition metal
  dichalcogenide moir{\'e} bilayers},\ }\href@noop {} {\bibfield  {journal}
  {\bibinfo  {journal} {Nature Communications}\ }\textbf {\bibinfo {volume}
  {14}},\ \bibinfo {pages} {2989} (\bibinfo {year} {2023})}\BibitemShut
  {NoStop}%
\bibitem [{\citenamefont {McHugh}\ \emph {et~al.}(2023)\citenamefont {McHugh},
  \citenamefont {Enaldiev},\ and\ \citenamefont {Fal'ko}}]{mchugh2023moire}%
  \BibitemOpen
  \bibfield  {author} {\bibinfo {author} {\bibfnamefont {J.~G.}\ \bibnamefont
  {McHugh}}, \bibinfo {author} {\bibfnamefont {V.~V.}\ \bibnamefont
  {Enaldiev}},\ and\ \bibinfo {author} {\bibfnamefont {V.~I.}\ \bibnamefont
  {Fal'ko}},\ }\bibfield  {title} {\bibinfo {title} {Moir{\'e} superstructures
  in marginally twisted nbse 2 bilayers},\ }\href@noop {} {\bibfield  {journal}
  {\bibinfo  {journal} {Physical Review B}\ }\textbf {\bibinfo {volume}
  {108}},\ \bibinfo {pages} {224111} (\bibinfo {year} {2023})}\BibitemShut
  {NoStop}%
\bibitem [{\citenamefont {Xie}\ and\ \citenamefont {Liu}(2023)}]{xie2023}%
  \BibitemOpen
  \bibfield  {author} {\bibinfo {author} {\bibfnamefont {B.}~\bibnamefont
  {Xie}}\ and\ \bibinfo {author} {\bibfnamefont {J.}~\bibnamefont {Liu}},\
  }\bibfield  {title} {\bibinfo {title} {Lattice distortions, moir{\'e}
  phonons, and relaxed electronic band structures in magic-angle twisted
  bilayer graphene},\ }\href@noop {} {\bibfield  {journal} {\bibinfo  {journal}
  {Physical Review B}\ }\textbf {\bibinfo {volume} {108}},\ \bibinfo {pages}
  {094115} (\bibinfo {year} {2023})}\BibitemShut {NoStop}%
\bibitem [{\citenamefont {Zhao}\ \emph {et~al.}(2023)\citenamefont {Zhao},
  \citenamefont {Li}, \citenamefont {Huang}, \citenamefont {Rupp},
  \citenamefont {G{\"o}ser}, \citenamefont {Vovk}, \citenamefont {Kruchinin},
  \citenamefont {Watanabe}, \citenamefont {Taniguchi}, \citenamefont {Bilgin}
  \emph {et~al.}}]{zhao2023}%
  \BibitemOpen
  \bibfield  {author} {\bibinfo {author} {\bibfnamefont {S.}~\bibnamefont
  {Zhao}}, \bibinfo {author} {\bibfnamefont {Z.}~\bibnamefont {Li}}, \bibinfo
  {author} {\bibfnamefont {X.}~\bibnamefont {Huang}}, \bibinfo {author}
  {\bibfnamefont {A.}~\bibnamefont {Rupp}}, \bibinfo {author} {\bibfnamefont
  {J.}~\bibnamefont {G{\"o}ser}}, \bibinfo {author} {\bibfnamefont {I.~A.}\
  \bibnamefont {Vovk}}, \bibinfo {author} {\bibfnamefont {S.~Y.}\ \bibnamefont
  {Kruchinin}}, \bibinfo {author} {\bibfnamefont {K.}~\bibnamefont {Watanabe}},
  \bibinfo {author} {\bibfnamefont {T.}~\bibnamefont {Taniguchi}}, \bibinfo
  {author} {\bibfnamefont {I.}~\bibnamefont {Bilgin}}, \emph {et~al.},\
  }\bibfield  {title} {\bibinfo {title} {Excitons in mesoscopically
  reconstructed moir{\'e} heterostructures},\ }\href@noop {} {\bibfield
  {journal} {\bibinfo  {journal} {Nature nanotechnology}\ }\textbf {\bibinfo
  {volume} {18}},\ \bibinfo {pages} {572} (\bibinfo {year} {2023})}\BibitemShut
  {NoStop}%
\bibitem [{\citenamefont {Halbertal}\ \emph {et~al.}(2023)\citenamefont
  {Halbertal}, \citenamefont {Klebl}, \citenamefont {Hsieh}, \citenamefont
  {Cook}, \citenamefont {Carr}, \citenamefont {Bian}, \citenamefont {Dean},
  \citenamefont {Kennes},\ and\ \citenamefont {Basov}}]{halbertal2023}%
  \BibitemOpen
  \bibfield  {author} {\bibinfo {author} {\bibfnamefont {D.}~\bibnamefont
  {Halbertal}}, \bibinfo {author} {\bibfnamefont {L.}~\bibnamefont {Klebl}},
  \bibinfo {author} {\bibfnamefont {V.}~\bibnamefont {Hsieh}}, \bibinfo
  {author} {\bibfnamefont {J.}~\bibnamefont {Cook}}, \bibinfo {author}
  {\bibfnamefont {S.}~\bibnamefont {Carr}}, \bibinfo {author} {\bibfnamefont
  {G.}~\bibnamefont {Bian}}, \bibinfo {author} {\bibfnamefont {C.~R.}\
  \bibnamefont {Dean}}, \bibinfo {author} {\bibfnamefont {D.~M.}\ \bibnamefont
  {Kennes}},\ and\ \bibinfo {author} {\bibfnamefont {D.~N.}\ \bibnamefont
  {Basov}},\ }\bibfield  {title} {\bibinfo {title} {Multilayered atomic
  relaxation in van der waals heterostructures},\ }\href
  {https://doi.org/10.1103/PhysRevX.13.011026} {\bibfield  {journal} {\bibinfo
  {journal} {Phys. Rev. X}\ }\textbf {\bibinfo {volume} {13}},\ \bibinfo
  {pages} {011026} (\bibinfo {year} {2023})}\BibitemShut {NoStop}%
\bibitem [{\citenamefont {Zhang}\ \emph {et~al.}(2024)\citenamefont {Zhang},
  \citenamefont {Wang}, \citenamefont {Liu}, \citenamefont {Fan}, \citenamefont
  {Cao},\ and\ \citenamefont {Xiao}}]{zhang2024polarization}%
  \BibitemOpen
  \bibfield  {author} {\bibinfo {author} {\bibfnamefont {X.-W.}\ \bibnamefont
  {Zhang}}, \bibinfo {author} {\bibfnamefont {C.}~\bibnamefont {Wang}},
  \bibinfo {author} {\bibfnamefont {X.}~\bibnamefont {Liu}}, \bibinfo {author}
  {\bibfnamefont {Y.}~\bibnamefont {Fan}}, \bibinfo {author} {\bibfnamefont
  {T.}~\bibnamefont {Cao}},\ and\ \bibinfo {author} {\bibfnamefont
  {D.}~\bibnamefont {Xiao}},\ }\bibfield  {title} {\bibinfo {title}
  {Polarization-driven band topology evolution in twisted mote2 and wse2},\
  }\href@noop {} {\bibfield  {journal} {\bibinfo  {journal} {Nature
  Communications}\ }\textbf {\bibinfo {volume} {15}},\ \bibinfo {pages} {4223}
  (\bibinfo {year} {2024})}\BibitemShut {NoStop}%
\bibitem [{\citenamefont {Ceferino}\ and\ \citenamefont
  {Guinea}(2024)}]{ceferino2024}%
  \BibitemOpen
  \bibfield  {author} {\bibinfo {author} {\bibfnamefont {A.}~\bibnamefont
  {Ceferino}}\ and\ \bibinfo {author} {\bibfnamefont {F.}~\bibnamefont
  {Guinea}},\ }\bibfield  {title} {\bibinfo {title} {Pseudomagnetic fields in
  fully relaxed twisted bilayer and trilayer graphene},\ }\href@noop {}
  {\bibfield  {journal} {\bibinfo  {journal} {2D Materials}\ }\textbf {\bibinfo
  {volume} {11}},\ \bibinfo {pages} {035015} (\bibinfo {year}
  {2024})}\BibitemShut {NoStop}%
\bibitem [{\citenamefont {Ezzi}\ \emph {et~al.}(2024)\citenamefont {Ezzi},
  \citenamefont {Pallewela}, \citenamefont {De~Beule}, \citenamefont {Mele},\
  and\ \citenamefont {Adam}}]{ezzi2024}%
  \BibitemOpen
  \bibfield  {author} {\bibinfo {author} {\bibfnamefont {M.~M.~A.}\
  \bibnamefont {Ezzi}}, \bibinfo {author} {\bibfnamefont {G.~N.}\ \bibnamefont
  {Pallewela}}, \bibinfo {author} {\bibfnamefont {C.}~\bibnamefont {De~Beule}},
  \bibinfo {author} {\bibfnamefont {E.}~\bibnamefont {Mele}},\ and\ \bibinfo
  {author} {\bibfnamefont {S.}~\bibnamefont {Adam}},\ }\bibfield  {title}
  {\bibinfo {title} {Analytical model for atomic relaxation in twisted
  moir{\'e} materials},\ }\href@noop {} {\bibfield  {journal} {\bibinfo
  {journal} {Physical Review Letters}\ }\textbf {\bibinfo {volume} {133}},\
  \bibinfo {pages} {266201} (\bibinfo {year} {2024})}\BibitemShut {NoStop}%
\bibitem [{\citenamefont {Hagel}\ \emph {et~al.}(2024)\citenamefont {Hagel},
  \citenamefont {Brem}, \citenamefont {Pineiro},\ and\ \citenamefont
  {Malic}}]{hagel2024}%
  \BibitemOpen
  \bibfield  {author} {\bibinfo {author} {\bibfnamefont {J.}~\bibnamefont
  {Hagel}}, \bibinfo {author} {\bibfnamefont {S.}~\bibnamefont {Brem}},
  \bibinfo {author} {\bibfnamefont {J.~A.}\ \bibnamefont {Pineiro}},\ and\
  \bibinfo {author} {\bibfnamefont {E.}~\bibnamefont {Malic}},\ }\bibfield
  {title} {\bibinfo {title} {Impact of atomic reconstruction on optical spectra
  of twisted tmd homobilayers},\ }\href@noop {} {\bibfield  {journal} {\bibinfo
   {journal} {Physical Review Materials}\ }\textbf {\bibinfo {volume} {8}},\
  \bibinfo {pages} {034001} (\bibinfo {year} {2024})}\BibitemShut {NoStop}%
\bibitem [{\citenamefont {Li}\ \emph {et~al.}(2024)\citenamefont {Li},
  \citenamefont {Brumme},\ and\ \citenamefont {Heine}}]{li2024}%
  \BibitemOpen
  \bibfield  {author} {\bibinfo {author} {\bibfnamefont {W.}~\bibnamefont
  {Li}}, \bibinfo {author} {\bibfnamefont {T.}~\bibnamefont {Brumme}},\ and\
  \bibinfo {author} {\bibfnamefont {T.}~\bibnamefont {Heine}},\ }\bibfield
  {title} {\bibinfo {title} {Relaxation effects in transition metal
  dichalcogenide bilayer heterostructures},\ }\href@noop {} {\bibfield
  {journal} {\bibinfo  {journal} {npj 2D Materials and Applications}\ }\textbf
  {\bibinfo {volume} {8}},\ \bibinfo {pages} {43} (\bibinfo {year}
  {2024})}\BibitemShut {NoStop}%
\bibitem [{\citenamefont {Dang}\ \emph {et~al.}(2025)\citenamefont {Dang},
  \citenamefont {Le},\ and\ \citenamefont {Woods}}]{dang2025}%
  \BibitemOpen
  \bibfield  {author} {\bibinfo {author} {\bibfnamefont {D.~T.-X.}\
  \bibnamefont {Dang}}, \bibinfo {author} {\bibfnamefont {D.-N.}\ \bibnamefont
  {Le}},\ and\ \bibinfo {author} {\bibfnamefont {L.~M.}\ \bibnamefont
  {Woods}},\ }\bibfield  {title} {\bibinfo {title} {Twisting in h-bn bilayers
  and their angle-dependent properties},\ }\href@noop {} {\bibfield  {journal}
  {\bibinfo  {journal} {Physical Review B}\ }\textbf {\bibinfo {volume}
  {112}},\ \bibinfo {pages} {085102} (\bibinfo {year} {2025})}\BibitemShut
  {NoStop}%
\bibitem [{\citenamefont {Yu}\ \emph {et~al.}(2025)\citenamefont {Yu},
  \citenamefont {Qian},\ and\ \citenamefont {Liu}}]{yu2025}%
  \BibitemOpen
  \bibfield  {author} {\bibinfo {author} {\bibfnamefont {J.}~\bibnamefont
  {Yu}}, \bibinfo {author} {\bibfnamefont {S.}~\bibnamefont {Qian}},\ and\
  \bibinfo {author} {\bibfnamefont {C.-C.}\ \bibnamefont {Liu}},\ }\bibfield
  {title} {\bibinfo {title} {General electronic structure calculation method
  for twisted systems},\ }\href@noop {} {\bibfield  {journal} {\bibinfo
  {journal} {Physical Review B}\ }\textbf {\bibinfo {volume} {111}},\ \bibinfo
  {pages} {075434} (\bibinfo {year} {2025})}\BibitemShut {NoStop}%
\bibitem [{\citenamefont {Fu}\ \emph {et~al.}(2025)\citenamefont {Fu},
  \citenamefont {Zhou},\ and\ \citenamefont {He}}]{fu2025review}%
  \BibitemOpen
  \bibfield  {author} {\bibinfo {author} {\bibfnamefont {Z.}~\bibnamefont
  {Fu}}, \bibinfo {author} {\bibfnamefont {X.}~\bibnamefont {Zhou}},\ and\
  \bibinfo {author} {\bibfnamefont {L.}~\bibnamefont {He}},\ }\bibfield
  {title} {\bibinfo {title} {Lattice reconstruction in twisted bilayer
  graphene},\ }\href@noop {} {\bibfield  {journal} {\bibinfo  {journal}
  {Journal of Physics: Condensed Matter}\ }\textbf {\bibinfo {volume} {37}},\
  \bibinfo {pages} {073001} (\bibinfo {year} {2025})}\BibitemShut {NoStop}%
\bibitem [{\citenamefont {Qian}\ \emph {et~al.}(2014)\citenamefont {Qian},
  \citenamefont {Liu}, \citenamefont {Fu},\ and\ \citenamefont
  {Li}}]{qian2014}%
  \BibitemOpen
  \bibfield  {author} {\bibinfo {author} {\bibfnamefont {X.}~\bibnamefont
  {Qian}}, \bibinfo {author} {\bibfnamefont {J.}~\bibnamefont {Liu}}, \bibinfo
  {author} {\bibfnamefont {L.}~\bibnamefont {Fu}},\ and\ \bibinfo {author}
  {\bibfnamefont {J.}~\bibnamefont {Li}},\ }\bibfield  {title} {\bibinfo
  {title} {Quantum spin hall effect in two-dimensional transition metal
  dichalcogenides},\ }\href@noop {} {\bibfield  {journal} {\bibinfo  {journal}
  {Science}\ }\textbf {\bibinfo {volume} {346}},\ \bibinfo {pages} {1344}
  (\bibinfo {year} {2014})}\BibitemShut {NoStop}%
\bibitem [{\citenamefont {Torun}\ \emph {et~al.}(2016)\citenamefont {Torun},
  \citenamefont {Sahin}, \citenamefont {Cahangirov}, \citenamefont {Rubio},\
  and\ \citenamefont {Peeters}}]{torun2016}%
  \BibitemOpen
  \bibfield  {author} {\bibinfo {author} {\bibfnamefont {E.}~\bibnamefont
  {Torun}}, \bibinfo {author} {\bibfnamefont {H.}~\bibnamefont {Sahin}},
  \bibinfo {author} {\bibfnamefont {S.}~\bibnamefont {Cahangirov}}, \bibinfo
  {author} {\bibfnamefont {A.}~\bibnamefont {Rubio}},\ and\ \bibinfo {author}
  {\bibfnamefont {F.}~\bibnamefont {Peeters}},\ }\bibfield  {title} {\bibinfo
  {title} {Anisotropic electronic, mechanical, and optical properties of
  monolayer {WTe$_2$}},\ }\href@noop {} {\bibfield  {journal} {\bibinfo
  {journal} {Journal of Applied Physics}\ }\textbf {\bibinfo {volume} {119}}
  (\bibinfo {year} {2016})}\BibitemShut {NoStop}%
\bibitem [{\citenamefont {Muechler}\ \emph {et~al.}(2016)\citenamefont
  {Muechler}, \citenamefont {Alexandradinata}, \citenamefont {Neupert},\ and\
  \citenamefont {Car}}]{muechler2016}%
  \BibitemOpen
  \bibfield  {author} {\bibinfo {author} {\bibfnamefont {L.}~\bibnamefont
  {Muechler}}, \bibinfo {author} {\bibfnamefont {A.}~\bibnamefont
  {Alexandradinata}}, \bibinfo {author} {\bibfnamefont {T.}~\bibnamefont
  {Neupert}},\ and\ \bibinfo {author} {\bibfnamefont {R.}~\bibnamefont {Car}},\
  }\bibfield  {title} {\bibinfo {title} {Topological nonsymmorphic metals from
  band inversion},\ }\href@noop {} {\bibfield  {journal} {\bibinfo  {journal}
  {Phys. Rev. X}\ }\textbf {\bibinfo {volume} {6}},\ \bibinfo {pages} {041069}
  (\bibinfo {year} {2016})}\BibitemShut {NoStop}%
\bibitem [{\citenamefont {Zheng}\ \emph {et~al.}(2016)\citenamefont {Zheng},
  \citenamefont {Cai}, \citenamefont {Ge}, \citenamefont {Zhang}, \citenamefont
  {Liu}, \citenamefont {Lu}, \citenamefont {Zhang}, \citenamefont {Qiu},
  \citenamefont {Taniguchi}, \citenamefont {Watanabe} \emph
  {et~al.}}]{fzheng2016}%
  \BibitemOpen
  \bibfield  {author} {\bibinfo {author} {\bibfnamefont {F.}~\bibnamefont
  {Zheng}}, \bibinfo {author} {\bibfnamefont {C.}~\bibnamefont {Cai}}, \bibinfo
  {author} {\bibfnamefont {S.}~\bibnamefont {Ge}}, \bibinfo {author}
  {\bibfnamefont {X.}~\bibnamefont {Zhang}}, \bibinfo {author} {\bibfnamefont
  {X.}~\bibnamefont {Liu}}, \bibinfo {author} {\bibfnamefont {H.}~\bibnamefont
  {Lu}}, \bibinfo {author} {\bibfnamefont {Y.}~\bibnamefont {Zhang}}, \bibinfo
  {author} {\bibfnamefont {J.}~\bibnamefont {Qiu}}, \bibinfo {author}
  {\bibfnamefont {T.}~\bibnamefont {Taniguchi}}, \bibinfo {author}
  {\bibfnamefont {K.}~\bibnamefont {Watanabe}}, \emph {et~al.},\ }\bibfield
  {title} {\bibinfo {title} {On the quantum spin hall gap of monolayer
  {1T$'$}-{WTe$_2$}},\ }\href@noop {} {\bibfield  {journal} {\bibinfo
  {journal} {Advanced Materials}\ }\textbf {\bibinfo {volume} {28}},\ \bibinfo
  {pages} {4845} (\bibinfo {year} {2016})}\BibitemShut {NoStop}%
\bibitem [{\citenamefont {Tang}\ \emph {et~al.}(2017)\citenamefont {Tang},
  \citenamefont {Zhang}, \citenamefont {Wong}, \citenamefont {Pedramrazi},
  \citenamefont {Tsai}, \citenamefont {Jia}, \citenamefont {Moritz},
  \citenamefont {Claassen}, \citenamefont {Ryu}, \citenamefont {Kahn} \emph
  {et~al.}}]{stang2017}%
  \BibitemOpen
  \bibfield  {author} {\bibinfo {author} {\bibfnamefont {S.}~\bibnamefont
  {Tang}}, \bibinfo {author} {\bibfnamefont {C.}~\bibnamefont {Zhang}},
  \bibinfo {author} {\bibfnamefont {D.}~\bibnamefont {Wong}}, \bibinfo {author}
  {\bibfnamefont {Z.}~\bibnamefont {Pedramrazi}}, \bibinfo {author}
  {\bibfnamefont {H.-Z.}\ \bibnamefont {Tsai}}, \bibinfo {author}
  {\bibfnamefont {C.}~\bibnamefont {Jia}}, \bibinfo {author} {\bibfnamefont
  {B.}~\bibnamefont {Moritz}}, \bibinfo {author} {\bibfnamefont
  {M.}~\bibnamefont {Claassen}}, \bibinfo {author} {\bibfnamefont
  {H.}~\bibnamefont {Ryu}}, \bibinfo {author} {\bibfnamefont {S.}~\bibnamefont
  {Kahn}}, \emph {et~al.},\ }\bibfield  {title} {\bibinfo {title} {Quantum spin
  hall state in monolayer {1T$'$}-{WTe$_2$}},\ }\href@noop {} {\bibfield
  {journal} {\bibinfo  {journal} {Nature Physics}\ }\textbf {\bibinfo {volume}
  {13}},\ \bibinfo {pages} {683} (\bibinfo {year} {2017})}\BibitemShut
  {NoStop}%
\bibitem [{\citenamefont {Fatemi}\ \emph {et~al.}(2018)\citenamefont {Fatemi},
  \citenamefont {Wu}, \citenamefont {Cao}, \citenamefont {Bretheau},
  \citenamefont {Gibson}, \citenamefont {Watanabe}, \citenamefont {Taniguchi},
  \citenamefont {Cava},\ and\ \citenamefont {Jarillo-Herrero}}]{fatemi2018}%
  \BibitemOpen
  \bibfield  {author} {\bibinfo {author} {\bibfnamefont {V.}~\bibnamefont
  {Fatemi}}, \bibinfo {author} {\bibfnamefont {S.}~\bibnamefont {Wu}}, \bibinfo
  {author} {\bibfnamefont {Y.}~\bibnamefont {Cao}}, \bibinfo {author}
  {\bibfnamefont {L.}~\bibnamefont {Bretheau}}, \bibinfo {author}
  {\bibfnamefont {Q.~D.}\ \bibnamefont {Gibson}}, \bibinfo {author}
  {\bibfnamefont {K.}~\bibnamefont {Watanabe}}, \bibinfo {author}
  {\bibfnamefont {T.}~\bibnamefont {Taniguchi}}, \bibinfo {author}
  {\bibfnamefont {R.~J.}\ \bibnamefont {Cava}},\ and\ \bibinfo {author}
  {\bibfnamefont {P.}~\bibnamefont {Jarillo-Herrero}},\ }\bibfield  {title}
  {\bibinfo {title} {Electrically tunable low-density superconductivity in a
  monolayer topological insulator},\ }\href@noop {} {\bibfield  {journal}
  {\bibinfo  {journal} {Science}\ }\textbf {\bibinfo {volume} {362}},\ \bibinfo
  {pages} {926} (\bibinfo {year} {2018})}\BibitemShut {NoStop}%
\bibitem [{\citenamefont {Xu}\ \emph {et~al.}(2018)\citenamefont {Xu},
  \citenamefont {Ma}, \citenamefont {Shen}, \citenamefont {Fatemi},
  \citenamefont {Wu}, \citenamefont {Chang}, \citenamefont {Chang},
  \citenamefont {Valdivia}, \citenamefont {Chan}, \citenamefont {Gibson} \emph
  {et~al.}}]{syxu2018}%
  \BibitemOpen
  \bibfield  {author} {\bibinfo {author} {\bibfnamefont {S.-Y.}\ \bibnamefont
  {Xu}}, \bibinfo {author} {\bibfnamefont {Q.}~\bibnamefont {Ma}}, \bibinfo
  {author} {\bibfnamefont {H.}~\bibnamefont {Shen}}, \bibinfo {author}
  {\bibfnamefont {V.}~\bibnamefont {Fatemi}}, \bibinfo {author} {\bibfnamefont
  {S.}~\bibnamefont {Wu}}, \bibinfo {author} {\bibfnamefont {T.-R.}\
  \bibnamefont {Chang}}, \bibinfo {author} {\bibfnamefont {G.}~\bibnamefont
  {Chang}}, \bibinfo {author} {\bibfnamefont {A.~M.~M.}\ \bibnamefont
  {Valdivia}}, \bibinfo {author} {\bibfnamefont {C.-K.}\ \bibnamefont {Chan}},
  \bibinfo {author} {\bibfnamefont {Q.~D.}\ \bibnamefont {Gibson}}, \emph
  {et~al.},\ }\bibfield  {title} {\bibinfo {title} {Electrically switchable
  berry curvature dipole in the monolayer topological insulator {WTe$_2$}},\
  }\href@noop {} {\bibfield  {journal} {\bibinfo  {journal} {Nature Physics}\
  }\textbf {\bibinfo {volume} {14}},\ \bibinfo {pages} {900} (\bibinfo {year}
  {2018})}\BibitemShut {NoStop}%
\bibitem [{\citenamefont {Yang}\ \emph {et~al.}(2018)\citenamefont {Yang},
  \citenamefont {Jin}, \citenamefont {Xu}, \citenamefont {Zheng}, \citenamefont
  {Wang},\ and\ \citenamefont {Xu}}]{jyang2018}%
  \BibitemOpen
  \bibfield  {author} {\bibinfo {author} {\bibfnamefont {J.}~\bibnamefont
  {Yang}}, \bibinfo {author} {\bibfnamefont {Y.}~\bibnamefont {Jin}}, \bibinfo
  {author} {\bibfnamefont {W.}~\bibnamefont {Xu}}, \bibinfo {author}
  {\bibfnamefont {B.}~\bibnamefont {Zheng}}, \bibinfo {author} {\bibfnamefont
  {R.}~\bibnamefont {Wang}},\ and\ \bibinfo {author} {\bibfnamefont
  {H.}~\bibnamefont {Xu}},\ }\bibfield  {title} {\bibinfo {title}
  {Oxidation-induced topological phase transition in monolayer
  {1T$'$}-{WTe$_2$}},\ }\href@noop {} {\bibfield  {journal} {\bibinfo
  {journal} {The journal of physical chemistry letters}\ }\textbf {\bibinfo
  {volume} {9}},\ \bibinfo {pages} {4783} (\bibinfo {year} {2018})}\BibitemShut
  {NoStop}%
\bibitem [{\citenamefont {Zhang}\ \emph {et~al.}(2019)\citenamefont {Zhang},
  \citenamefont {Zhang}, \citenamefont {Chen}, \citenamefont {Shen},
  \citenamefont {An}, \citenamefont {Hu}, \citenamefont {Dong}, \citenamefont
  {Liu},\ and\ \citenamefont {Zhu}}]{qzhang2019}%
  \BibitemOpen
  \bibfield  {author} {\bibinfo {author} {\bibfnamefont {Q.}~\bibnamefont
  {Zhang}}, \bibinfo {author} {\bibfnamefont {R.}~\bibnamefont {Zhang}},
  \bibinfo {author} {\bibfnamefont {J.}~\bibnamefont {Chen}}, \bibinfo {author}
  {\bibfnamefont {W.}~\bibnamefont {Shen}}, \bibinfo {author} {\bibfnamefont
  {C.}~\bibnamefont {An}}, \bibinfo {author} {\bibfnamefont {X.}~\bibnamefont
  {Hu}}, \bibinfo {author} {\bibfnamefont {M.}~\bibnamefont {Dong}}, \bibinfo
  {author} {\bibfnamefont {J.}~\bibnamefont {Liu}},\ and\ \bibinfo {author}
  {\bibfnamefont {L.}~\bibnamefont {Zhu}},\ }\bibfield  {title} {\bibinfo
  {title} {Remarkable electronic and optical anisotropy of layered
  1t’-{WTe$_2$} 2d materials},\ }\href@noop {} {\bibfield  {journal}
  {\bibinfo  {journal} {Beilstein Journal of Nanotechnology}\ }\textbf
  {\bibinfo {volume} {10}},\ \bibinfo {pages} {1745} (\bibinfo {year}
  {2019})}\BibitemShut {NoStop}%
\bibitem [{\citenamefont {Shi}\ \emph {et~al.}(2019)\citenamefont {Shi},
  \citenamefont {Kahn}, \citenamefont {Niu}, \citenamefont {Fei}, \citenamefont
  {Sun}, \citenamefont {Cai}, \citenamefont {Francisco}, \citenamefont {Wu},
  \citenamefont {Shen}, \citenamefont {Xu} \emph {et~al.}}]{shi2019}%
  \BibitemOpen
  \bibfield  {author} {\bibinfo {author} {\bibfnamefont {Y.}~\bibnamefont
  {Shi}}, \bibinfo {author} {\bibfnamefont {J.}~\bibnamefont {Kahn}}, \bibinfo
  {author} {\bibfnamefont {B.}~\bibnamefont {Niu}}, \bibinfo {author}
  {\bibfnamefont {Z.}~\bibnamefont {Fei}}, \bibinfo {author} {\bibfnamefont
  {B.}~\bibnamefont {Sun}}, \bibinfo {author} {\bibfnamefont {X.}~\bibnamefont
  {Cai}}, \bibinfo {author} {\bibfnamefont {B.~A.}\ \bibnamefont {Francisco}},
  \bibinfo {author} {\bibfnamefont {D.}~\bibnamefont {Wu}}, \bibinfo {author}
  {\bibfnamefont {Z.-X.}\ \bibnamefont {Shen}}, \bibinfo {author}
  {\bibfnamefont {X.}~\bibnamefont {Xu}}, \emph {et~al.},\ }\bibfield  {title}
  {\bibinfo {title} {Imaging quantum spin hall edges in monolayer {WTe$_2$}},\
  }\href@noop {} {\bibfield  {journal} {\bibinfo  {journal} {Science advances}\
  }\textbf {\bibinfo {volume} {5}},\ \bibinfo {pages} {eaat8799} (\bibinfo
  {year} {2019})}\BibitemShut {NoStop}%
\bibitem [{\citenamefont {Ok}\ \emph {et~al.}(2019)\citenamefont {Ok},
  \citenamefont {Muechler}, \citenamefont {Di~Sante}, \citenamefont
  {Sangiovanni}, \citenamefont {Thomale},\ and\ \citenamefont
  {Neupert}}]{ok2019}%
  \BibitemOpen
  \bibfield  {author} {\bibinfo {author} {\bibfnamefont {S.}~\bibnamefont
  {Ok}}, \bibinfo {author} {\bibfnamefont {L.}~\bibnamefont {Muechler}},
  \bibinfo {author} {\bibfnamefont {D.}~\bibnamefont {Di~Sante}}, \bibinfo
  {author} {\bibfnamefont {G.}~\bibnamefont {Sangiovanni}}, \bibinfo {author}
  {\bibfnamefont {R.}~\bibnamefont {Thomale}},\ and\ \bibinfo {author}
  {\bibfnamefont {T.}~\bibnamefont {Neupert}},\ }\bibfield  {title} {\bibinfo
  {title} {Custodial glide symmetry of quantum spin hall edge modes in
  monolayer wte 2},\ }\href@noop {} {\bibfield  {journal} {\bibinfo  {journal}
  {Phys. Rev. B}\ }\textbf {\bibinfo {volume} {99}},\ \bibinfo {pages} {121105}
  (\bibinfo {year} {2019})}\BibitemShut {NoStop}%
\bibitem [{\citenamefont {Zhao}\ \emph
  {et~al.}(2020{\natexlab{a}})\citenamefont {Zhao}, \citenamefont {Hu},
  \citenamefont {Qin}, \citenamefont {Xia}, \citenamefont {Liu}, \citenamefont
  {Wang}, \citenamefont {Guan}, \citenamefont {Li}, \citenamefont {Zheng},
  \citenamefont {Liu} \emph {et~al.}}]{czhao2020}%
  \BibitemOpen
  \bibfield  {author} {\bibinfo {author} {\bibfnamefont {C.}~\bibnamefont
  {Zhao}}, \bibinfo {author} {\bibfnamefont {M.}~\bibnamefont {Hu}}, \bibinfo
  {author} {\bibfnamefont {J.}~\bibnamefont {Qin}}, \bibinfo {author}
  {\bibfnamefont {B.}~\bibnamefont {Xia}}, \bibinfo {author} {\bibfnamefont
  {C.}~\bibnamefont {Liu}}, \bibinfo {author} {\bibfnamefont {S.}~\bibnamefont
  {Wang}}, \bibinfo {author} {\bibfnamefont {D.}~\bibnamefont {Guan}}, \bibinfo
  {author} {\bibfnamefont {Y.}~\bibnamefont {Li}}, \bibinfo {author}
  {\bibfnamefont {H.}~\bibnamefont {Zheng}}, \bibinfo {author} {\bibfnamefont
  {J.}~\bibnamefont {Liu}}, \emph {et~al.},\ }\bibfield  {title} {\bibinfo
  {title} {Strain tunable semimetal--topological-insulator transition in
  monolayer 1 t{$'$}-wte 2},\ }\href@noop {} {\bibfield  {journal} {\bibinfo
  {journal} {Phys. Rev. Lett.}\ }\textbf {\bibinfo {volume} {125}},\ \bibinfo
  {pages} {046801} (\bibinfo {year} {2020}{\natexlab{a}})}\BibitemShut
  {NoStop}%
\bibitem [{\citenamefont {Zhao}\ \emph
  {et~al.}(2020{\natexlab{b}})\citenamefont {Zhao}, \citenamefont {Fei},
  \citenamefont {Song}, \citenamefont {Choi}, \citenamefont {Palomaki},
  \citenamefont {Sun}, \citenamefont {Malinowski}, \citenamefont {McGuire},
  \citenamefont {Chu}, \citenamefont {Xu} \emph {et~al.}}]{wzhao2020}%
  \BibitemOpen
  \bibfield  {author} {\bibinfo {author} {\bibfnamefont {W.}~\bibnamefont
  {Zhao}}, \bibinfo {author} {\bibfnamefont {Z.}~\bibnamefont {Fei}}, \bibinfo
  {author} {\bibfnamefont {T.}~\bibnamefont {Song}}, \bibinfo {author}
  {\bibfnamefont {H.~K.}\ \bibnamefont {Choi}}, \bibinfo {author}
  {\bibfnamefont {T.}~\bibnamefont {Palomaki}}, \bibinfo {author}
  {\bibfnamefont {B.}~\bibnamefont {Sun}}, \bibinfo {author} {\bibfnamefont
  {P.}~\bibnamefont {Malinowski}}, \bibinfo {author} {\bibfnamefont {M.~A.}\
  \bibnamefont {McGuire}}, \bibinfo {author} {\bibfnamefont {J.-H.}\
  \bibnamefont {Chu}}, \bibinfo {author} {\bibfnamefont {X.}~\bibnamefont
  {Xu}}, \emph {et~al.},\ }\bibfield  {title} {\bibinfo {title} {Magnetic
  proximity and nonreciprocal current switching in a monolayer {WTe$_2$}
  helical edge},\ }\href@noop {} {\bibfield  {journal} {\bibinfo  {journal}
  {Nature Materials}\ }\textbf {\bibinfo {volume} {19}},\ \bibinfo {pages}
  {503} (\bibinfo {year} {2020}{\natexlab{b}})}\BibitemShut {NoStop}%
\bibitem [{\citenamefont {Hu}\ \emph {et~al.}(2021)\citenamefont {Hu},
  \citenamefont {Ma}, \citenamefont {Wan},\ and\ \citenamefont
  {Liu}}]{mhu2021}%
  \BibitemOpen
  \bibfield  {author} {\bibinfo {author} {\bibfnamefont {M.}~\bibnamefont
  {Hu}}, \bibinfo {author} {\bibfnamefont {G.}~\bibnamefont {Ma}}, \bibinfo
  {author} {\bibfnamefont {C.~Y.}\ \bibnamefont {Wan}},\ and\ \bibinfo {author}
  {\bibfnamefont {J.}~\bibnamefont {Liu}},\ }\bibfield  {title} {\bibinfo
  {title} {Realistic tight-binding model for monolayer transition metal
  dichalcogenides of {1T$'$} structure},\ }\href@noop {} {\bibfield  {journal}
  {\bibinfo  {journal} {Phys. Rev. B}\ }\textbf {\bibinfo {volume} {104}},\
  \bibinfo {pages} {035156} (\bibinfo {year} {2021})}\BibitemShut {NoStop}%
\bibitem [{\citenamefont {Zhao}\ \emph {et~al.}(2021)\citenamefont {Zhao},
  \citenamefont {Runburg}, \citenamefont {Fei}, \citenamefont {Mutch},
  \citenamefont {Malinowski}, \citenamefont {Sun}, \citenamefont {Huang},
  \citenamefont {Pesin}, \citenamefont {Cui}, \citenamefont {Xu} \emph
  {et~al.}}]{wzhao2021}%
  \BibitemOpen
  \bibfield  {author} {\bibinfo {author} {\bibfnamefont {W.}~\bibnamefont
  {Zhao}}, \bibinfo {author} {\bibfnamefont {E.}~\bibnamefont {Runburg}},
  \bibinfo {author} {\bibfnamefont {Z.}~\bibnamefont {Fei}}, \bibinfo {author}
  {\bibfnamefont {J.}~\bibnamefont {Mutch}}, \bibinfo {author} {\bibfnamefont
  {P.}~\bibnamefont {Malinowski}}, \bibinfo {author} {\bibfnamefont
  {B.}~\bibnamefont {Sun}}, \bibinfo {author} {\bibfnamefont {X.}~\bibnamefont
  {Huang}}, \bibinfo {author} {\bibfnamefont {D.}~\bibnamefont {Pesin}},
  \bibinfo {author} {\bibfnamefont {Y.-T.}\ \bibnamefont {Cui}}, \bibinfo
  {author} {\bibfnamefont {X.}~\bibnamefont {Xu}}, \emph {et~al.},\ }\bibfield
  {title} {\bibinfo {title} {Determination of the spin axis in quantum spin
  hall insulator candidate monolayer wte 2},\ }\href@noop {} {\bibfield
  {journal} {\bibinfo  {journal} {Phys. Rev. X}\ }\textbf {\bibinfo {volume}
  {11}},\ \bibinfo {pages} {041034} (\bibinfo {year} {2021})}\BibitemShut
  {NoStop}%
\bibitem [{\citenamefont {Sun}\ \emph {et~al.}(2022)\citenamefont {Sun},
  \citenamefont {Zhao}, \citenamefont {Palomaki}, \citenamefont {Fei},
  \citenamefont {Runburg}, \citenamefont {Malinowski}, \citenamefont {Huang},
  \citenamefont {Cenker}, \citenamefont {Cui}, \citenamefont {Chu} \emph
  {et~al.}}]{bsun2022}%
  \BibitemOpen
  \bibfield  {author} {\bibinfo {author} {\bibfnamefont {B.}~\bibnamefont
  {Sun}}, \bibinfo {author} {\bibfnamefont {W.}~\bibnamefont {Zhao}}, \bibinfo
  {author} {\bibfnamefont {T.}~\bibnamefont {Palomaki}}, \bibinfo {author}
  {\bibfnamefont {Z.}~\bibnamefont {Fei}}, \bibinfo {author} {\bibfnamefont
  {E.}~\bibnamefont {Runburg}}, \bibinfo {author} {\bibfnamefont
  {P.}~\bibnamefont {Malinowski}}, \bibinfo {author} {\bibfnamefont
  {X.}~\bibnamefont {Huang}}, \bibinfo {author} {\bibfnamefont
  {J.}~\bibnamefont {Cenker}}, \bibinfo {author} {\bibfnamefont {Y.-T.}\
  \bibnamefont {Cui}}, \bibinfo {author} {\bibfnamefont {J.-H.}\ \bibnamefont
  {Chu}}, \emph {et~al.},\ }\bibfield  {title} {\bibinfo {title} {Evidence for
  equilibrium exciton condensation in monolayer {WTe$_2$}},\ }\href@noop {}
  {\bibfield  {journal} {\bibinfo  {journal} {Nature Physics}\ }\textbf
  {\bibinfo {volume} {18}},\ \bibinfo {pages} {94} (\bibinfo {year}
  {2022})}\BibitemShut {NoStop}%
\bibitem [{\citenamefont {Maximenko}\ \emph {et~al.}(2022)\citenamefont
  {Maximenko}, \citenamefont {Chang}, \citenamefont {Chen}, \citenamefont
  {Hirsbrunner}, \citenamefont {Swiech}, \citenamefont {Hughes}, \citenamefont
  {Wagner},\ and\ \citenamefont {Madhavan}}]{maximenko2022}%
  \BibitemOpen
  \bibfield  {author} {\bibinfo {author} {\bibfnamefont {Y.}~\bibnamefont
  {Maximenko}}, \bibinfo {author} {\bibfnamefont {Y.}~\bibnamefont {Chang}},
  \bibinfo {author} {\bibfnamefont {G.}~\bibnamefont {Chen}}, \bibinfo {author}
  {\bibfnamefont {M.~R.}\ \bibnamefont {Hirsbrunner}}, \bibinfo {author}
  {\bibfnamefont {W.}~\bibnamefont {Swiech}}, \bibinfo {author} {\bibfnamefont
  {T.~L.}\ \bibnamefont {Hughes}}, \bibinfo {author} {\bibfnamefont {L.~K.}\
  \bibnamefont {Wagner}},\ and\ \bibinfo {author} {\bibfnamefont
  {V.}~\bibnamefont {Madhavan}},\ }\bibfield  {title} {\bibinfo {title}
  {Nanoscale studies of electric field effects on monolayer
  {1T$'$}-{WTe$_2$}},\ }\href@noop {} {\bibfield  {journal} {\bibinfo
  {journal} {npj Quantum Materials}\ }\textbf {\bibinfo {volume} {7}},\
  \bibinfo {pages} {29} (\bibinfo {year} {2022})}\BibitemShut {NoStop}%
\bibitem [{\citenamefont {Lee}\ \emph {et~al.}(2023)\citenamefont {Lee},
  \citenamefont {Kwon}, \citenamefont {Lee}, \citenamefont {Park},
  \citenamefont {Cha}, \citenamefont {Watanabe}, \citenamefont {Taniguchi},
  \citenamefont {Jo},\ and\ \citenamefont {Choi}}]{jlee2023}%
  \BibitemOpen
  \bibfield  {author} {\bibinfo {author} {\bibfnamefont {J.}~\bibnamefont
  {Lee}}, \bibinfo {author} {\bibfnamefont {J.}~\bibnamefont {Kwon}}, \bibinfo
  {author} {\bibfnamefont {E.}~\bibnamefont {Lee}}, \bibinfo {author}
  {\bibfnamefont {J.}~\bibnamefont {Park}}, \bibinfo {author} {\bibfnamefont
  {S.}~\bibnamefont {Cha}}, \bibinfo {author} {\bibfnamefont {K.}~\bibnamefont
  {Watanabe}}, \bibinfo {author} {\bibfnamefont {T.}~\bibnamefont {Taniguchi}},
  \bibinfo {author} {\bibfnamefont {M.-H.}\ \bibnamefont {Jo}},\ and\ \bibinfo
  {author} {\bibfnamefont {H.}~\bibnamefont {Choi}},\ }\bibfield  {title}
  {\bibinfo {title} {Spinful hinge states in the higher-order topological
  insulators {WTe$_2$}},\ }\href@noop {} {\bibfield  {journal} {\bibinfo
  {journal} {Nature Communications}\ }\textbf {\bibinfo {volume} {14}},\
  \bibinfo {pages} {1801} (\bibinfo {year} {2023})}\BibitemShut {NoStop}%
\bibitem [{\citenamefont {Watson}\ \emph {et~al.}(2025)\citenamefont {Watson},
  \citenamefont {Ripoll}, \citenamefont {Tong}, \citenamefont {Kumar},
  \citenamefont {Que}, \citenamefont {Chan}, \citenamefont {Lin}, \citenamefont
  {Mukherjee}, \citenamefont {Garnica}, \citenamefont {Edmonds} \emph
  {et~al.}}]{lwatson2025}%
  \BibitemOpen
  \bibfield  {author} {\bibinfo {author} {\bibfnamefont {L.}~\bibnamefont
  {Watson}}, \bibinfo {author} {\bibfnamefont {J.}~\bibnamefont {Ripoll}},
  \bibinfo {author} {\bibfnamefont {Z.}~\bibnamefont {Tong}}, \bibinfo {author}
  {\bibfnamefont {A.}~\bibnamefont {Kumar}}, \bibinfo {author} {\bibfnamefont
  {Y.}~\bibnamefont {Que}}, \bibinfo {author} {\bibfnamefont {Y.-H.}\
  \bibnamefont {Chan}}, \bibinfo {author} {\bibfnamefont {H.}~\bibnamefont
  {Lin}}, \bibinfo {author} {\bibfnamefont {S.}~\bibnamefont {Mukherjee}},
  \bibinfo {author} {\bibfnamefont {M.}~\bibnamefont {Garnica}}, \bibinfo
  {author} {\bibfnamefont {M.~T.}\ \bibnamefont {Edmonds}}, \emph {et~al.},\
  }\bibfield  {title} {\bibinfo {title} {Observation of the charge density wave
  excitonic order parameter in topological insulator monolayer {WTe$_2$}},\
  }\href@noop {} {\bibfield  {journal} {\bibinfo  {journal} {ACS nano}\
  }\textbf {\bibinfo {volume} {19}},\ \bibinfo {pages} {32374} (\bibinfo {year}
  {2025})}\BibitemShut {NoStop}%
\bibitem [{\citenamefont {Kresse}\ and\ \citenamefont
  {Furthmüller}(1996)}]{KRESSE199615}%
  \BibitemOpen
  \bibfield  {author} {\bibinfo {author} {\bibfnamefont {G.}~\bibnamefont
  {Kresse}}\ and\ \bibinfo {author} {\bibfnamefont {J.}~\bibnamefont
  {Furthmüller}},\ }\bibfield  {title} {\bibinfo {title} {Efficiency of
  ab-initio total energy calculations for metals and semiconductors using a
  plane-wave basis set},\ }\href
  {https://doi.org/https://doi.org/10.1016/0927-0256(96)00008-0} {\bibfield
  {journal} {\bibinfo  {journal} {Computational Materials Science}\ }\textbf
  {\bibinfo {volume} {6}},\ \bibinfo {pages} {15} (\bibinfo {year}
  {1996})}\BibitemShut {NoStop}%
\bibitem [{\citenamefont {Kresse}\ and\ \citenamefont
  {Furthm\"uller}(1996)}]{PhysRevB.54.11169}%
  \BibitemOpen
  \bibfield  {author} {\bibinfo {author} {\bibfnamefont {G.}~\bibnamefont
  {Kresse}}\ and\ \bibinfo {author} {\bibfnamefont {J.}~\bibnamefont
  {Furthm\"uller}},\ }\bibfield  {title} {\bibinfo {title} {Efficient iterative
  schemes for ab initio total-energy calculations using a plane-wave basis
  set},\ }\href {https://doi.org/10.1103/PhysRevB.54.11169} {\bibfield
  {journal} {\bibinfo  {journal} {Phys. Rev. B}\ }\textbf {\bibinfo {volume}
  {54}},\ \bibinfo {pages} {11169} (\bibinfo {year} {1996})}\BibitemShut
  {NoStop}%
\bibitem [{\citenamefont {Ozaki}(2003)}]{PhysRevB.67.155108}%
  \BibitemOpen
  \bibfield  {author} {\bibinfo {author} {\bibfnamefont {T.}~\bibnamefont
  {Ozaki}},\ }\bibfield  {title} {\bibinfo {title} {Variationally optimized
  atomic orbitals for large-scale electronic structures},\ }\href
  {https://doi.org/10.1103/PhysRevB.67.155108} {\bibfield  {journal} {\bibinfo
  {journal} {Phys. Rev. B}\ }\textbf {\bibinfo {volume} {67}},\ \bibinfo
  {pages} {155108} (\bibinfo {year} {2003})}\BibitemShut {NoStop}%
\bibitem [{\citenamefont {Ozaki}\ and\ \citenamefont
  {Kino}(2004)}]{PhysRevB.69.195113}%
  \BibitemOpen
  \bibfield  {author} {\bibinfo {author} {\bibfnamefont {T.}~\bibnamefont
  {Ozaki}}\ and\ \bibinfo {author} {\bibfnamefont {H.}~\bibnamefont {Kino}},\
  }\bibfield  {title} {\bibinfo {title} {Numerical atomic basis orbitals from h
  to kr},\ }\href {https://doi.org/10.1103/PhysRevB.69.195113} {\bibfield
  {journal} {\bibinfo  {journal} {Phys. Rev. B}\ }\textbf {\bibinfo {volume}
  {69}},\ \bibinfo {pages} {195113} (\bibinfo {year} {2004})}\BibitemShut
  {NoStop}%
\bibitem [{\citenamefont {Ozaki}\ and\ \citenamefont
  {Kino}(2005)}]{PhysRevB.72.045121}%
  \BibitemOpen
  \bibfield  {author} {\bibinfo {author} {\bibfnamefont {T.}~\bibnamefont
  {Ozaki}}\ and\ \bibinfo {author} {\bibfnamefont {H.}~\bibnamefont {Kino}},\
  }\bibfield  {title} {\bibinfo {title} {Efficient projector expansion for the
  ab initio lcao method},\ }\href {https://doi.org/10.1103/PhysRevB.72.045121}
  {\bibfield  {journal} {\bibinfo  {journal} {Phys. Rev. B}\ }\textbf {\bibinfo
  {volume} {72}},\ \bibinfo {pages} {045121} (\bibinfo {year}
  {2005})}\BibitemShut {NoStop}%
\bibitem [{\citenamefont {Perdew}\ \emph {et~al.}(1996)\citenamefont {Perdew},
  \citenamefont {Burke},\ and\ \citenamefont
  {Ernzerhof}}]{PhysRevLett.77.3865}%
  \BibitemOpen
  \bibfield  {author} {\bibinfo {author} {\bibfnamefont {J.~P.}\ \bibnamefont
  {Perdew}}, \bibinfo {author} {\bibfnamefont {K.}~\bibnamefont {Burke}},\ and\
  \bibinfo {author} {\bibfnamefont {M.}~\bibnamefont {Ernzerhof}},\ }\bibfield
  {title} {\bibinfo {title} {Generalized gradient approximation made simple},\
  }\href {https://doi.org/10.1103/PhysRevLett.77.3865} {\bibfield  {journal}
  {\bibinfo  {journal} {Phys. Rev. Lett.}\ }\textbf {\bibinfo {volume} {77}},\
  \bibinfo {pages} {3865} (\bibinfo {year} {1996})}\BibitemShut {NoStop}%
\bibitem [{\citenamefont {Grimme}\ \emph {et~al.}(2010)\citenamefont {Grimme},
  \citenamefont {Antony}, \citenamefont {Ehrlich},\ and\ \citenamefont
  {Krieg}}]{Grimme2010}%
  \BibitemOpen
  \bibfield  {author} {\bibinfo {author} {\bibfnamefont {S.}~\bibnamefont
  {Grimme}}, \bibinfo {author} {\bibfnamefont {J.}~\bibnamefont {Antony}},
  \bibinfo {author} {\bibfnamefont {S.}~\bibnamefont {Ehrlich}},\ and\ \bibinfo
  {author} {\bibfnamefont {H.}~\bibnamefont {Krieg}},\ }\bibfield  {title}
  {\bibinfo {title} {A consistent and accurateab initioparametrization of
  density functional dispersion correction (dft-d) for the 94 elements h-pu},\
  }\bibfield  {journal} {\bibinfo  {journal} {The Journal of Chemical Physics}\
  }\textbf {\bibinfo {volume} {132}},\ \href
  {https://doi.org/10.1063/1.3382344} {10.1063/1.3382344} (\bibinfo {year}
  {2010})\BibitemShut {NoStop}%
\bibitem [{\citenamefont {Bl{\"o}chl}(1994)}]{PhysRevB.50.17953}%
  \BibitemOpen
  \bibfield  {author} {\bibinfo {author} {\bibfnamefont {P.~E.}\ \bibnamefont
  {Bl{\"o}chl}},\ }\bibfield  {title} {\bibinfo {title} {Projector
  augmented-wave method},\ }\href {https://doi.org/10.1103/PhysRevB.50.17953}
  {\bibfield  {journal} {\bibinfo  {journal} {Phys. Rev. B}\ }\textbf {\bibinfo
  {volume} {50}},\ \bibinfo {pages} {17953} (\bibinfo {year}
  {1994})}\BibitemShut {NoStop}%
\bibitem [{\citenamefont {Morrison}\ \emph {et~al.}(1993)\citenamefont
  {Morrison}, \citenamefont {Bylander},\ and\ \citenamefont
  {Kleinman}}]{PhysRevB.47.6728}%
  \BibitemOpen
  \bibfield  {author} {\bibinfo {author} {\bibfnamefont {I.}~\bibnamefont
  {Morrison}}, \bibinfo {author} {\bibfnamefont {D.~M.}\ \bibnamefont
  {Bylander}},\ and\ \bibinfo {author} {\bibfnamefont {L.}~\bibnamefont
  {Kleinman}},\ }\bibfield  {title} {\bibinfo {title} {Nonlocal hermitian
  norm-conserving vanderbilt pseudopotential},\ }\href
  {https://doi.org/10.1103/PhysRevB.47.6728} {\bibfield  {journal} {\bibinfo
  {journal} {Phys. Rev. B}\ }\textbf {\bibinfo {volume} {47}},\ \bibinfo
  {pages} {6728} (\bibinfo {year} {1993})}\BibitemShut {NoStop}%
\bibitem [{\citenamefont {Monkhorst}\ and\ \citenamefont
  {Pack}(1976)}]{PhysRevB.13.5188}%
  \BibitemOpen
  \bibfield  {author} {\bibinfo {author} {\bibfnamefont {H.~J.}\ \bibnamefont
  {Monkhorst}}\ and\ \bibinfo {author} {\bibfnamefont {J.~D.}\ \bibnamefont
  {Pack}},\ }\bibfield  {title} {\bibinfo {title} {Special points for
  brillouin-zone integrations},\ }\href
  {https://doi.org/10.1103/PhysRevB.13.5188} {\bibfield  {journal} {\bibinfo
  {journal} {Phys. Rev. B}\ }\textbf {\bibinfo {volume} {13}},\ \bibinfo
  {pages} {5188} (\bibinfo {year} {1976})}\BibitemShut {NoStop}%
\bibitem [{\citenamefont {He}\ \emph {et~al.}(2014)\citenamefont {He},
  \citenamefont {Hummer},\ and\ \citenamefont {Franchini}}]{he2014}%
  \BibitemOpen
  \bibfield  {author} {\bibinfo {author} {\bibfnamefont {J.}~\bibnamefont
  {He}}, \bibinfo {author} {\bibfnamefont {K.}~\bibnamefont {Hummer}},\ and\
  \bibinfo {author} {\bibfnamefont {C.}~\bibnamefont {Franchini}},\ }\bibfield
  {title} {\bibinfo {title} {Stacking effects on the electronic and optical
  properties of bilayer transition metal dichalcogenides {MoS$_2$}, {MoSe$_2$},
  {WS$_2$}, and {WSe$_2$}},\ }\href@noop {} {\bibfield  {journal} {\bibinfo
  {journal} {Phys. Rev. B}\ }\textbf {\bibinfo {volume} {89}},\ \bibinfo
  {pages} {075409} (\bibinfo {year} {2014})}\BibitemShut {NoStop}%
\bibitem [{\citenamefont {Zhai}\ \emph {et~al.}(2024)\citenamefont {Zhai},
  \citenamefont {Li}, \citenamefont {Wang}, \citenamefont {Zhai}, \citenamefont
  {Yao}, \citenamefont {Li}, \citenamefont {Wang}, \citenamefont {Yang},
  \citenamefont {Chi}, \citenamefont {Liang} \emph {et~al.}}]{zhai2024review}%
  \BibitemOpen
  \bibfield  {author} {\bibinfo {author} {\bibfnamefont {W.}~\bibnamefont
  {Zhai}}, \bibinfo {author} {\bibfnamefont {Z.}~\bibnamefont {Li}}, \bibinfo
  {author} {\bibfnamefont {Y.}~\bibnamefont {Wang}}, \bibinfo {author}
  {\bibfnamefont {L.}~\bibnamefont {Zhai}}, \bibinfo {author} {\bibfnamefont
  {Y.}~\bibnamefont {Yao}}, \bibinfo {author} {\bibfnamefont {S.}~\bibnamefont
  {Li}}, \bibinfo {author} {\bibfnamefont {L.}~\bibnamefont {Wang}}, \bibinfo
  {author} {\bibfnamefont {H.}~\bibnamefont {Yang}}, \bibinfo {author}
  {\bibfnamefont {B.}~\bibnamefont {Chi}}, \bibinfo {author} {\bibfnamefont
  {J.}~\bibnamefont {Liang}}, \emph {et~al.},\ }\bibfield  {title} {\bibinfo
  {title} {Phase engineering of nanomaterials: transition metal
  dichalcogenides},\ }\href@noop {} {\bibfield  {journal} {\bibinfo  {journal}
  {Chemical Reviews}\ }\textbf {\bibinfo {volume} {124}},\ \bibinfo {pages}
  {4479} (\bibinfo {year} {2024})}\BibitemShut {NoStop}%
\bibitem [{\citenamefont {Chaikin}\ and\ \citenamefont
  {Lubensky}(1995)}]{chaikinbooksec6}%
  \BibitemOpen
  \bibfield  {author} {\bibinfo {author} {\bibfnamefont {P.~M.}\ \bibnamefont
  {Chaikin}}\ and\ \bibinfo {author} {\bibfnamefont {T.~C.}\ \bibnamefont
  {Lubensky}},\ }\href@noop {} {\emph {\bibinfo {title} {Principles of
  Condensed Matter Physics, {{\rm Sec.6.4}}}}}\ (\bibinfo  {publisher}
  {Cambridge University Press},\ \bibinfo {address} {Cambridge},\ \bibinfo
  {year} {1995})\BibitemShut {NoStop}%
\bibitem [{\citenamefont {Steiner}\ \emph {et~al.}(2016)\citenamefont
  {Steiner}, \citenamefont {Khmelevskyi}, \citenamefont {Marsmann},\ and\
  \citenamefont {Kresse}}]{PhysRevB.93.224425}%
  \BibitemOpen
  \bibfield  {author} {\bibinfo {author} {\bibfnamefont {S.}~\bibnamefont
  {Steiner}}, \bibinfo {author} {\bibfnamefont {S.}~\bibnamefont
  {Khmelevskyi}}, \bibinfo {author} {\bibfnamefont {M.}~\bibnamefont
  {Marsmann}},\ and\ \bibinfo {author} {\bibfnamefont {G.}~\bibnamefont
  {Kresse}},\ }\bibfield  {title} {\bibinfo {title} {Calculation of the
  magnetic anisotropy with projected-augmented-wave methodology and the case
  study of disordered {${\mathrm{Fe}}_{1\ensuremath{-}x}{\mathrm{Co}}_{x}$}
  alloys},\ }\href {https://doi.org/10.1103/PhysRevB.93.224425} {\bibfield
  {journal} {\bibinfo  {journal} {Phys. Rev. B}\ }\textbf {\bibinfo {volume}
  {93}},\ \bibinfo {pages} {224425} (\bibinfo {year} {2016})}\BibitemShut
  {NoStop}%
\bibitem [{\citenamefont {San-Jose}\ and\ \citenamefont
  {Prada}(2013)}]{sanjose2013}%
  \BibitemOpen
  \bibfield  {author} {\bibinfo {author} {\bibfnamefont {P.}~\bibnamefont
  {San-Jose}}\ and\ \bibinfo {author} {\bibfnamefont {E.}~\bibnamefont
  {Prada}},\ }\bibfield  {title} {\bibinfo {title} {Helical networks in twisted
  bilayer graphene under interlayer bias},\ }\href
  {https://doi.org/10.1103/PhysRevB.88.121408} {\bibfield  {journal} {\bibinfo
  {journal} {Phys. Rev. B}\ }\textbf {\bibinfo {volume} {88}},\ \bibinfo
  {pages} {121408} (\bibinfo {year} {2013})}\BibitemShut {NoStop}%
\bibitem [{\citenamefont {Nam}\ and\ \citenamefont
  {Koshino}(2017{\natexlab{b}})}]{nam2017lattice}%
  \BibitemOpen
  \bibfield  {author} {\bibinfo {author} {\bibfnamefont {N.~N.~T.}\
  \bibnamefont {Nam}}\ and\ \bibinfo {author} {\bibfnamefont {M.}~\bibnamefont
  {Koshino}},\ }\bibfield  {title} {\bibinfo {title} {Lattice relaxation and
  energy band modulation in twisted bilayer graphene},\ }\href@noop {}
  {\bibfield  {journal} {\bibinfo  {journal} {Phys. Rev. B}\ }\textbf {\bibinfo
  {volume} {96}},\ \bibinfo {pages} {075311} (\bibinfo {year}
  {2017}{\natexlab{b}})},\ \bibinfo {note} {errata ibid {\bf 101}, 099901
  (2020)}\BibitemShut {NoStop}%
\bibitem [{\citenamefont {Efimkin}\ and\ \citenamefont
  {MacDonald}(2018)}]{efimkin2018}%
  \BibitemOpen
  \bibfield  {author} {\bibinfo {author} {\bibfnamefont {D.~K.}\ \bibnamefont
  {Efimkin}}\ and\ \bibinfo {author} {\bibfnamefont {A.~H.}\ \bibnamefont
  {MacDonald}},\ }\bibfield  {title} {\bibinfo {title} {Helical network model
  for twisted bilayer graphene},\ }\href
  {https://doi.org/10.1103/PhysRevB.98.035404} {\bibfield  {journal} {\bibinfo
  {journal} {Phys. Rev. B}\ }\textbf {\bibinfo {volume} {98}},\ \bibinfo
  {pages} {035404} (\bibinfo {year} {2018})}\BibitemShut {NoStop}%
\bibitem [{\citenamefont {Walet}\ and\ \citenamefont
  {Guinea}(2019)}]{walet2019emergence}%
  \BibitemOpen
  \bibfield  {author} {\bibinfo {author} {\bibfnamefont {N.~R.}\ \bibnamefont
  {Walet}}\ and\ \bibinfo {author} {\bibfnamefont {F.}~\bibnamefont {Guinea}},\
  }\bibfield  {title} {\bibinfo {title} {The emergence of one-dimensional
  channels in marginal-angle twisted bilayer graphene},\ }\href@noop {}
  {\bibfield  {journal} {\bibinfo  {journal} {2D Materials}\ }\textbf {\bibinfo
  {volume} {7}},\ \bibinfo {pages} {015023} (\bibinfo {year}
  {2019})}\BibitemShut {NoStop}%
\bibitem [{\citenamefont {Hsu}\ \emph {et~al.}(2023)\citenamefont {Hsu},
  \citenamefont {Loss},\ and\ \citenamefont {Klinovaja}}]{hsu2023}%
  \BibitemOpen
  \bibfield  {author} {\bibinfo {author} {\bibfnamefont {C.-H.}\ \bibnamefont
  {Hsu}}, \bibinfo {author} {\bibfnamefont {D.}~\bibnamefont {Loss}},\ and\
  \bibinfo {author} {\bibfnamefont {J.}~\bibnamefont {Klinovaja}},\ }\bibfield
  {title} {\bibinfo {title} {General scattering and electronic states in a
  quantum-wire network of moir{\'e} systems},\ }\href@noop {} {\bibfield
  {journal} {\bibinfo  {journal} {Physical Review B}\ }\textbf {\bibinfo
  {volume} {108}},\ \bibinfo {pages} {L121409} (\bibinfo {year}
  {2023})}\BibitemShut {NoStop}%
\bibitem [{\citenamefont {Nakatsuji}\ \emph {et~al.}(2023)\citenamefont
  {Nakatsuji}, \citenamefont {Kawakami},\ and\ \citenamefont
  {Koshino}}]{nakatsuji2023}%
  \BibitemOpen
  \bibfield  {author} {\bibinfo {author} {\bibfnamefont {N.}~\bibnamefont
  {Nakatsuji}}, \bibinfo {author} {\bibfnamefont {T.}~\bibnamefont
  {Kawakami}},\ and\ \bibinfo {author} {\bibfnamefont {M.}~\bibnamefont
  {Koshino}},\ }\bibfield  {title} {\bibinfo {title} {Multiscale lattice
  relaxation in general twisted trilayer graphenes},\ }\href
  {https://doi.org/10.1103/PhysRevX.13.041007} {\bibfield  {journal} {\bibinfo
  {journal} {Phys. Rev. X}\ }\textbf {\bibinfo {volume} {13}},\ \bibinfo
  {pages} {041007} (\bibinfo {year} {2023})}\BibitemShut {NoStop}%
\bibitem [{\citenamefont {Wang}\ and\ \citenamefont {Hsu}(2024)}]{hcwang2024}%
  \BibitemOpen
  \bibfield  {author} {\bibinfo {author} {\bibfnamefont {H.-C.}\ \bibnamefont
  {Wang}}\ and\ \bibinfo {author} {\bibfnamefont {C.-H.}\ \bibnamefont {Hsu}},\
  }\bibfield  {title} {\bibinfo {title} {Electrically tunable correlated domain
  wall network in twisted bilayer graphene},\ }\href@noop {} {\bibfield
  {journal} {\bibinfo  {journal} {2D Materials}\ }\textbf {\bibinfo {volume}
  {11}},\ \bibinfo {pages} {035007} (\bibinfo {year} {2024})}\BibitemShut
  {NoStop}%
\bibitem [{\citenamefont {Chang}\ \emph {et~al.}(2025)\citenamefont {Chang},
  \citenamefont {Saito},\ and\ \citenamefont {Hsu}}]{yychang2025}%
  \BibitemOpen
  \bibfield  {author} {\bibinfo {author} {\bibfnamefont {Y.-Y.}\ \bibnamefont
  {Chang}}, \bibinfo {author} {\bibfnamefont {K.}~\bibnamefont {Saito}},\ and\
  \bibinfo {author} {\bibfnamefont {C.-H.}\ \bibnamefont {Hsu}},\ }\bibfield
  {title} {\bibinfo {title} {Quasi-two-dimensional spin helix and
  magnon-induced singularity in twisted bilayer graphene},\ }\href@noop {}
  {\bibfield  {journal} {\bibinfo  {journal} {Physical Review Research}\
  }\textbf {\bibinfo {volume} {7}},\ \bibinfo {pages} {L032066} (\bibinfo
  {year} {2025})}\BibitemShut {NoStop}%
\bibitem [{\citenamefont {Fujimoto}\ \emph {et~al.}(2022)\citenamefont
  {Fujimoto}, \citenamefont {Kawakami},\ and\ \citenamefont
  {Koshino}}]{fujimoto2022}%
  \BibitemOpen
  \bibfield  {author} {\bibinfo {author} {\bibfnamefont {M.}~\bibnamefont
  {Fujimoto}}, \bibinfo {author} {\bibfnamefont {T.}~\bibnamefont {Kawakami}},\
  and\ \bibinfo {author} {\bibfnamefont {M.}~\bibnamefont {Koshino}},\
  }\bibfield  {title} {\bibinfo {title} {Perfect one-dimensional interface
  states in a twisted stack of three-dimensional topological insulators},\
  }\href@noop {} {\bibfield  {journal} {\bibinfo  {journal} {Physical Review
  Research}\ }\textbf {\bibinfo {volume} {4}},\ \bibinfo {pages} {043209}
  (\bibinfo {year} {2022})}\BibitemShut {NoStop}%
\bibitem [{\citenamefont {C{\u{a}}lug{\u{a}}ru}\ \emph
  {et~al.}(2025)\citenamefont {C{\u{a}}lug{\u{a}}ru}, \citenamefont {Jiang},
  \citenamefont {Hu}, \citenamefont {Pi}, \citenamefont {Yu}, \citenamefont
  {Vergniory}, \citenamefont {Shan}, \citenamefont {Felser}, \citenamefont
  {Schoop}, \citenamefont {Efetov} \emph {et~al.}}]{cualuguaru2025}%
  \BibitemOpen
  \bibfield  {author} {\bibinfo {author} {\bibfnamefont {D.}~\bibnamefont
  {C{\u{a}}lug{\u{a}}ru}}, \bibinfo {author} {\bibfnamefont {Y.}~\bibnamefont
  {Jiang}}, \bibinfo {author} {\bibfnamefont {H.}~\bibnamefont {Hu}}, \bibinfo
  {author} {\bibfnamefont {H.}~\bibnamefont {Pi}}, \bibinfo {author}
  {\bibfnamefont {J.}~\bibnamefont {Yu}}, \bibinfo {author} {\bibfnamefont
  {M.~G.}\ \bibnamefont {Vergniory}}, \bibinfo {author} {\bibfnamefont
  {J.}~\bibnamefont {Shan}}, \bibinfo {author} {\bibfnamefont {C.}~\bibnamefont
  {Felser}}, \bibinfo {author} {\bibfnamefont {L.~M.}\ \bibnamefont {Schoop}},
  \bibinfo {author} {\bibfnamefont {D.~K.}\ \bibnamefont {Efetov}}, \emph
  {et~al.},\ }\bibfield  {title} {\bibinfo {title} {Moir{\'e} materials based
  on m-point twisting},\ }\href@noop {} {\bibfield  {journal} {\bibinfo
  {journal} {Nature}\ }\textbf {\bibinfo {volume} {643}},\ \bibinfo {pages}
  {376} (\bibinfo {year} {2025})}\BibitemShut {NoStop}%
\bibitem [{\citenamefont {Haldane}(1981)}]{haldane1981}%
  \BibitemOpen
  \bibfield  {author} {\bibinfo {author} {\bibfnamefont {F.~D.~M.}\
  \bibnamefont {Haldane}},\ }\bibfield  {title} {\bibinfo {title} {{`Luttinger
  liquid theory'} of one-dimensional quantum fluids. {I.} properties of the
  luttinger model and their extension to the general 1d interacting spinless
  fermi gas},\ }\href@noop {} {\bibfield  {journal} {\bibinfo  {journal}
  {Journal of Physics C: Solid State Physics}\ }\textbf {\bibinfo {volume}
  {14}},\ \bibinfo {pages} {2585} (\bibinfo {year} {1981})}\BibitemShut
  {NoStop}%
\bibitem [{\citenamefont {Egger}\ and\ \citenamefont
  {Gogolin}(1997)}]{egger1997}%
  \BibitemOpen
  \bibfield  {author} {\bibinfo {author} {\bibfnamefont {R.}~\bibnamefont
  {Egger}}\ and\ \bibinfo {author} {\bibfnamefont {A.~O.}\ \bibnamefont
  {Gogolin}},\ }\bibfield  {title} {\bibinfo {title} {Effective low-energy
  theory for correlated carbon nanotubes},\ }\href@noop {} {\bibfield
  {journal} {\bibinfo  {journal} {Physical review letters}\ }\textbf {\bibinfo
  {volume} {79}},\ \bibinfo {pages} {5082} (\bibinfo {year}
  {1997})}\BibitemShut {NoStop}%
\bibitem [{\citenamefont {Bockrath}\ \emph {et~al.}(1999)\citenamefont
  {Bockrath}, \citenamefont {Cobden}, \citenamefont {Lu}, \citenamefont
  {Rinzler}, \citenamefont {Smalley}, \citenamefont {Balents},\ and\
  \citenamefont {McEuen}}]{bockrath1999}%
  \BibitemOpen
  \bibfield  {author} {\bibinfo {author} {\bibfnamefont {M.}~\bibnamefont
  {Bockrath}}, \bibinfo {author} {\bibfnamefont {D.~H.}\ \bibnamefont
  {Cobden}}, \bibinfo {author} {\bibfnamefont {J.}~\bibnamefont {Lu}}, \bibinfo
  {author} {\bibfnamefont {A.~G.}\ \bibnamefont {Rinzler}}, \bibinfo {author}
  {\bibfnamefont {R.~E.}\ \bibnamefont {Smalley}}, \bibinfo {author}
  {\bibfnamefont {L.}~\bibnamefont {Balents}},\ and\ \bibinfo {author}
  {\bibfnamefont {P.~L.}\ \bibnamefont {McEuen}},\ }\bibfield  {title}
  {\bibinfo {title} {Luttinger-liquid behaviour in carbon nanotubes},\
  }\href@noop {} {\bibfield  {journal} {\bibinfo  {journal} {Nature}\ }\textbf
  {\bibinfo {volume} {397}},\ \bibinfo {pages} {598} (\bibinfo {year}
  {1999})}\BibitemShut {NoStop}%
\bibitem [{\citenamefont {Yao}\ \emph {et~al.}(1999)\citenamefont {Yao},
  \citenamefont {Postma}, \citenamefont {Balents},\ and\ \citenamefont
  {Dekker}}]{yao1999}%
  \BibitemOpen
  \bibfield  {author} {\bibinfo {author} {\bibfnamefont {Z.}~\bibnamefont
  {Yao}}, \bibinfo {author} {\bibfnamefont {H.~W.~C.}\ \bibnamefont {Postma}},
  \bibinfo {author} {\bibfnamefont {L.}~\bibnamefont {Balents}},\ and\ \bibinfo
  {author} {\bibfnamefont {C.}~\bibnamefont {Dekker}},\ }\bibfield  {title}
  {\bibinfo {title} {Carbon nanotube intramolecular junctions},\ }\href@noop {}
  {\bibfield  {journal} {\bibinfo  {journal} {Nature}\ }\textbf {\bibinfo
  {volume} {402}},\ \bibinfo {pages} {273} (\bibinfo {year}
  {1999})}\BibitemShut {NoStop}%
\bibitem [{\citenamefont {Emery}\ \emph {et~al.}(2000)\citenamefont {Emery},
  \citenamefont {Fradkin}, \citenamefont {Kivelson},\ and\ \citenamefont
  {Lubensky}}]{emery2000}%
  \BibitemOpen
  \bibfield  {author} {\bibinfo {author} {\bibfnamefont {V.~J.}\ \bibnamefont
  {Emery}}, \bibinfo {author} {\bibfnamefont {E.}~\bibnamefont {Fradkin}},
  \bibinfo {author} {\bibfnamefont {S.~A.}\ \bibnamefont {Kivelson}},\ and\
  \bibinfo {author} {\bibfnamefont {T.~C.}\ \bibnamefont {Lubensky}},\
  }\bibfield  {title} {\bibinfo {title} {Quantum theory of the smectic metal
  state in stripe phases},\ }\href@noop {} {\bibfield  {journal} {\bibinfo
  {journal} {Physical review letters}\ }\textbf {\bibinfo {volume} {85}},\
  \bibinfo {pages} {2160} (\bibinfo {year} {2000})}\BibitemShut {NoStop}%
\bibitem [{\citenamefont {Vishwanath}\ and\ \citenamefont
  {Carpentier}(2001)}]{vishwanath2001}%
  \BibitemOpen
  \bibfield  {author} {\bibinfo {author} {\bibfnamefont {A.}~\bibnamefont
  {Vishwanath}}\ and\ \bibinfo {author} {\bibfnamefont {D.}~\bibnamefont
  {Carpentier}},\ }\bibfield  {title} {\bibinfo {title} {Two-dimensional
  anisotropic non-fermi-liquid phase of coupled luttinger liquids},\
  }\href@noop {} {\bibfield  {journal} {\bibinfo  {journal} {Physical review
  letters}\ }\textbf {\bibinfo {volume} {86}},\ \bibinfo {pages} {676}
  (\bibinfo {year} {2001})}\BibitemShut {NoStop}%
\bibitem [{\citenamefont {Yang}\ \emph {et~al.}(2026)\citenamefont {Yang},
  \citenamefont {Zhang}, \citenamefont {Sato}, \citenamefont {Nishimura},
  \citenamefont {Nakashima}, \citenamefont {Chen}, \citenamefont {Aso},
  \citenamefont {Onodera}, \citenamefont {Moriya}, \citenamefont {Watanabe}
  \emph {et~al.}}]{yang2026}%
  \BibitemOpen
  \bibfield  {author} {\bibinfo {author} {\bibfnamefont {X.}~\bibnamefont
  {Yang}}, \bibinfo {author} {\bibfnamefont {Y.}~\bibnamefont {Zhang}},
  \bibinfo {author} {\bibfnamefont {K.}~\bibnamefont {Sato}}, \bibinfo {author}
  {\bibfnamefont {A.}~\bibnamefont {Nishimura}}, \bibinfo {author}
  {\bibfnamefont {M.}~\bibnamefont {Nakashima}}, \bibinfo {author}
  {\bibfnamefont {L.}~\bibnamefont {Chen}}, \bibinfo {author} {\bibfnamefont
  {K.}~\bibnamefont {Aso}}, \bibinfo {author} {\bibfnamefont {M.}~\bibnamefont
  {Onodera}}, \bibinfo {author} {\bibfnamefont {R.}~\bibnamefont {Moriya}},
  \bibinfo {author} {\bibfnamefont {K.}~\bibnamefont {Watanabe}}, \emph
  {et~al.},\ }\bibfield  {title} {\bibinfo {title} {Racemic van der waals
  assembly of planar-chiral 2d materials for intrinsic one-dimensional
  moir{\'e} superlattices},\ }\href@noop {} {\bibfield  {journal} {\bibinfo
  {journal} {ACS Nano}\ } (\bibinfo {year} {2026})}\BibitemShut {NoStop}%
\end{thebibliography}%
\end{document}